\documentclass[twocolumn]{aastex631}

\usepackage{amsmath,mathtools,mathrsfs,amssymb}
\usepackage{hyperref}
\usepackage{verbatim}
\usepackage{xspace}
\usepackage{CJK}

 \received{ ***, ***, 2021}
 \revised{*** ***, ***}
 \accepted{***}

\shorttitle{
general relativistic, multi-wavelength radiative transfer}
\shortauthors{Kawashima et al.}
\graphicspath{{./}{figures/}}
\begin{document}
\begin{CJK*}{UTF8}{gbsn}

\title{
\texttt{RAIKOU} (来光): A General Relativistic, Multi-wavelength Radiative Transfer Code
}

        \correspondingauthor{Tomohisa Kawashima}
        \email{kawshm@icrr.u-tokyo.ac.jp}
        
        \author[0000-0001-8527-0496]{Tomohisa Kawashima}
        \affiliation{Institute for Cosmic Ray Research, The University of Tokyo, 5-1-5 Kashiwanoha, Kashiwa, Chiba 277-8582, Japan}

\author[0000-0002-2309-3639]{Ken Ohsuga}
\affiliation{Center for Computational Sciences, University of Tsukuba, Ten-nodai, 1-1-1 Tsukuba, Ibaraki 305-8577, Japan}

\author[0000-0003-0114-5378]{Hiroyuki R. Takahashi}
\affiliation{
Department of Natural Sciences, Faculty of Arts and Sciences, 
Komazawa University, 1-23-1 Komazawa, Setagaya, Tokyo 154-8525, Japan}

\begin{abstract}
We present a general relativistic, ray-tracing radiative transfer code \texttt{RAIKOU} (来光) for  multi-wavlength studies of spectra and images including the black hole shadows around Kerr black holes.
Important radiative processes in hot plasmas around black holes, i.e., (cyclo-)synchrotron, bremsstrahlung emission/absorption and Compton/inverse-Compton scattering, are incorporated.
The Maxwell-J\"uttner and single/broken power-law electron distribution functions are implemented to calculate the radiative transfer via both of the thermal and the nonthermal electrons. 
Two calculation algorithms
are implemented for studies of both the images and broadband spectra.
An observer-to-emitter algorithm, which inversely solve the radiative transfer equation from the observer screen to  emitting plasmas, 
is suitable for efficient calculations of the
images, e.g., the black hole shadows, and spectra without the Compton effects.
On the other hand, 
an emitter-to-observer algorithm, by which photons are transported with a Monte-Carlo method including the effects of Compton/inverse-Compton scatterings, 
enables us to compute multi-wavelength spectra with their energy bands broadly ranging from radio to very-high-energy gamma-ray. 
The code is generally applicable to accretion flows around Kerr black holes with relativistic jets and winds/coronae with various mass accretion rate (i.e., radiatively inefficient accretion flows, super-Eddington accretion flows, and others). 
We demonstrate an application of the code to a radiatively innefficent accretion flow onto a supermassive black hole.
\end{abstract}



\section{Introduction} \label{sec:intro}


Astrophysical black holes (BHs) are thought to be powered by the gravitational energy release via the accretion flows. The energy release in the accretion flows will result in the emission of an enormous amount of electromagnetic radiation \citep[e.g.,][]{Shakura_Sunyaev_1973, Abramowicz_etal_1988, Narayan_Yi_1994} and the ejection of powerful outflows (i.e., relativistic jets and mildly relativistic winds), which are theoretically suggested to be accelerated 
by magnetic processes \citep{Blandford_Znajek_1977, Blandford_Payne_1982, Lynden-Bell_1996}, 
and radiative force \citep{Sikora_1981}.
In order to study BH spacetimes and physics of these plasma dynamics,
computations of general relativistic radiative transfer (GRRT) will be a powerful tool,
because GRRT calculations of images and spectra will enable us to directly compare the theoretical models with the observed data.

One of the most important observable features of  BHs is their shadows or silhouettes \citep{Bardeen_1973, Luminet_1979, Falcke_etal_2000, Takahashi_2004, Broderick_Loeb_2006, Noble_etal_2007, Dexter_etal_2012, Moscibrodzka_etal_2016, Chael_etal_2018}, 
because their images are formed as a consequence of the significant light-bending effect due to the extremely strong gravity 
of the BHs.
Recently, the Event Horizon Telescope (EHT) detected the image of BH shadow in the elliptical galaxy M87* \citep{EHTC2019_1,EHTC2019_2, EHTC2019_3, EHTC2019_4, EHTC2019_5, EHTC2019_6} and it provided us a strong and important evidence of the supermassive BHs (SMBHs), whose formation scenario is still controversial \citep{Rees_1984review,Volonteri_2012Science}. 

The diameter of the observed photon ring, which is a bright ring image formed by the photons passing through near the unstable photon orbit around the BHs, is $\sim 40 \mu$as \citep{EHTC2019_4}.
The BH mass can be estimated to be $6.5 \times 10^{9}M_{\odot}$ \citep{EHTC2019_6} 
because the diameter of the photon ring is roughly proportional to the BH mass, i.e., ${\sim} 10 r_{\rm g}$, where $r_{\rm g}$ is the gravitational radius defined as $r_{\rm g} \equiv GM/c^2$, $G$ is the gravitaional constant, $M$ is the BH mass, and $c$ is the speed of light.
However, the constrain of the BH spin is still grand challenge because the diameter of the photon ring depends weakly on the BH spin, in practice the diameter changes within $\pm 5 \%$ \citep[e.g.,][]{Psaltis_etal_2015} and it would be difficult to be identified using the current EHT.
In addition, the dynamics of accretion flows and launching mechanism of jets are still enigmatic.
It would be, therefore, important to explore these problem using the additional feature, e.g., polarization \citep{Broderick_Loeb_2006, Shcherbakov_etal_2012, Dexter_2016_GRTRANS, Moscibrodzka_etal_2017, Gold_etal_2017, Tsunetoe_etal_2020, Tsunetoe_etal_2021, EHTC_2021b} and/or multi-wavelength spectra \citep{Ohsuga_etal_2005_SED, Moscibrodzka_etal_2009,Moscibrodzka_etal_2016, Chael_etal_2016, Chael_etal_2018}, where the latter  approach is the motivation of the code development in this paper.

The multi-wevelength radiative transfer calculations, whose energy band ranging from radio to very-high-energy (VHE) gamma-ray, potentially constrain the various parameters with respect to BH spacetimes and theoretical models of the accretion flows and jets via the simultaneous comparison between theoretical and observed radiative flux in the various energy bands.
 In order to calculate the spectral energy distributions (SEDs) of accretion flows and jets around BHs above the X-ray band, it is necessary to incorporate the effects of the Compton and inverse-Compton scattering \citep[e.g.,][]{Rybicki_Lightman_1979}.
For the calculations of the Comptonized SEDs, Monte-Carlo radiative transfer \citep[e.g.,][]{Pozdnyakov_etal_1977,Pozdnyakov_etal_1983,Gorecki_Wilczewski_1984,Canfield_etal_1987,Dolence_etal_2009} are suitable for accurate computations of the Compton/inverse-Compton scattering without introducing approximations (e.g., the Kompaneets approximation, which is applicable when the radiative field are isotropic in the fluid-rest frame \citep{Kompaneets_1957}), although the Monte-Calro calculations require high computational costs.

Despite the most of the previous works on GRRT calculations focused on the thermal electrons, the observations of BHs strongly demonstrates the evidences of the acceleration of electrons. 
For example, the VHE gamma-ray observation of M87 by MAGIC Telescope indicates the generation of nonthermal electrons at the distance ${\sim} 100r_{\rm g}$ from the BH in M87 \citep{MAGIC_2020_M87}. 
Recent simultaneous multi-wavelength observation campaign for EHT 2017 found that the gamma-ray emission is not generated at the same place with the ring-emission in the vicinity of the BH  if the isotropic distribution of the nonthermal electrons is assumed \citep{EHT_MWL_2021}. 
On the other hand, IC 310 shows the sub-day scale, rapidly varying VHE gamma-ray emission, which is interpreted as the emission orginated from the nonthermal electrons accelerated from the gap electric field in the vinicity of the BH \citep{MAGIC_2014Science_IC310}.
In the lower frequency bands, the synchrotron emission via nonthermal electrons can reproduce the radio emission at $\nu < 100$ GHz \citep{Yuan_etal_2003, Davelaar_etal_2018}
and it will be important to consider the images and spectra at 230 GHz, infrared , and X-ray bands in Sgr A* \citep[e.g.,][]{Ball_etal_2016, Chael_etal_2017, Mao_etal_2017, Chatterjee_2020_arxiv, Scepi_etal_2021_arxiv}. 

Calculations of the Comptonized SEDs are important for the general studies on BHs, whose mass ranging from stellar-mass BHs to SMBHs, with the mass accretion rate ranging from sub-Eddington  to super-Eddington accretion-rate.
For example, in the super-Eddington accretion flows, the Compton/inverse-Compton scattering in jets and winds drastically modifies the SEDs \citep{Kawashima_etal_2012, Kitaki_etal_2017, Narayan_etal_2017,Curd_Narayan_2019, Kawanaka_Mineshige_2021}, which will be important to explain the SEDs of ultraluminous X-ray sources (ULXs) and tidal disruption events (TDEs).
The origin of the hard X-ray photons from the BH accretion flows with marginally Eddington accretion rate, which is an open question and is proposed to be Comptonized in hot coronae formed above the standard disks or inside the truncated accretion disks, can be also approached \citep{Kawanaka_etal_2008, Davis_etal_2009, Schnittman_etal_2013}.

In this paper, we describe our newly developed general relativistic, multi-wavelength radiative transfer code \texttt{RAIKOU} (来光), by which the multi-wavelength SEDs from radio to VHE gamma-ray bands and images of the BH shadow, accretion flows and outflows can be calculated.
The electron distribution functions of the thermal electrons with the Maxwell J\"uttner distribution and the nonthermal electrons with single/broken power-law distributions are incorporated.
The VHE gamma-ray emission above TeV energy bands can be computed because of the implement of the  inverse-Compton scattering processes via nonthermal electrons. 
The code is applicable to calculate the images and spectra of accreting BHs by post-processing general relativisitic (radiation-)magnetohydrodynamic simulations, hereafter GR(R)MHD,  as well as by using semi-analytic models. 
In \S \ref{sec:method}---\ref{sec:method_E2O}, the formalism and numerical method are described.
In \S \ref{sec:test}, numerical tests are shown.
In \S \ref{sec:App}, we demonstrate an application of the code to calculations of multi-wavelength images and spectra of a radiatively inefficient accretion flow onto Sgr A* based on GRMHD simulations.
We devote \S \ref{sec:summary} as the summary and discussion.

\section{Overview of Method} \label{sec:method}
We describe the formalism of equations and algorithms implemented to  \texttt{RAIKOU}.
For the various purposes of the radiative transfer calculations, 
two types of calculation algorithm are employed.
One is the observer-to-emitter algorithm, by which the radiative transfer equations are integrated from the observer-screen to the emitting plasma. This method enables us to efficiently compute the accurate images and spectra in the limit of plasmas being optically thin against the inverse-Compton scattering.
The other is the emitter-to-observer algorithm, by which the superphotons are transported from the emitting plasma to the observer with a Monte-Carlo method.
By using the emitter-to-observer algorithm, one can compute broadband spectra from radio to VHE gamma-rays which is significantly affected by Compton/inverse-Compoton scattering.

The same formalism and algorithm of the ray-tracing part, i.e., the integration module of the geodesic equations of photons, are used for both of the oberver-to-emitter and the emitter-to-observer algorithm. 
The details of the radiative transfer part including various radiative processes, i.e., the emission, abosorption, and scattering, are presented in the subsequent subsections.

\section{Geodesics}

We integrate the Hamilton's canonical equations of motion, which describe time evolution of the position and momentum of photons $r$, $\theta$, $\varphi$, $p_{r}$ and $p_{\theta}$ in Boyer-Lindquist (BL) coordinates, to solve the geodesic equations for photons \citep{Schnittman_Krolik_2013}.
 We do not solve the time evolution of conserved variable of photons, i.e., time and azimuthal components of photons $p_t$ and $p_\varphi$ during the calculation of the geodesics, while these are updated when the photons are scattered via the Compton process.

The metric of the BL coordinate is written as follows:
\begin{equation}
g_{\mu \nu} = 
\begin{pmatrix}
-\alpha^2 + \omega^2 \varpi^2 & 0 & 0 & -\omega \varpi^2 \\
0 & \rho^2/\Delta & 0 & 0 \\
0 & 0 & \rho^2 & 0 \\
-\omega \varpi^2 & 0 & 0 & \varpi^2
\end{pmatrix},
\end{equation}
where, in the geometrical unit $G=c=1$, 
\begin{eqnarray}
   \rho^2 &=& r^2 + a^2 \cos ^2 \theta, \\
   \Delta &=& r^2 - 2Mr + a^2, \\
   \alpha^2 &=& \frac{\rho^2 \Delta}{\rho^2 \Delta + 2Mr\left(a^2 + r^2\right)}, \\         
   \omega   &=& \frac{2Mar}{\rho^2 \Delta + 2Mr\left(a^2 + r^2\right)}, \\   
   \varpi^2 &=& \frac{\rho^2 \Delta + 2Mr\left(a^2 + r^2\right)}{\rho^2} \sin^2 \theta.
\end{eqnarray}

The Hamiltonian in the BL coordinate is described as
\begin{eqnarray}
   & & H(r, \theta, \varphi, p_r, p_\theta, p_\varphi)
   =  -p_t \nonumber \\ 
   &=& \omega p_\varphi 
      + \alpha \left( \frac{\Delta}{\rho ^2} p_r^2 
               + \frac{1}{\rho ^2} p_\theta^2
               + \frac{1}{\varpi ^2} p_\varphi^2 + m^2 
               \right)^{1/2}. 
\end{eqnarray}
The Hamilton equations $dx^{i}/dt = \partial H/ \partial p_i$ and  $dp_{i}/dt = -\partial H/ \partial x^{i}$ are then written as follows:
\begin{eqnarray}
 \frac{dr}{dt} &=& - \frac{p_r}{p_t + \omega p_\varphi} \frac{\alpha^2 \Delta}{\rho^2}, \\
 \frac{d\theta}{dt} &=& - \frac{p_\theta}{p_t + \omega p_\varphi} \frac{\alpha^2}{\rho^2}, \\
 \frac{d\varphi}{dt} &=& \omega - \frac{p_\varphi}{p_t + \omega p_\varphi} \frac{\alpha^2}{\varpi^2}, \\
  \frac{dp_r}{dt} &=& - \frac{\partial \omega}{\partial r} p_\varphi 
                     +  \frac{p_t + \omega p_\varphi} {\alpha} \frac{\partial \alpha}{\partial r}
                       \nonumber \\
                & &   +  \frac{1}{2} \frac{\alpha^2}{p_t + \omega p_\varphi} 
                       \frac{\partial}{\partial r}
                       \left(  \frac{\Delta}{\rho^2}p_r^2 
                             + \frac{1}{\rho^2}p_\theta^2
                             + \frac{1}{\varpi^2}p_\varphi^2
                       \right), \\
 \frac{dp_\theta}{dt} &=& - \frac{\partial \omega}{\partial \theta} p_\varphi 
                     +  \frac{p_t + \omega p_\varphi}{\alpha} \frac{\partial \alpha}{\partial \theta}
                       \nonumber \\
                & &   + \frac{1}{2} \frac{\alpha^2}{p_t + \omega p_\varphi} 
                       \frac{\partial}{\partial \theta}
                       \left(  \frac{\Delta}{\rho^2}p_r^2 
                             + \frac{1}{\rho^2}p_\theta^2
                             + \frac{1}{\varpi^2}p_\varphi^2
                       \right).
 \end{eqnarray}
As is mentioned above,  $dp_\varphi/dt$ is not solved during the calculation of geodesics as $p_{\varphi}$ is conserved variable in the axisymmetric spacetime.
The coefficients in the above Hamilton equations are described as follows:
\begin{eqnarray}
 \frac{\partial \omega}{\partial r} &=& 
 - \frac{\omega^2}{2Ma} \left[3r^2 + a^2\left(1 + \cos^2 \theta \right) - \frac{a^4}{r^2} \cos^2 \theta \right], \\
\frac{\partial \omega}{\partial \theta} &=& 
 - \frac{\omega^2 \sin \theta \cos \theta}{2Ma} \left(2Ma^2 - a^2r - \frac{a^4}{r} \right) , \\
\frac{\partial \alpha}{\partial r} &=& 
  \frac{1}{2\alpha} \frac{\partial \alpha^2}{\partial r}, \\
 \frac{\partial \alpha}{\partial \theta} &=& 
  \frac{1}{2\alpha} \frac{\partial \alpha^2}{\partial \theta}, 
\end{eqnarray}
\begin{eqnarray}
 \frac{\partial \alpha^2}{\partial r} &=& 
 - 
\alpha^4 \left(\frac{2M}{\Delta \rho^2}\right)
\left(\frac{a^4 - r^4}{\Delta} -  \frac{2r^2 a^2 \sin^2 \theta}{\rho^2} \right)
 , \\
\frac{\partial \alpha^2}{\partial \theta} &=& 
 - \alpha^4 \left[ \frac{4Ma^2 r \sin \theta \cos \theta\left(r^2 + a^2 \right)}
  {\Delta \rho^4}\right], 
 \end{eqnarray}
 \begin{eqnarray}
\frac{\partial }{\partial r}\left(\frac{1}{\varpi^2}\right) &=& 
 - \frac{2\sin^2 \theta }{\varpi^4} \nonumber \\
 && \left[r + \frac{M a^2 \sin^2 \theta 
  \left(a^2 \cos^2 \theta - r^2 \right)}{\rho^4} \right], \\
\frac{\partial }{\partial \theta}\left(\frac{1}{\varpi^2}\right) &=& 
 - \frac{2\sin \theta \cos \theta}{\varpi^4} \nonumber \\
 &&  \left[2M a^2 r \sin^2 \theta 
    \left(\frac{r^2 + a^2}{\rho^4}+\frac{1}{\rho^2} \right) 
  +  \left(r^2 + a^2 \right) 
 \right], \\
\frac{\partial }{\partial r}\left(\frac{\Delta}{\rho^2} \right) &=& 
  \frac{2}{\rho^2} \left( r - M - \frac{r \Delta}{\rho^2}\right), \\
\frac{\partial }{\partial \theta}\left(\frac{\Delta}{\rho^2} \right) &=& 
  \frac{2}{\rho^4}  a^2 \Delta \sin \theta \cos \theta, \\
\frac{\partial }{\partial r}\left(\frac{1}{\rho^2} \right) &=& 
 - \frac{2r}{\rho^4}, \\
\frac{\partial }{\partial \theta}\left(\frac{1}{\rho^2} \right) &=& 
  \frac{2}{\rho^4}  a^2 \sin \theta \cos \theta. 
 \end{eqnarray}

The above equations are integrated using an 8th-order embedded Runge-Kutta method with the adaptive step-size control \citep[e.g.,][]{Numerical_Recipes}.

\section{Radiative transfer (I): Observer-to-Emitter algorithmm} \label{sec:method_O2E}




The covariant radiative transfer equations, 
which are expressed by a suitable formula for the time-reversed calculations,
are solved when we use the  observer-to-emitter algorithm. 

The covariant radiative transfer equations without scatterings are described as following:
\begin{equation}
    \frac{d \mathcal{I}}{d \tau_{\nu}} =  - \mathcal{I} + \mathcal{S}, \label{eq:RTcov}
\end{equation}
where ${\tau}_{\nu}$ is the optical depth for photons with their frequency $\nu$ measured in the observer frame.

The invariant specific intensity $\mathcal{I}$ and source function $\mathcal{S}$ is wrtten as
\begin{eqnarray}
\mathcal{I} = \frac {I_{\nu}}{\nu^3}, \\
\mathcal{S} = \frac {\mathcal{J}}{\mathcal{A}}, \label{eq:S_cov}\\
\mathcal{J} = \frac {j_{{\nu}{\rm (tot)}}}{\nu^2}, \label{eq:J_cov}\\
\mathcal{A} = \nu \alpha_{{\nu}{\rm (tot)}} \label{eq:A_cov},
\end{eqnarray}
where $\mathcal{J}$ and $\mathcal{A}$ are the invariant emissivity and absorption coefficient, respectively.

The covariant radiative transfer equation (\ref{eq:RTcov}) leads to the equation which is convenient to be integrated from the observer to emitter \citep{Younsi_etal_2012, Pu_etal_2016_Odyssey}:
\begin{equation}
    \frac{d \mathcal{I}}{d \tau_{\nu {\rm (a)}}} = \mathcal{S}
    e^{-\tau_{\nu {\rm (a)}}} \label{eq:RTcov_o2e},
\end{equation}
where $\tau_{\nu {\rm (a)}}$ is the optical depth for absorption calculated from the observer to emitter. 
In practice, the infinitesimal changes of the invariant quantity $\tau_{\nu {\rm (a)}}$ calculated in our code such as 
\begin{eqnarray}
d \tau_{\nu {\rm (a)}} = {\alpha}^{\rm (z)}_{\nu_{\rm z}} dl = 
\left(\frac{\nu_{\rm f}}{\nu_{\rm z}}\right){\alpha}^{\rm (f)}_{\nu_{\rm f}} dl, \label{eq:dtau_a}
\end{eqnarray}
where $dl$ is the distance of photon propagation measured in the zero angular momentum observer \citep[ZAMO;][]{Bardeen_etal_1972} frame.
In the ZAMO frame, the infinitesimal $dx^{\hat \mu}$ can be transformed from that in the BL coordinate $dx^{\mu}$ as
\begin{eqnarray}
dx^{\hat{\mu}} = e^{\hat {\mu}}_{\mu} dx^{\mu},
\end{eqnarray}
where
\begin{equation}
    e^{\hat {\mu}}_{\mu} = 
    \begin{pmatrix}
    \alpha & 0 & 0 & 0 \\
    0 & {\rho}/\sqrt{{\Delta}} & 0 & 0 \\
    0 & 0 & {\rho} & 0 \\
    -{\omega}{\Delta}\sin\theta /{\alpha} & 0 & 0 & 
    {\varpi}
    \end{pmatrix}.
\end{equation}
Then, $dl$ is obtained by $dl = \sqrt{(dx^{\hat r})^2 + (dx^{\hat \theta})^2  + (dx^{\hat \varphi})^2}$.
Since the radiative transfer equation is not scale free, we should calculate $dl$ in the physical unit, i.e., cgs unit in the code, while the geodesic equations can be calculated with being normalized by the gravitational radius.
We note that $\nu_{\rm f}$ and $\nu_{\rm z}$ are the photon frequency observed in fluid-rest and ZAMO frame, respectively.
The covariant 4-momentum of photon in the ZAMO frame is calculated as
\begin{eqnarray}
p_{\hat{\mu}} = e^{\mu}_{\hat {\mu}} p_{\mu},
\end{eqnarray}
where $p_{\mu}$ is the covariant 4-momentum in the BL coordinates and 
\begin{equation}
    e^{\mu}_{\hat {\mu}} = 
    \begin{pmatrix}
    1/\alpha & 0 & 0 & \omega \alpha \\
    0 & \sqrt{{\Delta}}/{\rho} & 0 & 0 \\
    0 & 0 & 1/{\rho} & 0 \\
    0 & 0 & 0 & 
    1/{\varpi}
    \end{pmatrix}.
\end{equation}

The covariant source function Eq. (\ref{eq:S_cov}) is calculated by using total emissivity and absorption coefficient described in the following subsection.

\subsection{Emission and Absorption}

Various emission mechanism can be simultaneously incorporated by summing up each emissivity in the fluid-rest frame:
\begin{eqnarray}
j^{\rm (f)}_{{\nu_{\rm f}}{\rm (tot)}} = \sum j^{\rm (f)}_{\nu_{\rm f}},
\end{eqnarray}
where the detailed formulation of the emissivity of each radiative processes via thermal and nonthermal electrons are described in Appendix \ref{app:emission_form}


The absorption coefficient can be obtained by summing up those for each radiative process: 
\begin{eqnarray}
\alpha^{\rm (f)}_{\nu_{\rm f} ({\rm tot})} = \sum \alpha^{\rm (f)}_{\nu_{\rm f}}, \label{eq:alpha_tot}
\end{eqnarray}
where the detailed description of the absorption coefficients for the radiative processes via thermal and nonthermal electrons  are shown in Appendix \ref{app:absorb_form}.

We integrate the covariant radiative transfer equation from the observer to the emitter described as Eq (\ref{eq:RTcov_o2e}), by substituting the total emissivity and absorption coefficient to the covariant form of radiative quantities desribed in Eq (\ref{eq:S_cov}), (\ref{eq:J_cov}), and (\ref{eq:A_cov}).



\subsection{Observer Screen}
The observer screen is located at the distance far from the black hole, e.g., $10^4 r_{\rm g}$ in default. 
Each ray incoming to the screen is assumed to be perpendicular to the surface of the screen. 
We inversely trace the photon which arises from the center of the pixel of the screen.
We solve the radiative transfer with various photon frequency at observer screen simultaneously, during the ray-tracing calculations.

\section{Radiative Transfer (II): Emitter-to-Observer algorithm} \label{sec:method_E2O}

The emitter-to-observer algorithm with a Monte Carlo method is used to transport the photons taking into account the effects of Compton/inverse-Compton scattering.
For the Monte Carlo calculations, pseudo random numbers are generated by using Mersenne-Twister method. 

The four reference frames are used in this method: (i) observer frame (BL frame) is used for the ray tracing, (ii) ZAMO frame is basically used for analysis, (iii) fluid-rest frame is used for the calculation of emission and absorption coefficient, and (iv) electron-rest frame is used for computation of Compton scattering. 
For the transformation among (ii), (iii) and (iv), the Lorentz transformation is used because these are the tetrad frames. \\

\subsection{Emission}

To transport the photons with the Monte Carlo method, computational particles representing the  photons so called superphotons, are generated in the computation box.
The weight of superphotons $w$, which is a dimensionless quantity, represents the 
ratio of number density of real photons to superphotons in the phasespace generated in the unit time  \citep[e.g., ][]{Dolence_etal_2009}:
 \begin{eqnarray}
 \frac{1}{\sqrt{-g}}\frac{d N_{\rm real}}{dt d^3x d\nu d\Omega} = w \frac{1}{\sqrt{-g}}\frac{d N_{\rm super}}{dt d^3x d\nu d\Omega}
 =\left(\frac{\nu}{\nu_{\rm f}}\right)\frac{j^{\rm (f)}_{\nu_{\rm f}}}{h\nu_{\rm f}}, \label{eq:SP}
 \end{eqnarray}
where $N_{\rm real}$ and $N_{\rm super}$ are the number of real photons and superphotons, respectively.
Therefore, $w$ can be described as 
 \begin{eqnarray}
  w = \frac{\sqrt{-g} \Delta t \Delta^3 x \Delta \ln {\nu}}{\bar{N}_{\rm super} h} 
  \int  j^{\rm (f)}_{\nu_{\rm f}} d \Omega^{\rm (f)} \label{eq:weight},
 \end{eqnarray}
where $\bar{N}_{\rm super}$ is the number of superphotons per photon-frequency bin per computational cell.
It should be noted that $\sqrt{-g}$, $\Delta t$, $\Delta^3 x$ $d \nu$ are defined in the observer frame, while $d \Omega^{\rm (f)}$ is defined in the fluid-rest frame\footnote{This equation is the same as that in \citep{Dolence_etal_2009}, because $j_{\nu} d\Omega = j_{{\nu}_{\rm f}}^{(\rm f)} d\Omega^{\rm (f)}$, where $j_{\nu}$ is the emissivity in the observer frame.}.

\subsection{Absorption}
The absorption processes reduce the weight of the superphoton as
\begin{eqnarray}
w_{\rm new} = w \exp\left[-\Delta \tau_{\nu {\rm (a)}}\right],
\end{eqnarray}
where $w_{\rm new}$ is the updated value of superphoton weight $w$ after one tiemstep, and $\Delta \tau_{\nu {\rm (a)}}$ is the optical depth for absorption per timestep of ray tracing.
The total absorption coefficient described by (\ref{eq:alpha_tot}) and formulation of the optical depth (\ref{eq:dtau_a}) are used to evaluate $\Delta \tau_{\nu {\rm (a)}}$.

\subsection{Scattering}

The overview of the procedure for calculating the Compton scattering is as follows: (i) we calculate the optical depth for the Compton scattering $d\tau_{\nu ({\rm s})}$ between the current and previous position of the superphoton, (ii) we evaluate the probability of Compton scattering from $d\tau_{\nu ({\rm s})}$, and examine whether the scattering takes place at the current position by using a Monte Carlo method, (iii) the Compton scattering with the Monte Carlo method is calculated if the scattering occurs, and (iv) we go back to (i) after integrating the geodesic equation for one timestep.
Our procedure for the solving Compton scattering with a Monte Carlo method is similar to that of \cite{Dolence_etal_2009}, except that we incorporate the effect of the nonthermal electrons in addition to the thermal electrons.


\subsubsection{Optical Depth for Compton Scattering}
The optical depth for the Compton scattering is computed taking into account the fluid motion and kinetic motion of thermal and nonthemal electrons:
\begin{eqnarray}
d\tau_{\nu {\rm (s)}} =  \frac{\nu_{\rm f}}{\nu_{\rm z}} \left(n_{{\rm e,  th}} \sigma_{\rm KN, th} + n_{{\rm e, nth}} \sigma_{\rm KN, nth} \right)  dl,
\end{eqnarray}
where $\sigma_{\rm KN, th}$ and $\sigma_{\rm KN, nth}$ are the Klein-Nishina cross section modified by taking into account the effect of kinetic motion of thermal and nonthermal electrons in the fluid-rest frame, respectively.
We note that the former is the same as the hot cross section in \cite{Dolence_etal_2009}.
These modified cross section is formulated as:
\begin{eqnarray}
\sigma_{\rm KN, th} &=& \frac{1}{n_{{\rm e, th}}} \int d^3p_{\rm e} \frac{dn_{{\rm e, th}}}{d^3 p_{\rm e}} \left(1- \mu_{\rm e} \beta_{\rm e} \right) \sigma_{\rm KN}, \nonumber \\
&=& \frac{1}{2n_{{\rm e, th}}} \int d\gamma_{\rm e} \int d\mu_{\rm e} \frac{dn_{{\rm e, th}}}{d\gamma_{\rm e}}  \nonumber \\
&&\left(1- \mu_{\rm e} \sqrt{1 -{\gamma_{\rm e}}^{-2}} \right) \sigma_{\rm KN}, \\
\sigma_{\rm KN, nth} 
&=& \frac{1}{2n_{{\rm e, nth}}} \int d\gamma_{\rm e} \int d\mu_{\rm e} \frac{dn_{{\rm e, nth}}}{d\gamma_{\rm e}}  \nonumber \\
&&\left(1- \mu_{\rm e} \sqrt{1 -{\gamma_{\rm e}}^{-2}} \right) \sigma_{\rm KN},
\end{eqnarray}
where $dn_{{\rm e, th}}/d\gamma_{\rm e}$ is the Maxwell-J\"uttner distribution function described in eq. (\ref{eq:MJdist}), $dn_{{\rm e, nth}}/d\gamma_{\rm e}$ is the distribution function of nonthermal electron assumed to be isotropic in the momentum space, and $\sigma_{\rm KN}$ is the Klein-Nishina cross section defined in the electron-rest frame,
\begin{eqnarray}
\sigma_{\rm KN} &=& \sigma_{\rm T} \frac{3}{4\epsilon_{\rm e}^2} \nonumber \\
&& \left[2 + \frac{\epsilon_{\rm e}^2 (1 + \epsilon_{\rm e})}{(1 + 2 \epsilon_{\rm e})^2} + \frac{\epsilon_{\rm e}^2 - 2\epsilon_{\rm e} -2}{2\epsilon_{\rm e}} \ln (1 + 2\epsilon_{\rm e})\right] \label{eq:KN_crosssection},
\end{eqnarray}
where $\epsilon_{\rm e}$ is the photon energy, which is normalized by the rest-mass energy of electrons, in the electron-rest frame.

We tabulate the modified cross section for thermal and nonthermal electrons before the GRRT calculations. 
It should be noted that the nonthermal electrons are divided to subgroups by their Lorentz factor in the fluid rest frame, which is explained in the next subsection.


\subsubsection{Scattering Probability}
It should be determined whether the Compton scattering occurs at each timestep of integration of geodesic equation.  
In order to accelerate the calculation, it is useful to introduce the the bias parameter $b$ \citep{Dolence_etal_2009}:
\begin{eqnarray}
P_{\rm (s)} = 1 - \exp \left(- b \Delta \tau_{\nu {\rm (s)}}\right). \label{eq:scatprob2}
\end{eqnarray}

Due to the introduction of the bias parameter $b$,  each superphoton is split into two superphotons with dividing the weight $w$, when it is determined to be scattered.
The weight of the scattered photons $w_{\rm s}$ and the the other one are described as follows: 
\begin{eqnarray}
&w_{\rm s}& = w \left[ \frac{1 - \exp \left(\Delta \tau_{\nu {\rm (s)}}\right)}{1 - \exp \left(- b \Delta \tau_{\nu {\rm (s)}}\right)} \right] \label{eq:w_s}, \\
&w_{\rm nos}& = w - w_{\rm s} 
\end{eqnarray}
so that the number of photons is conserved via the scattering processes with the split of superphotons\footnote{We note that Eq.(\ref{eq:w_s}) leads to $w_{\rm s} = {w}/{b}$
for the limit $\Delta \tau_{\nu {\rm (s)}} << 1$, which is consistent with \cite{Dolence_etal_2009}.}.

When we include the nonthermal electrons, we further extend this method.
At first, we set the bias parameter $b$ for nonthermal electrons as follows:
\begin{eqnarray}
 b_{\rm nth, i} = b_{\rm th} \frac{\Delta \tau_{\nu {\rm (s) th}}}{\Delta \tau_{\nu {\rm (s) nth, i}}},
\end{eqnarray}
where $b_{\rm th}$/$b_{\rm nth, i}$ and ${\Delta \tau_{\nu {\rm (s) th}}}$/${\Delta \tau_{\nu {\rm (s) nth, i}}}$  are the bias parameter and the optical depth for the Compton scattering for thermal/nonthermal electrons, respectively.
The subscript "i"
identifies a subgroup of nonthermal electrons, which is introduced to accelerate the calculation.
In astrophysical plasmas, the nonthermal electrons with power-law distribution $\gamma_{\rm e}^{-p}$ usually has the power-law index with $p > 0$, i.e., the number density of nonthermal electorns with high Lorentz factor is very small. 
It requires, therefore, a huge computational time to sample the nonthermal electrons with high Lorentz factor with Monte-Carlo method.
In order to avoid this problem, we divide the energy distribution of nonthermal electrons into subgroups with their Lorentz factor. 

 We summarize our practical method to determine whether the scattering occur at each timestep and which of the thermal and nonthermal electrons are sampled if the scattering takes place.
We tabulate the modified cross section for each subgroup of the nonthermal electrons:
\begin{eqnarray}
\sigma_{\rm KN, nth, i} 
&=& \frac{1}{2n_{\rm e,  nth}} \int_{\gamma_{{\rm e,  min, i}}}^{\gamma_{{\rm e, max, i}}} d\gamma_{\rm e} \int_{-1}^{1} d\mu_{\rm e} \frac{dn_{\rm e,  th}}{d\gamma_{\rm e}}  \nonumber \\
&&\left(1- \mu_{\rm e} \sqrt{1 -{\gamma_{\rm e}}^{-2}} \right) \sigma_{\rm KN},
\end{eqnarray}
as well as that for thermal electrons (which are not divided to subgroups).
Then the optical depth for the Compton scattering is calculated as 
\begin{eqnarray}
\Delta \tau_{\nu {\rm (s) th}} =   \left(\frac{\nu_{\rm f}}{\nu_{\rm z}}\right)n_{\rm e,  th} \sigma_{\rm KN, th} \Delta l, \\
\Delta \tau_{\nu {\rm (s) nth, i}} =   \left(\frac{\nu_{\rm f}}{\nu_{\rm z}}\right)n_{\rm e,  nth, i} \sigma_{\rm KN, nth, i} \Delta l,
\end{eqnarray}
The scattering probability is calculated as 
\begin{eqnarray}
P_{\rm (s)} = 1 - \exp \left[- \left(b_{\rm th} \Delta \tau_{\nu {\rm (s) th}}
 + \sum_{\rm i = 1}^{\rm i_{max}} b_{\rm nth, i} \Delta \tau_{\nu {\rm (s) nth. i}} \right)  \right]. \label{eq:scatprob3}
\end{eqnarray}

We generate a pseudo-random number ${\xi}_1$ to determine whether the superphotons are scattered at each timestep. If the following condition is satisfied, the superphoton is scattered:
\begin{eqnarray}
 P_{\rm (s)} > {\xi}_1.
\end{eqnarray}
When this condition is satisfied, we examine which of the thermal and nonthermal electrons scatters the superphoton.
We assume that the thermal electrons scatter the superphotons if the following relation is satisfied:
\begin{eqnarray}
\frac{b_{\rm th} \Delta \tau_{\nu {\rm (s) th}}}{b_{\rm th} \Delta \tau_{\nu {\rm (s) th}}+ \sum_{\rm i = 1}^{\rm i_{max}} b_{\rm nth, i} \Delta \tau_{\nu {\rm (s) nth. i}}}  
\ge \xi_2,
\end{eqnarray}
where, $\xi_2$ is a newly generated random number.
The nonthermal electrons in the subgroup $i = i^{\prime} (< i_{\rm max})$ is selected when the following condition is satisfied:
\begin{eqnarray}
\frac{b_{\rm th} \Delta \tau_{\nu {\rm (s) th}} + \sum_{\rm i = 1}^{\rm i^{\prime}} b_{\rm nth, i} \Delta \tau_{\nu {\rm (s) nth. i}}}{b_{\rm th} \Delta \tau_{\nu {\rm (s) th}}+ \sum_{\rm i = 1}^{\rm i_{max}} b_{\rm nth, i} \Delta \tau_{\nu {\rm (s) nth. i}}}  
< \xi_2 \\ 
\le 
\frac{b_{\rm th} \Delta \tau_{\nu {\rm (s) th}}  + \sum_{\rm i = 1}^{i^{\prime} + 1} b_{\rm nth, i \Delta \tau_{\nu {\rm (s) nth. i}}}}
{b_{\rm th} \Delta \tau_{\nu {\rm (s) th}}+ \sum_{\rm i = 1}^{\rm i_{max}} b_{\rm nth, i} \Delta \tau_{\nu {\rm (s) nth. i}}}.
\end{eqnarray}
In the other case, the nonthermal electrons in the subgroup $i = i_{\rm max}$ is selected.

\subsubsection{Solving Compton Scattering}

When the condition for the scattering is satisfied, we solve the Compton scattering by using a Monte Carlo method.
First of all, (i) a scattering electron in the fluid-rest frame is sampled, and then,
(ii) the four-momentum of the scattered photon is determined using another pseudo-random number. 

Before the Compton scattering, the scattering electrons are sampled as follows:
\begin{enumerate}
    \item The Lorentz factor of thermal or nonthermal electrons in the fluid-rest frame is determined.
    For thermal electrons, the Lorentz factor of the electrons is sampled to reproduce the Maxwell-J\"utter distribution (eq. \ref{eq:MJdist}). We employ a procedure based on a rejection method described in \cite{Canfield_etal_1987}, which is summarized in Appendix \ref{sec:eDF_MJ_MC}. On the other hand, for nonthermal electrons with a single/broken power law distribution, the Lorentz factor of the scattering electrons can be sampled  much easier by the inverse function method descibed in Appendix \ref{sec:eDF_sPL_MC} and \ref{sec:eDF_brPL_MC}.

\item The direction of the velocity of the scattering electron, i.e., the angle between the momenta of scattering electron and the incident photon, is determined.
When we sample the scattering electrons, 
the relativistic motion of electron should be taken into consideration during the MC sampling.
This is because (i) the scattering cross section in the fluid-rest frame is $(1- \mu_{\rm e} \beta_{\rm e}) \sigma_{\rm KN}$ and (ii) $\sigma_{\rm KN}$ is significantly reduced when the Doppler-boosted photon energy in the electron-rest frame is significantly larger than the rest-mass  energy of electrons \citep{Gorecki_Wilczewski_1984}.
In order to take into account these effects, we follow the algorithm in \cite{Canfield_etal_1987} as follows. The probability distribution function is anisotropic by the factor $(1- \mu_{\rm e} \beta_{\rm e})$.
The cosine of the angle between the incident photon and the scattering electron $\mu_{\rm e}$ is then sampled by the inverse function method using a pseudo random number $\xi_3$:
\begin{eqnarray}
\mu_{\rm e} = \frac{1 - \sqrt{1+ \beta_{\rm e}^2 + 2\beta_{\rm e} -4 \beta_{\rm e}\xi_3}}{\beta_{\rm e}}.
\end{eqnarray}
There still remains the degree of freedom of the  other angle of four-momentum of the electron, and this angle is sampled by a uniform random number. 
At this step, the four-momentum of the electron  is tentatively determined.

\item Finally, we examine the validity of the sampled electrons with their four-momentum using a rejection method.
We accept the sampled electron if it satisfies
\begin{eqnarray}
\frac{\sigma_{\rm KN}}{\sigma_{\rm T}} \ge \xi_4,
\end{eqnarray}
where $\sigma_{\rm KN}$ shown in equation (\ref{eq:KN_crosssection}) is evaluated by 
the photon energy in the electron-rest frame $\epsilon_{\rm e}$, which is calculated by the Lorentz transformation of the four-momentum of the incident photon from the fluid-rest frame to the electron-rest frame of the sampled electron.
\end{enumerate}

Next, we sample the four-momentum of the scattered photons.
As described above, the four-momentum vector of the incident photons is Lorentz transformed from the fluid-rest frame to the electron-rest frame.
Using a pseudo-random number $\xi_5$, a tentative value of the scattered photon energy can be represented as 
\begin{eqnarray}
\epsilon_{\rm e}^{\prime} = (1-\xi_5) \epsilon_{\rm e,  min}^{\prime} + \xi_5 \epsilon_{\rm e,  max}^{\prime},
\end{eqnarray}
where ${\epsilon^{\prime}_{\rm e}}$ is related to
${\epsilon_{\rm e}}$ and the cosine of the angle between the electron and the scattered photon 
$\mu_{\rm e}^{\prime} = 1 + 1/{\epsilon_{\rm e}} + 1/{\epsilon^{\prime}_{\rm e}}$, 
so that 
$\epsilon_{{\rm e, min}}^{\prime} = (1+2\epsilon_{\rm e})/\epsilon_{\rm e}$ 
and $\epsilon_{{\rm e,  max}}^{\prime} = \epsilon_{\rm e}$.
Once $\epsilon_e$ is determined, we obtain $\mu_e^{\prime}$.
The remaining angle, whose probability distribution function uniformly distributes $[0, 2 \pi)$, is determined by generating another pseudo-random number.

Finally, in order to examine whether the tentative value of $\epsilon_{\rm e}^{\prime}$ is presumable, we  accept the value which satisfies the following relation:
\begin{eqnarray}
\frac{{\rm d}\sigma_{\rm KN}/{\rm d}\epsilon_{\rm e}^{\prime}}{{\rm max}({\rm d}\sigma_{\rm KN}/{\rm d}\epsilon_{\rm e}^{\prime})} \ge \xi_6,
\end{eqnarray}
where
\begin{eqnarray}
\frac{{\rm d}\sigma_{\rm KN}}{{\rm d}\epsilon_{\rm e}^{\prime}} = \frac{3 \sigma_{\rm T}}{8} \frac{1}{\epsilon_{\rm e}^2} \left( \frac{\epsilon_{\rm e}}{\epsilon_{\rm e}^{\prime}} + \frac{\epsilon_{\rm e}^{\prime}}{\epsilon_{\rm e}} - 1 + {\mu^{\prime}_{\rm e}}^{2} \right), \\
{\rm max}\left(\frac{{\rm d}\sigma_{\rm KN}}{{\rm d}\epsilon_{\rm e}^{\prime}}\right) = \left.\frac{{\rm d}\sigma_{\rm KN}}{{\rm d}\epsilon_{\rm e}^{\prime}}\right|
_{\epsilon_{\rm e}^{\prime} = \epsilon_{\rm e}} = \frac{3 \sigma_{\rm T}}{4} \frac{1}{\epsilon_{\rm e}^2}.
\end{eqnarray}
If the tentative $\epsilon_{\rm e}^{\prime}$ is rejected, the procedure for sampling the scattered photon is repeated until it is accepted through the rejection method.
Once the sampled scattered photon is accepted, the four-momentum of the scattered photon in the electron-rest frame
is converted to that in the fluid-rest frame and in the ZAMO frame via the Lorentz transformation. 
The covariant four-momentum of photon in the BL frame is also calculated via the coordinate transformation
in order to solve the geodesic equations
at the next time step.


\subsection{Calculation of SED}
The SED is calculated after the superphoton escape to the outer boundary of the computational domain. 
We divide the bins of the photon-frequency in the observer frame $\nu$ and the viewing-angle $\theta_{\rm ob}$ in such a way that $\Delta \ln \nu$  and $\Delta \cos \theta_{\rm ob}$ are constant.
Here viewing angle $\theta_{\rm ob}$ 
is defined as the 
angle between the rotation axis of the Kerr BH and the line-of-sight of  observers.

For an observer at $j$th viewing-angle bin, the value of $\nu L_{\nu}$ at $i$th frequency bin can be calculated as follows:
\begin{eqnarray}
\nu L_{\nu}{\rm (i,j)}  = \frac{4\pi}{\Delta \ln \nu \Delta \Omega \Delta t} \sum_{\rm n} w_{\rm n}{\rm (i,j)} h \nu_{\rm n}{\rm (i,j)},
\end{eqnarray}
where $w_{\rm n} {\rm (i,j)}$ is the weight of $\rm n$-th photons leaving the computational domain from the outer boundary with belonging to the $\rm i$-th frequency bin and the $\rm j$-th viewing-angle bin. 
We set $\Delta \Omega = 2\pi \Delta \cos \theta_{\rm ob}$ and $\Delta t = 1$ in default.



\section{Numerical Tests} \label{sec:test}

\subsection{Trajectory of photon} \label{sec:test-trajectory}
We test the geodesics solver by comparing the numerical results and the analytic solution on the equatorial plane of a Kerr BH for the case with $p_{\varphi} = -a p_t$ \citep{Chandrasekahr_1983}:
\begin{eqnarray}
\varphi = \frac{a}{r_{+} - r_{-}} \left[\ln \left(\frac{r}{r_+} - 1\right) - 
\ln \left(\frac{r}{r_-} - 1\right) \right], \label{eq:RT_Chandra}
\end{eqnarray}
where $r_+$ and $r_-$ are the outer and inner horizon of the Kerr BHs $r_{\pm} = M \pm \sqrt{M^2 - a^2}$.

\begin{figure}[ht!]
\includegraphics[width=0.95\columnwidth]{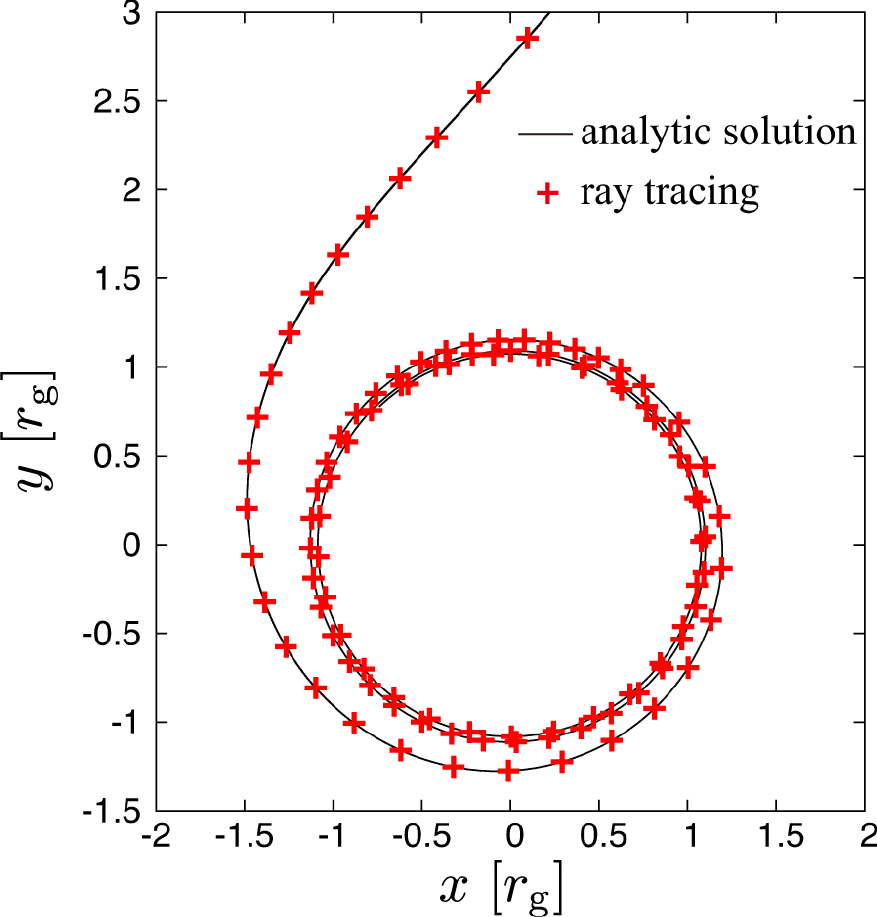}
\caption{
Comparison between the ray-tracing calculation (the red points) and the analytic solution (the black line) of a trajectory of a photon on the equatorial plane of a Kerr BH with $a_* = 0.998$.  \label{fig:RT_Chandra}}.
\end{figure}

Figure \ref{fig:RT_Chandra} shows the result with setting $a_* = 0.998$, where $a_* \equiv a/M$. 
The BH rotates in the counterclockwise direction.
The photons are initiated at $r = 10r_{\rm g}$, $\theta = \pi/2$, and $\varphi$ given by eq. (\ref{eq:RT_Chandra}).
The four-momentum of photon is $p_{\rm t} = -E$, $p_{\theta} = 0$, $p_{\varphi} = a_* E$, and $p_r$ is derived from the the constraint $p^\mu p_\mu = 0$.
As the photon gets close to the BH, its trajectory is strongly affected by the BH gravity and the frame-dragging effect.
The result computed by our code (red points) well reproduces the analytic solution (black line). 


\subsection{Black Hole Shadow} \label{sec:test-photonring}

\begin{figure*}[ht!]
\centering
\includegraphics[width=2.\columnwidth]{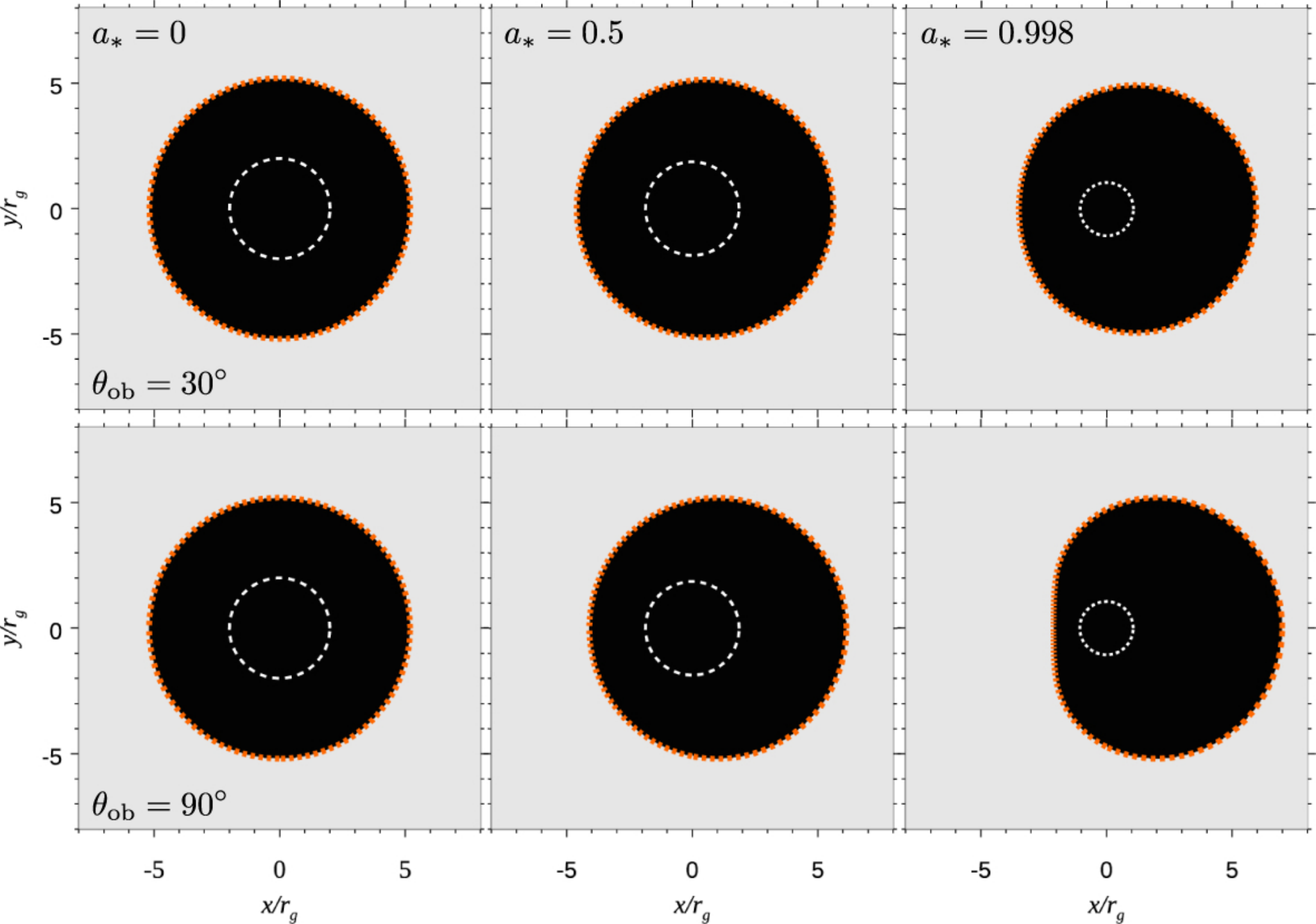}
\caption{The computed BH shadows for the case with $a_* = 0$ (left), 0.5 (center), and 0.998 (right).
The top and bottom panels displays the images with the viewing angle $\theta_{\rm ob} = 30^{\circ}$  and $90^{\circ}$, respectively.
The black color displays the region where the photon emitted from the observer screen with time-reversing way is captured by the BH, while the light gray represents the area in which the photons escape to infinity.
The dotted, orange lines are the edge of BH shadow obtained by the analytic solution.
The dashed white lines represent the radius of the outer horizon of BHs in the BL coordinate.  \label{fig:RT_BHshadow}}.
\end{figure*}

As another tests of the geodesic equation solver, 
the shape of the BH shadows are calculated with the viewing angle $30^{\circ}$ and $90^{\circ}$ for $a_*= 0$, 0.5, and 0.998, and shown in Figure \ref{fig:RT_BHshadow}.
The photon trajectory is solved from the observer screen with the time-reversing way.
The black color displays the region where the photon 
is captured by the black hole, while the light-gray color represents the area where the photons escape to infinity.
The dotted, orange lines represent the shape of black-hole shadow given by an analytic solution, whose radius and angle in the 
observer screen  is described as follows \citep{Bardeen_1973, Johannsen_2013, Wong_2021}:
\begin{eqnarray}
x &=& -\frac{\xi}{\sin \theta_{\rm ob}}, \\
y &=& \pm \sqrt{\eta + a^2 \cos^2 \theta_{\rm ob} - \xi^2\cot^2 \theta_{\rm ob}}, \\
\end{eqnarray}
where $x$ and $y$ are the cartesian coordinate on the observer screen in which the BH spin vector projected onto the screen is on the $y$-axis, and $\theta_{\rm ob}$ is the viewing angle of the observer.
The variables $\xi$ and $\eta$ are parameterized by the radius of the corresponding photon orbit $r_{\rm ph}$ as follows:
\begin{eqnarray}
\xi &=& -\frac{r_{\rm ph}^2 (r_{\rm ph} - 3M) + a^2 (r_{\rm ph}+M)}{a(r_{\rm ph} - M)}, \\
\eta &=& \frac{r_{\rm ph}^3 [4a^2M - r_{\rm ph}(r_{\rm ph} - 3M)^2 ] }{a^2 (r_{\rm ph} -M)^2}\\
\end{eqnarray}
We note that $r_{\rm ph}$ continuously exists between their minimum and maximum vales $r_{\rm ph, min}$ and $r_{\rm ph, max}$, which are the roots of the equation 
\begin{eqnarray}
y^2 &=& a^2 \cos^2 \theta_{\rm ob} + \frac{1}{a^2 (r_{\rm ph}-M)^2} \nonumber\\
&&[a^2 (r_{\rm ph}+M) + r_{\rm ph}^2(r_{\rm ph}-3M)]^2 \cot^2 \theta_{\rm ob} \nonumber \\
&&+ r_{\rm ph}^3[r_{\rm ph}(r_{\rm ph} - 3M)^2 - 4 a^2] =0,
\end{eqnarray}
where the region formed by these unstable circular photon orbits is so called photon shell \citep{Johnson_etal_2020_ScienceAdv, EHTC2019_5, Kawashima_etal_2021}.

Our numerical calculations successfully reproduce the analytical shape of the BH shadow for the various inclination angle and BH spin. 
We note that, as the BH spin increases, the location of the BH shadow shifts in the direction perpendicular to the BH spin vector projected onto the observer screen, because of the frame-dragging effect \citep[for the application to a possible estimate of the BH spin,  see][]{Kawashima_etal_2019,Chael_etal_2021_arXiv}.

More tests of computations of geodesic equations and black hole shadows using \texttt{RAIKOU} are also carried out by GRRT code comparison projects of EHT \citep{GRRT_code_comparison_2020}.


\subsection{Compton scattering via thermal electron} \label{sec:test-Compton scattering_th}

In this section, we test the Compton scattering of blackbody photons via thermal electrons in a uniform plasma sphere.
For the sake of the simplicity, the spacetime is assumed to be flat (i.e., the Minkowski spacetime) and the absorption processes are ignored here.
The radiative temperature of blackbody photons $T_{\rm rad}$ is set to be $k_{\rm B} T_{\rm rad}/m_{\rm e}c^2 = 10^{-8}$, where $k_{\rm B}$, $m_{\rm e}$ and $c$ are the Boltzmann constant, electron rest mass, and speed of light, respectively.
The blackbody photons are injected from the inner sphere at $r = r_{\rm in}=1$ cm. The outer boundary of the computational domain locates at $r=r_{\rm out} = 10^5$ cm, so that the finite radius of $r_{\rm in}$ negligibly affect the results.

\begin{figure}[ht!]
\centering
\includegraphics[width=0.98\columnwidth]{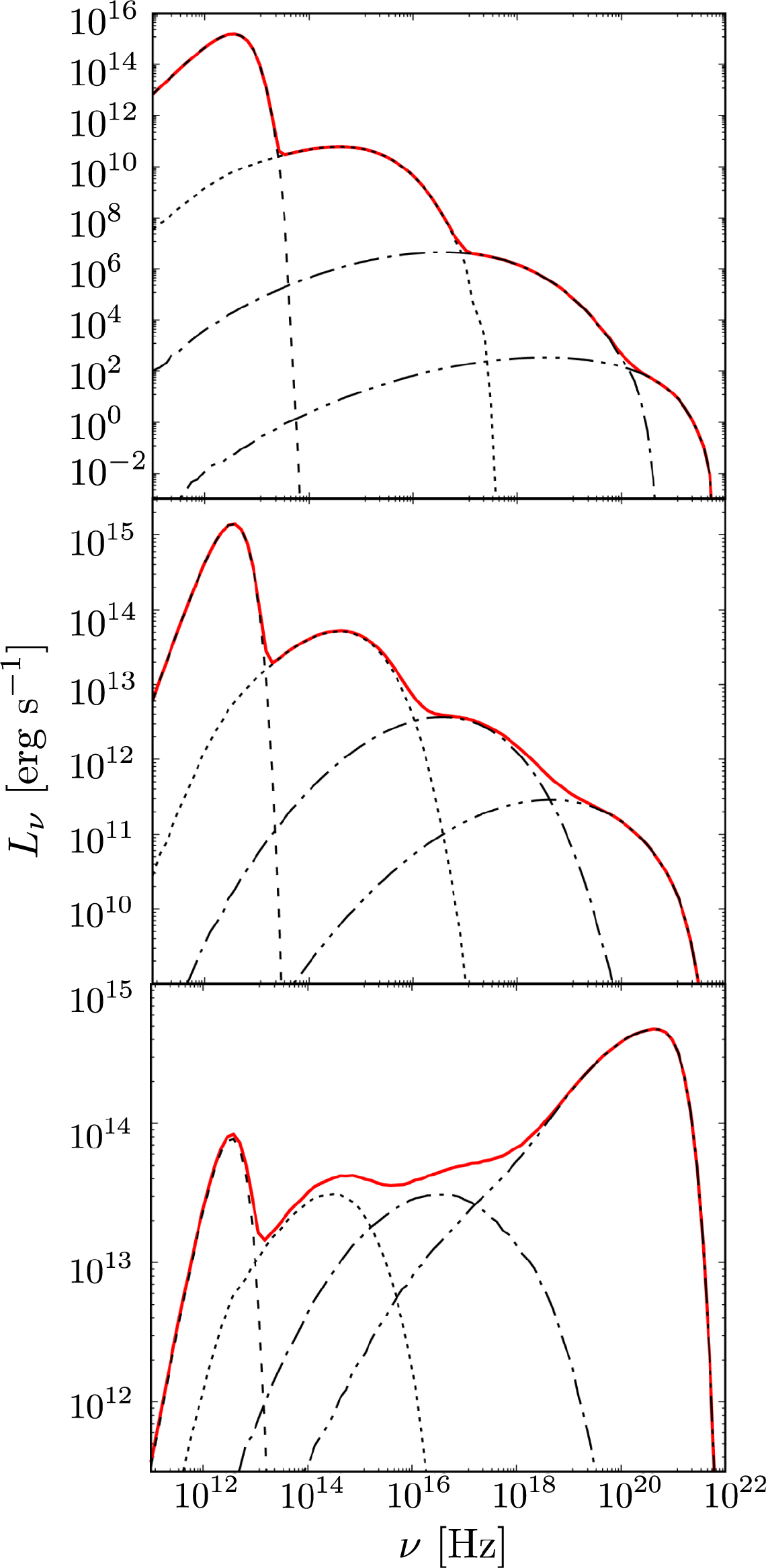}
\caption{Comptonized SEDs of blackbody photons, which are scattered by the hot electrons $\Theta_{\rm e} = 4$ in the plasma sphere with optical depth for Thomson scattering $\tau_{\rm (s)} = 10^{-4}$(top), $10^{-1}$(middle), and 3(bottom). 
The resulting SEDs are displayed with the red solid lines.
Black lines represent the decomposed SEDs, whose photons are escaped towards the observer without being scattered (dashed), with being scattered once (dotted), twice (dash-dot-dash), and more than three times (dash-dot-dot-dash).}
\label{fig:SED_Comp_th}
\end{figure}

We examine the Compton scattering via thermal electrons as in \cite{Dolence_etal_2009}.
The temperature and number density of electrons are set to be $\Theta_{\rm e} = k_{\rm B}T_{\rm e}/m_{\rm e}c^2 = 4$ and $n_{\rm e,  th} = \tau_{\rm (s)}/\sigma_{\rm T}(r_{\rm out}-r_{\rm in})$, respectively. Here, $\tau_{\rm (s)}$ is the optical depth for Thomson scattering and we set $\tau_{\rm (s)} = 10^{-4}$, $10^{-1}$, and $3$.
Figure \ref{fig:SED_Comp_th} displays the resulting SEDs, which reproduces the results shown in \cite{Dolence_etal_2009}.

In Fig \ref{fig:SED_Comp_th}, the SEDs are also decomposed into those of photons without being scattered, with being scattered once, twice, and three or more times.
The optically thin cases (i.e., $\tau_{(s)} = 10^{-4}$ and $10^{-1}$) more clearly show the feature of individual Comptonization processes. 
Spectral bumps at $\nu \sim 10^{15}, 10^{17}$, and $\sim 10^{19}$ Hz are formed via the inverse-Compton scattering, which is consistent with the fact that the change of photon frequency  per scattering $\Delta \nu$ is $\Delta \nu / \nu \sim 16\Theta_{\rm e}^2 \sim 10^2$ \citep[e.g.,][]{Rybicki_Lightman_1979}.
One can also find that the $L_{\nu}$ at each spectral bump reduces by $\sim 10^{-4}$ and $\sim 10^{-1}$ for the case with ${\tau}_{\rm (s)} = 10^{-4}$ and $10^{-1}$, respectively, since the $L_{\nu}$ at the frequency of scattered photon reduced by the scattaring probability $1 - \exp(-{\tau}_{\rm (s)}) \sim \tau_{(\rm s)}$ in the optically thin limit.
In the optically thick case $\tau_{\rm (s)} = 3$, the number of scattering is $\sim$ 10 and the Compton {\it y}-parameter  is significantly greater than unity ${\it y} = 16\Theta_{\rm e}^2 \max(\tau_{\rm (s)}, \tau_{\rm (s)}^2) \sim 2\times 10^3 \gg 1$, which results in the formation of the Wien-like peak around $3k_{\rm B}T_{\rm e}/h \sim 10^{21}$ Hz.
The resulting SEDs therefore, well reproduce the spectral features of Comptonization both in the opticallty thin and thick plasmas.

\subsection{Compton scattering via nonthermal electron} \label{sec:test-Compton scattering_nth}

We examine the Comptonization of blackbody photons via nonthermal electrons with single and broken power-law distribution functions in the uniform spherical plasma. 
The radial optical depth against the Thomson scattering is $10^{-4}$.
In this subsection, we show the SEDs of $\nu L_{\nu}$ rather than $L_{\nu}$ to clearly present the  high energy photons Compton-upscattered by electrons with a high Lorentz factor.

\begin{figure}[ht!]
\centering
\includegraphics[width=0.98\columnwidth]{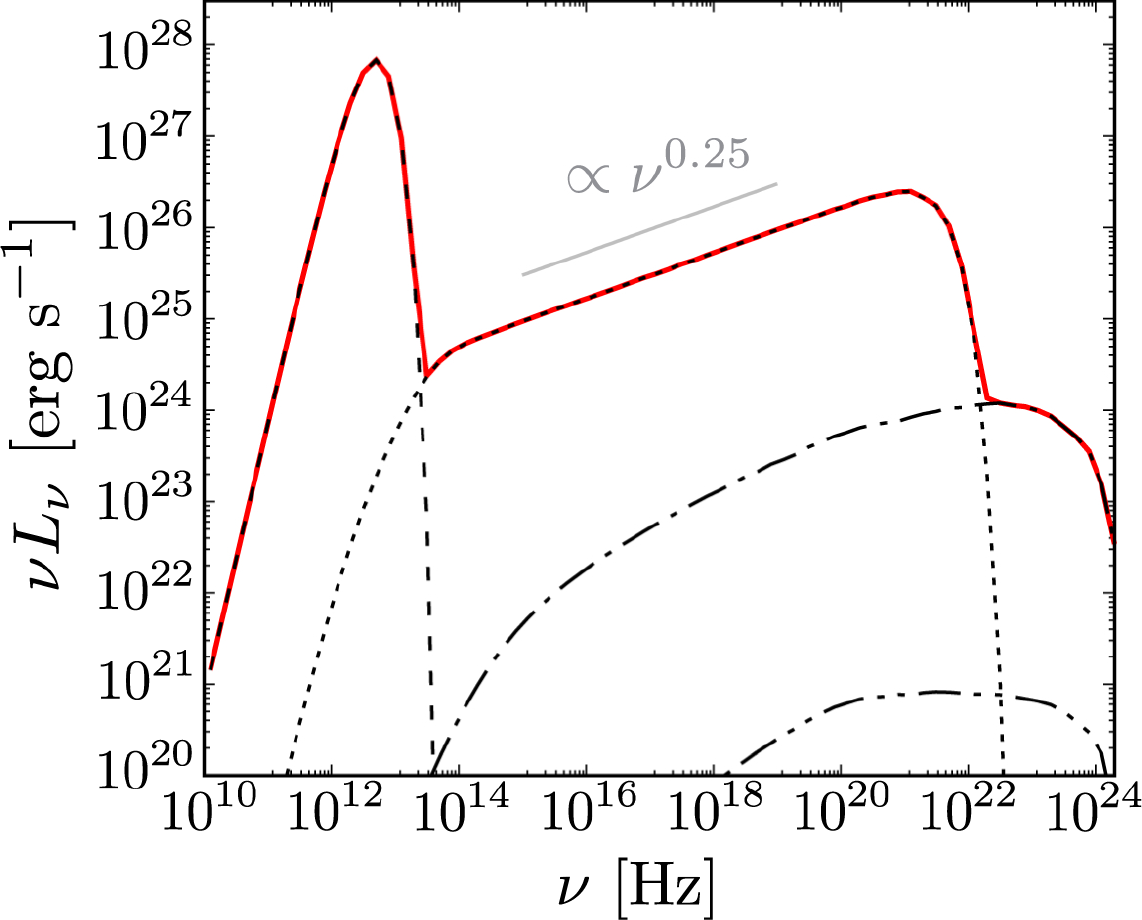}

\caption{Comptonized SEDs of blackbody photons, which are scattered by nonthermal electrons with power-law index $p=2.5$ in the plasma sphere.
The minimum and maximum Lorentz factor of nonthermal electrons are $\gamma_{\rm e,  min} = 2$ and $\gamma_{\rm e,  max}=2 \times 10^4$, respectively.
The optical depth for Thomson scattering is $\tau_{\rm (s)} = 10^{-4}$. 
The resulting SEDs are displayed with the red solid lines.
Black lines represent the decomposed SEDs, whose photons are escaped towards the observer without being scattered (dashed), with being scattered once (dotted), twice (dash-dot-dash), and more than three times (dash-dot-dot-dash).}
\label{fig:SED_Comp_nth_SP1}
\end{figure}

Figure \ref{fig:SED_Comp_nth_SP1} displays the SEDs of photons Comptonized via nonthermal electrons with a single power-law distribution $n_{\rm e,  nth} \propto \gamma_{\rm e}^{-p}$ for $\gamma_{\rm e,  min} \le \gamma_{\rm e} \le \gamma_{\rm e,  max}$, where $\gamma_{\rm min} = 2$ and $\gamma_{\rm max} = 2\times10^{4}$.
The SED peak of photons, which experienced one scattering event, appears at $\sim 10^{21}$ Hz. 
This peak is formed by inverse-Compton scattering with electrons whose Lorentz factor $\gamma_e \, {\simeq} \, \gamma_{\rm e,  max}$, i.e., the photons with the peak frequency of injected blackbody photons ${\nu}_{\rm bb} \simeq 4 \times 10^{12}$ Hz are upscattered to the photon frequency $ \gamma_{\rm e,  max}^2 {\nu}_{\rm bb}$ $\simeq 10^{21}$ Hz.
The additional peak at $\nu \sim 10^{23}$ Hz is formed by two scattering events, where the peak frequency appears at  
$\lesssim \gamma_{\rm e,  max} m_{\rm e} c^2/h \sim 10^{24}$ Hz because of the reduce of the scattering cross section (i.e., Klein-Nishina effect).

\begin{figure}[ht!]
\centering
\includegraphics[width=0.98\columnwidth]{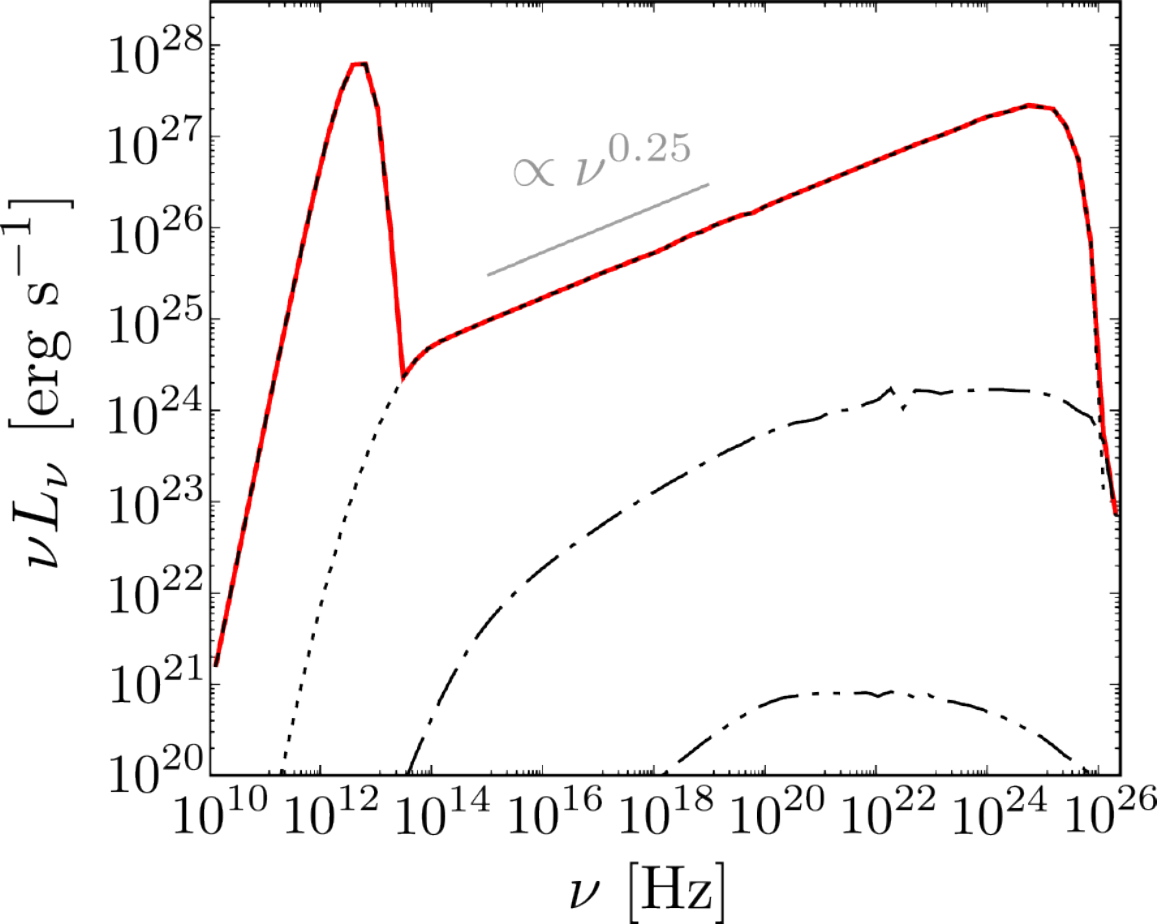}
\caption{The same as Fig. \ref{fig:SED_Comp_nth_SP1} except that $\gamma_{\rm max} = 2\times10^6$. 
}
\label{fig:SED_Comp_nth_SP2}
\end{figure}

\begin{figure}[ht!]
\centering
\includegraphics[width=0.98\columnwidth]{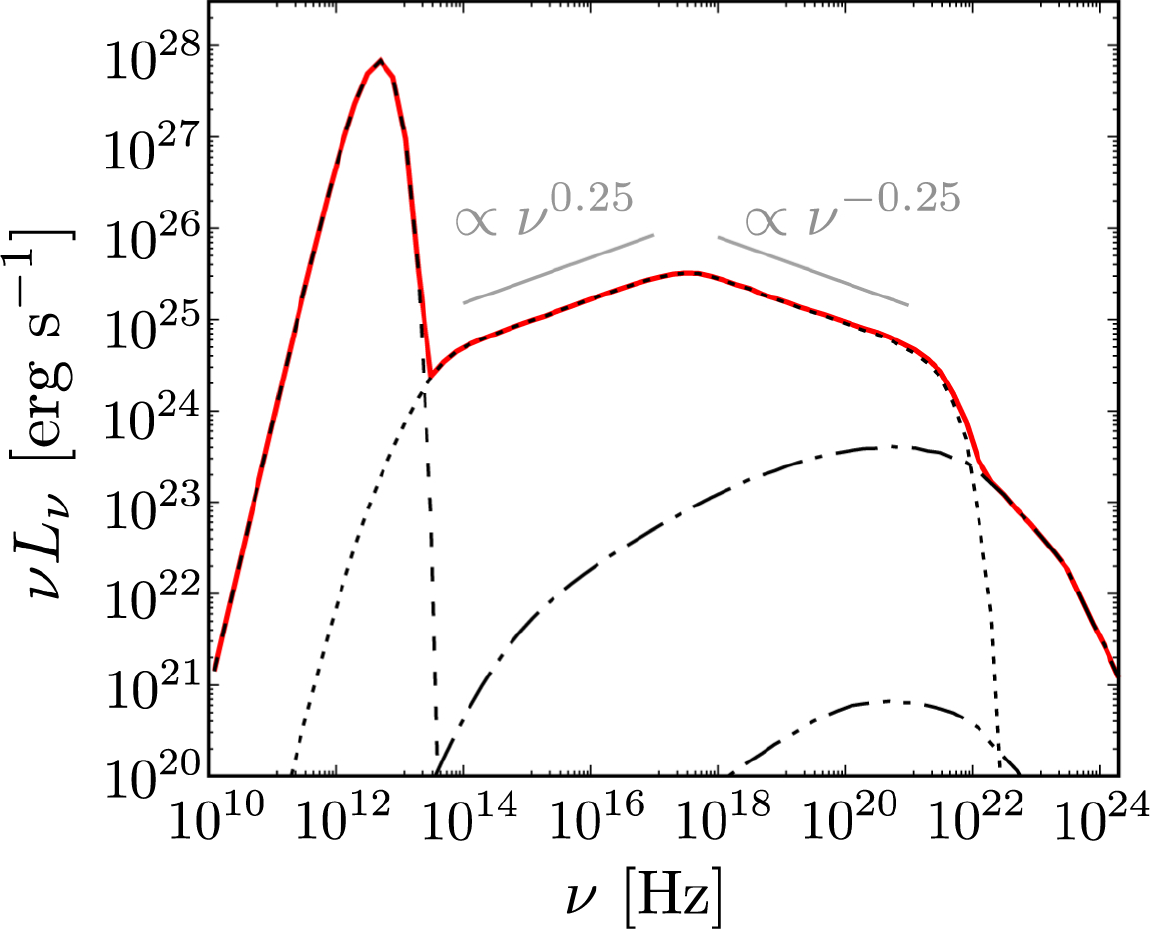}
\caption{The same as Fig. \ref{fig:SED_Comp_nth_SP1}, except  for broken power-law distribution of nonthermal electrons.
The power-law indices are  $p_1=2.5$  and $p_2 = 3.5$ between 
$2 \le \gamma_{\rm e} \le 2 \times 10^2$ and $200 < \gamma_{\rm e} \le 2 \times 10^{4}$, respectively.}
\label{fig:SED_Comp_nth_BP1}
\end{figure}

Figure \ref{fig:SED_Comp_nth_SP2} shows the Comptonized SEDs with the same model setup except for the higher maximum Lorentz factor of nonthermal electrons $\gamma_{\rm e,  max} = 2\times10^{6}$. 
In the gamma-ray energy bands, only one remarkable spectral peak appears at $\nu \sim 10^{25}$ Hz, although two peaks appear in the gamma-ray energy band for the model with the lower maximum Lorentz factor 
(see Fig. \ref{fig:SED_Comp_nth_SP1}).
This is because the spectral peak formed by the one-scattering component $\gamma_{\rm e,  max}^2  \nu_{\rm bb}  \sim 10^{25}$ Hz and the possible maximum peak of Comptonized SEDs $\gamma_{\rm e,   max} m_{\rm e} c^2/h \sim 10^{26}$ Hz are similar. For the component with scattering twice or more, the Compton scattering with high Lorentz factor of electrons are highly suppressed by the Klein-Nishina effect and the spectral peak appears at $\nu \lesssim \gamma_{\rm e,  max} m_{\rm e}c^2/h \sim 10^{26}$ Hz.

In figure \ref{fig:SED_Comp_nth_BP1}, we show a result of Comptonized SEDs calculated for broken power-law model of nonthermal electrons. 
The power-law distribution extends from $\gamma_{\rm e,  min} = 2\times 10^2$ to $\gamma_{\rm e,  max} = 2\times 10^4$  and the power-law index is broken at $\gamma_{\rm br} =  2\times10^2$.
The power-law indices are $p_1 = 2.5$ and $p_2 = 3.5$ in $\gamma_{\rm e,  min} \le \gamma_{\rm e} \le \gamma_{\rm e,  br}$ and $\gamma_{\rm e,  br} < \gamma_{\rm e} \le \gamma_{\rm e,  max}$, respectively.
As a consequence of the break of power-law electron destribution function, the power-law SED of photons also breaks at $\nu_{\rm br} \sim (4k_{\rm B}T_{\rm e}/h) \gamma_{\rm br}^2 \sim 10^{17}$ Hz. 
The power-law indices in the SED is $(3-p_{\rm 1})/2 = 0.25$ and  $(3-p_{\rm 2})/2 = -0.25$ below and above $\nu_{\rm br}$, respectively.

\subsection{Synchrotron Self-Compton} \label{sec:test-SSC}

In this subsection, the synchrotron self-Compton processes via nonthermal electrons are tested. 
We set a plasma sphere with uniform spatial distribution of number density of nonthermal electrons $n_{\rm e, nth} = 10^3 {\rm cm^{-3}}$ and magnetic-field strength $B=10^{-2}$ G, where the orientation of the magnetic field is isotropic.
The radius of the plasma sphere is set to be $10^{15}$ cm. 
The bulk speed of the plasma sphere is assumed to be zero. 
We examine the models with single- and broken-power law distribution function of nonthermal electrons.

\begin{figure}[ht!]
\centering
\includegraphics[width=0.98\columnwidth]{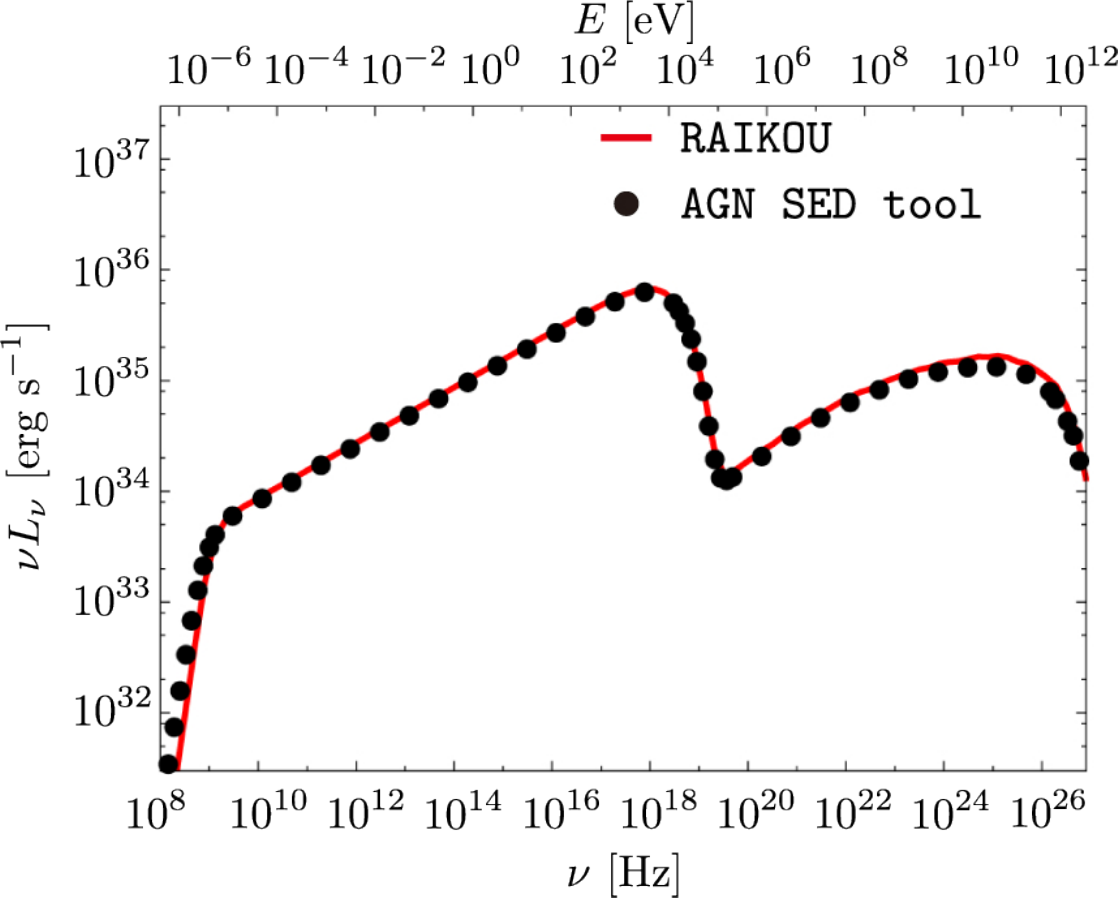}
\caption{SEDs of synchrotron Self-Compton via nonthermal electrons with a single power-law distribution. The power-law indices are $p = 2.5$ for the electrons with the Lorentz factor $\gamma_{\rm e, min} \le \gamma_{\rm e} \le \gamma_{\rm e, max}$. 
Here, we set $\gamma_{\rm e, min} = 10^2$  and $\gamma_{\rm e, max} = 10^7$. The red line and the black circles present the results obtained by \texttt{RAIKOU} and a public code \texttt{AGN SED tool}, respectively.}
\label{fig:SED_SSC_nth_SPL}
\end{figure}

\begin{figure}[ht!]
\centering
\includegraphics[width=0.98\columnwidth]{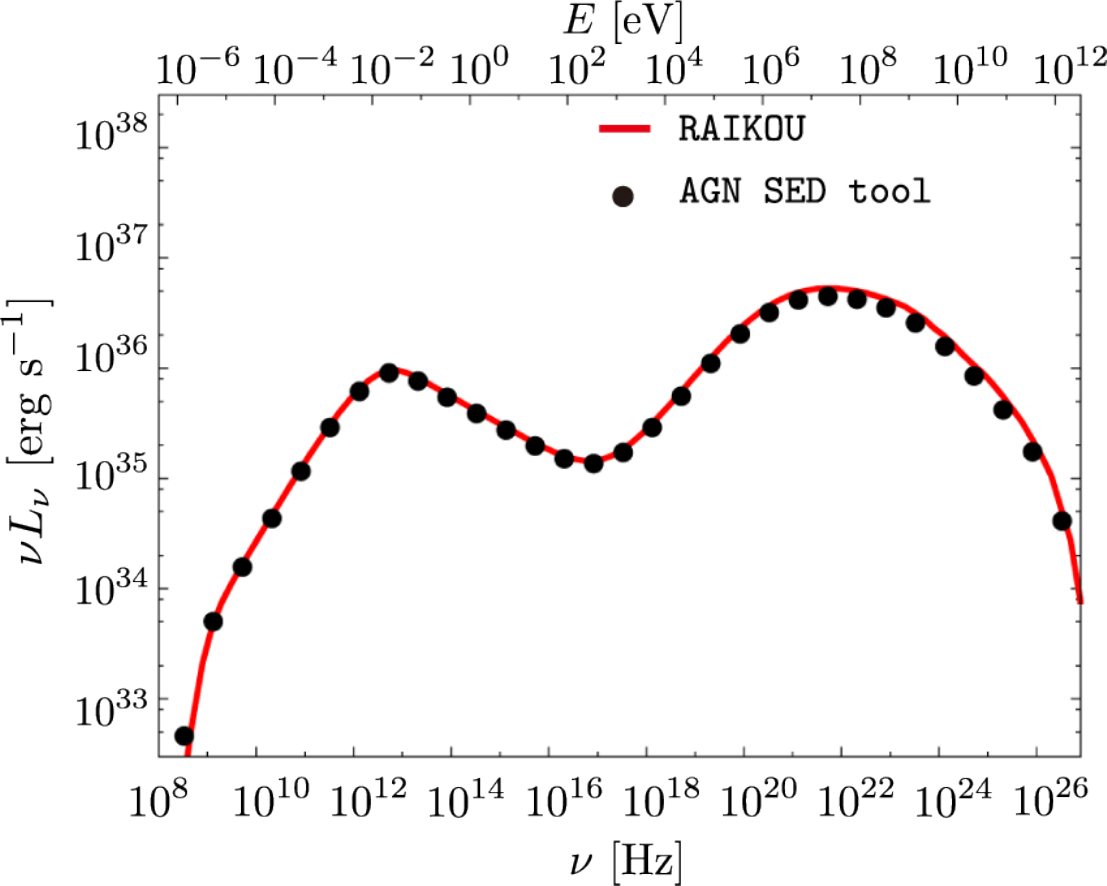}
\caption{Same as Fig. \ref{fig:SED_SSC_nth_SPL}, but for a broken power-law distribution. The power-law indices are $p_1 = 1.5$ and $p_2 = 3.5$ for the electrons with the Lorentz factor $\gamma_{\rm e, min} \le \gamma_{\rm e} \le \gamma_{\rm e, br}$ and $\gamma_{\rm e, br} < \gamma_{\rm e} \le \gamma_{\rm e, max}$, respectively. Here, we set $\gamma_{\rm e, min} = 10^2$, $\gamma_{\rm e, br} = 10^4$, and $\gamma_{\rm e, max} = 10^7$.}
\label{fig:SED_SSC_nth_BPL}
\end{figure}

First, we present the results for the model with a single power-law distribution of electrons, where the distribution function is shown in eq. \ref{eq:eDF_sPL_pgen}.
The power-law indices is $p = 2.5$ for the Lorentz factor of nonthermal electrons $\gamma_{\rm e, min} \le \gamma_{\rm e} \le \gamma_{\rm e,  max}$,
where $\gamma_{\rm e, min} = 10^2$ and $\gamma_{\rm e,  max} = 10^7$.
Figure \ref{fig:SED_SSC_nth_SPL} shows the resulting SED calculated by \texttt{RAIKOU} (red line). 
As a reference, the SED calculated by a numerical code without based on MC methods \citep[\texttt{AGN SED tool,}][]{Massaro_etal_2006, Tramacere_etal_2009, Tramacere_etal_2011}, in which one-zone models are assumed, is also plotted as black filled circles.
It is found that our result well agrees with the reference. 

The spectral features in Fig. \ref{fig:SED_SSC_nth_SPL} is briefly mentioned here. 
The SED is steep in $\nu \lesssim 10^9$ Hz because of the synchrotron self-absorbed process. 
In $10^{9} \lesssim \nu \lesssim 10^{17.5}$ Hz, the SED formed by the optically-thin, synchrotron emission via the nonthermal electrons with $p=2.5$, i.e., $\nu L_{\nu} \propto (3-p)/2 = 0.25$ appears. 
As is well known, the SEDs due to the synchrotron emission via electrons with a power-law distribution is the same as those due to the inverse Compton scattering of photons via nonthermal electrons with the same distribution function \citep[e.g.,][]{Rybicki_Lightman_1979}. 
In $\nu \gtrsim 10^{19}$ Hz, the spectral bump resulted by the inverse Compton scattering of synchrotron photons via nonthermal electrons (i.e., synchrotron self-Compton) appears. 
Here, it should be noted that the SED formed by the synchrotron self-Compton processes show a bump rather than the simple power-law shape, because the seed photons (i.e., synchrotron photons) have the broadband SEDs which consequently result in the deviation from the simple power-law SED of the Comptonized photons.


Next, we present the results for the model with a broken power-law distribution of electrons, 
where the distribution function is shown in eqs. (\ref{eq:BrPL1}) and (\ref{eq:BrPL2}).
The power-law indices are $p_1 = 1.5$ and $p_2 = 3.5$ for $\gamma_{\rm e,   min} \le \gamma_{\rm e} \le \gamma_{\rm e,  br}$ and $\gamma_{\rm e,  br} \le \gamma_{\rm e} \le \gamma_{\rm e,  max}$, respectively,
where $\gamma_{\rm e,  min} = 10^2$, $\gamma_{\rm e,  br} = 10^4$, and $\gamma_{\rm e,  max} = 10^7$.
Figure \ref{fig:SED_SSC_nth_BPL} shows the resulting SED. As is the calculation with single power-law model, our computed SED well reproduce that calculated by \texttt{AGN SED tool}.
Here, we briefly note the spectral features of calculated SED in Figure \ref{fig:SED_SSC_nth_BPL} . 
As is the same as the single power-law model,
the SED is steep in $\lesssim 10^9$ Hz because of the synchrotron self-absorbed process. 
In $10^{9} \lesssim \nu \lesssim 10^{12.5}$ Hz and $10^{12.5} \lesssim \nu \lesssim 10^{17}$ Hz, the optically-thin, synchrotron emission via the nonthermal electrons with $p_1$ and $p_2$ appears, respectively. 
The spectral indices of $\nu L_{\nu}$ are $(3-p_1)/2 = 0.75$ and $(3-p_2)/2 = -0.25$, respectively.
In $\nu \gtrsim 10^{17}$ Hz, the spectral bump resulted by the synchrotron self-Compton via the nonthermal electrons appears. 

\section{Application to Accreting Kerr  Black Hole based on GRMHD simulations} \label{sec:App}

We show examples of full code tests with applying \texttt{RAIKOU} to the calculation of images and SEDs of the accretion flows, wind, and jets of GRMHD 
simulation data.
We demonstrate the results of a radiatively insufficient accretion flow around a SMBH  where the accretion flows are simulated with by using a GRRMHD code \texttt{UWABAMI}  \citep{Takahashi_etal_2016}.
 The key parameters for the GRRT calculations are described bellow and summarized in Table \ref{tab:GRRT_parameter}.
The GRMHD simulation data are briefly summarized in appendix \ref{appendix:simulation}.


\begin{figure*}[ht!]
\centering
\includegraphics[width=0.95\textwidth]{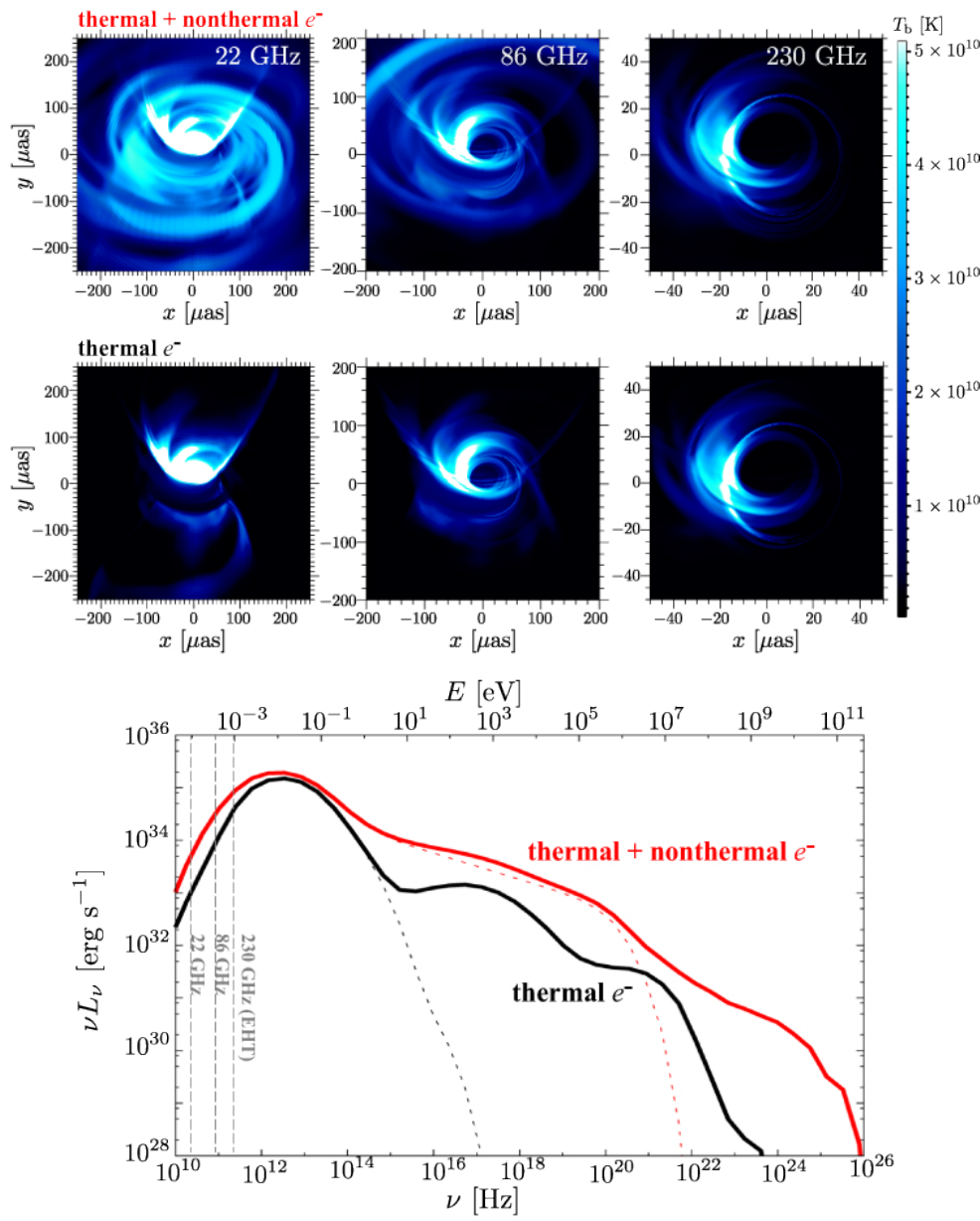}
\caption{(Top): Images of RIAF calculated using GRMHD simulation data  taking into account thermal and nonthermal electrons.
The image is shown in units of brightness temperature.
(Middle): The same as top panels except that the thermal electrons are only taken into account. 
(Bottom): The SEDs with taking into account both of the thermal and nonthermal electorns (red) and thermal electrons only (black). The solid and the dashed lines show the SED with and without the inverse-Compton scattering, respectively. The vertical, gray dashed lines represent the photon frequencies at 22, 86, and 230 GHz.}
\label{fig:image_SED_RIAF}
\end{figure*}




We consider the synchrotron 
emission/abosrption and Compton/inverse-Compton scattering processes via thermal and nonthermal electrons.
The dimensionless spin parameter $a_*$ is set to be 0.9375. 
We need to give the scale of the system, because GRRT calculations are not scale free.
Here, we set the parameters suitable for one of the main targets of EHT, Sgr A*:
the BH mass $M_{\rm BH} = 4.1 \times 10^6 M_{\odot}$, the mass accretion rate ${\dot M} \sim 10^{-8} M_{\odot} ~{\rm y}^{-1}$, and Distance of the observer screen for image calculation $D = 8.1$ kpc.
We set the black hole mass, accretion rate, and distance suitable for one of the main targets of EHT, Sgr A*:
The field-of view (FoV) of the screen is assumed to be $(10^3 \mu{\rm as}, 10^3 \mu{\rm as})$, which is divided by $(2.5 \times 10^3, \, 2.5 \times 10^3)$ pixels.

Because of the uncertainty of the electron temperature in the GRMHD simulations,  
we use a formula describing $T_{p} / T_e$ as a function of the plasma $\beta$, which is the same as \cite{Moscibrodzka_etal_2016,EHTC2019_5}:
\begin{eqnarray}
    \frac{T_p}{T_e} = \frac{1}{1+\beta^2}R_{\rm high} + \frac{\beta^2}{1+\beta^2}R_{\rm low},
\end{eqnarray}
where $R_\mathrm{high} = 5$ and $R_\mathrm{low}=1$.
The distribution function of nonthermal electrons is assumed to be a single power-law  with a power-law index $p = 3.5$. 
The minimum and maximum Lorentz factor of the nonthermal electrons are $\gamma_{\rm e,  min} = 30$ and $\gamma_{\rm e,  max} = 10^6$, respectively.
The number density of nonthermal electrons are set in such a way that their energy density $u_{\rm e,nth}$ is proportional to that of the magnetic energy density $u_B$ at each computational cell:
\begin{eqnarray}
u_{\rm e, nth} = \eta_{\rm nth} u_B,
\end{eqnarray}
where a efficiency parameter ${\eta}_{\rm nth}$ is set to be  0.03 and to be uniform in space in this work.

Figure \ref{fig:image_SED_RIAF} demonstrates images of accretion flow and jet at 22 (left), 86 (center), and 230 GHz (right),
and the broadband SEDs from radio to VHE gamma-ray.
The top/middle panels show the images with/without taking into account the effects of nonthermal electrons.
The size of bright region becomes large
as a consequence of the inclusion of nonthermal electrons, which is consistent with the previous interpretation of the Sgr A* using simple toy models \citep[e.g.,][]{Ozel_2000,Cho_2021}. This is because the emissivity via the synchrotron processes is still high in the outer region, while the plasma temperature is low and the thermal synchrotron emission is very faint there. 
Although the effect of nonthermal electrons is less significant at higher frequency in the radio band, the image taking into account the nonthermal electorns still shows a slightly extended emission image including compared with that without nonthermal electrons at 230 GHz (i.e., at the photon frequency observing black hole shadows by EHT)\footnote{We simply assumed that the energy density of the nonthermal electrons is proportional to the magnetic field energy. If we change the amount of the nonthermal electrons based on the local Maxwell  J\"uttner distribution, the morphology of the resulting image may change (Davelaar et al., private communication)}.

\begin{deluxetable*}{cccccccccccc}
\tablecaption{A set of parameters for the GRRT calculation post-processing the GRMHD simulation.}\label{tab:GRRT_parameter}
\tablewidth{10pc}
\tablehead{
\colhead{$a_*$} & 
\colhead{$M_{\rm BH} [M_{\odot}]$} & \colhead{$D$ [kpc]} & 
\colhead{${\dot M}$ [$M_{\odot} {\rm y}^{-1}$]} &
\colhead{$R_{\rm high}$} & 
\colhead{$R_{\rm low}$} & 
\colhead{$p$} & 
\colhead{$\gamma_{\rm e,  min}$} & \colhead{$\gamma_{\rm e,  max}$}   & \colhead{$\eta_{\rm nth}$}  
& \colhead{$\theta_{\rm ob}$}  & 
\colhead{FoV [$\mu{\rm as} \times \mu {\rm as}$]}
}
\startdata
0.9375 & $4.1 \times 10^6$ & 8.3  &
$\sim 1 \times 10^{-8}$ & 5 & 1 & 
3.5 & 30 & $10^6$ & 0.03 & $45^{\circ}$ &
$10^3 \times 10^3$
 \\
\enddata
 \tablecomments{
 A set of parameters for GRMHD simulations is summarized 
 in Appendix \ref{appendix:simulation}.}
\end{deluxetable*}

The broadband SEDs shown in the bottom panel in Figure \ref{fig:image_SED_RIAF} present significant differences between the models with and without nonthermal electrons. 
As is described above, the luminosity in the lower radio band ${\lesssim}$  100 GHz is increased via the nonthermal synchrotron emission.
Importantly, the significant amount of X-ray photons are generated by the synchrtron emission via nonthermal electrons in addition to the inverse Compton scattering via the thermal electrons.
The SED with thermal and nonthermal electrons (red solid line) at $\sim 10^{17-20}$ Hz is the superposition of the nonthermal synchrotron (red dashed line) plus the SSC via thermal electrons (black solid line).
In the gamma-rays band $\gtrsim 10^{20}$, the significant amount of the photons up to ${\sim}$ TeV are generated by the inverse-Compton scattering due to the nonthermal electrons (red solid line), while a poor amount of gamma-ray photons with $\nu > 10^{22}$ are emitted in the model with thermal electrons only (black solid line).
The power-law shaped SED with the spectral index $\sim (3-p)/2 = -0.25$ expected for synchrotron emmission and inverse-Compton scattering via nonthermal electron  appears in the photon-frequency range $10^{16} \, \lesssim \, \nu \,  \lesssim \, 10^{22}$ Hz and $10^{23} \, \lesssim \, \nu \,  \lesssim \, 10^{25}$ Hz, respectively, while the SED shows slightly complicated structure because of the contribution of the thermal electron processes.

\begin{figure}
\centering
\includegraphics[width=0.98\columnwidth]{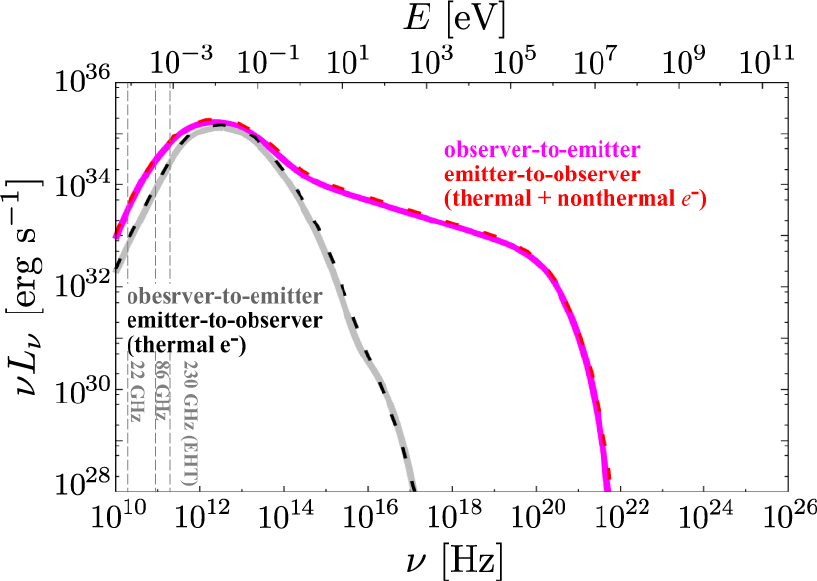}
\caption{Comparison of SED calculated by using the observer-to-emitter (solid lines) and emitter-to-observer algorithm (dashed lines). The results with thermal plus nonthermal electrons are shown in magenta and red color, while the those with thermal electron only are colored in gray and black.} 
\label{fig:SED_e2o_o2e_compare}
\end{figure}

We check the consistency between the two algorithms, i.e., observer-to-emitter and emitter-to-observer algorithms. 
Figure \ref{fig:SED_e2o_o2e_compare} shows the SEDs calculated by using the observer-to-emitter and emitter-to-observer algorithms,
where the Compton effects are not included for the comparison because the observer-to-emitter algorithm cannot solve the Compton scattering. 
The SEDs with the former algorithms are averaged in the azimuthal direction to compare with the latter ones, since our broadband SEDs are computed by summing up over the azimuthal angle $\varphi$ and averaged as in \texttt{grmonty} \citep{Dolence_etal_2009}.
One may find that the SEDs computed by the different methods are mutually-consistent, except a small difference in $\nu \sim 10^{15-16}$ Hz, i.e., the SED are slightly deviated due to the relativistic Doppler effects caused by the relativistic bulk motion in the jet between $10 r_{\rm g} \le r < 30 r_{\rm g}$.
It should be also noted that the deviation is still  $\lesssim (\rm several) \times 10\%$ at highest in the all frequency range of the SEDs, which will be enough to compare the theoretical SEDs with observation data. 
The radio emission, which is crucial for the comparison with the EHT data, is nearly identical between the two different calculation algorithms.

\begin{figure}
\centering
\includegraphics[width=0.98\columnwidth]{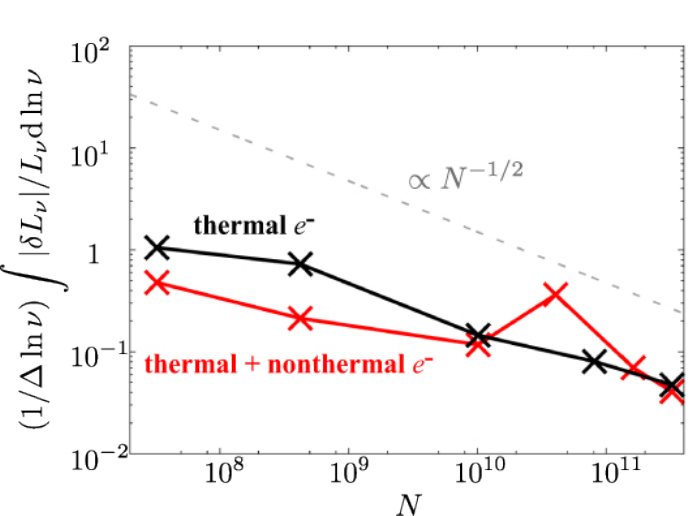}
\caption{Integrated fractional error as a function of the number of generated superphotons for the model with thermal and nonthermal electrons (red) and thermal electrons only (black), respectively. 
The gray dashed line indicates the error proportional to $N^{-1/2}$ as a reference.}
\label{fig:convergence_RIAF}
\end{figure}

The results of the self-convergence test for our broadband SEDs calculations are shown in Figure \ref{fig:convergence_RIAF}. 
We used the SEDs with $N \simeq 4.9\times 10^{11}$ shown in Figure \ref{fig:image_SED_RIAF} for the references, where  $N$ is the total number of superphotons.
As is expected, the relative errors accumulated over the photon-frequency bins decrease roughly as $N^{-1/2}$ for both of the models with and without the nonthermal electrons.


\section{Summary and Discussion} \label{sec:summary}

We have developed a multi-wavelength, general relativistic radiative transfer code \texttt{RAIKOU} (来光) to compute the multi-wavelength images and SEDs of accreting Kerr BHs in the broadband energy bands from radio to VHE gamma-ray, based on a ray-tracing method. 
The two different algorithms are implemented to \texttt{RAIKOU}; an observer-to-emitter algorithm for  efficient calculations of images and an emitter-to-observer algorithm based on a Monte-Carlo method for computations of broadband SEDs by incorporating the inverse-Compton scattering of soft photons via themal and nonthermal electrons.

The radiative processes of cyclo-synchrotron emission/absorption via thermal electrons, synchrotron emission/absorption via thermal and nonthermal electrons, bremsstrahlung emission/absorption via thermal electrons, and inverser-Compton/Compton scattering via thermal and nonthermal electrons are incorporated. 
The implementation of these important radiative processes to the code enabled us to calculate the detailed images and broadband SEDs with the broad energy band ranging from radio to VHE gamma-ray.
The nonthermal synchrotron emission will enlarge the bright region in the resulting images, and X-ray and gamma-ray emissions are significantly increased due to the combination of the nonthermal synchrotron emission and the inverse-Compton scattering via the nonthermal electrons, respectively. 
These will be useful to constrain the physical parameters in  the main target of EHT, i.e., Sgr A* and M87, and other accreting black holes. 
Although we focused on the application to the calculations for radiatively inefficient accretion flow in this paper, our code can be applicable to the studies on the super-Eddington accretion flow around black holes and neutron stars, which will be realized in ultraluminous X-ray sources and tidal disruption events which we leave as future works.

In the current version of our code, the $e^{\pm}$ creation and annihilation via the $\gamma$-$\gamma$ collision, which will be important to study the formation of the $e^{\pm}$ jet \citep{Wong_etal_2021} and the SEDs in VHE gamma-ray bands absorbed by the intrinsic photons and the extragalactic background light (EBL). 
While the latter effect, i.e., the absorption of VHE gamma-ray via the pair creation due to the interaction with EBL, is negligible to study the Galactic Center and nearby low luminosity AGN, which are the main targets of EHT (e.g., Sgr A* and M87), it is important to be incorporated when we extend our study to the SEDs of distant AGNs \citep[e.g.,][]{MAGIC_EBL_2008_Science}.
It should be also mentioned that we focused on the leptonic radiative mechanism. 
The incorporation of the hadronic processes, e.g., the $p$-$p$ and $p$-$\gamma$ collision and successive production of gamma-ray photons via the decay of pions, as well as the production of  high energy neutrinos via the decay of the charged pions.
The direct synchrotron emission via accelerated protons will also contribute to the VHE gamma-ray emissions. 
In our demonstration, the fast-light approximation (i.e., the snapshot data is fixed during the radiative transfer calculation) is used. 
For more precise calculations and the study of the time-variable images and SEDs, it will be important to use the snapshot depending on the physical time of during the GRRT computation.
The inclusion of these effects will be addressed in the near future.

\begin{acknowledgements}
 We thank  M. Kino, I. Cho, K. Kawaguchi, and K. Asano for fruitful discussion.
 The numerical simulations were carried out on the XC50 at the Center for  Computational Astrophysics, National Astronomical Observatory of Japan. This work also used computational resources of the supercomputer Fugaku provided by RIKEN through the HPCI System Research Project (Project ID: hp120286).
 This work was supported by JSPS KAKENHI grant Nos. JP18K13594, 19H01908, 19H01906 (T.K.), 18K03710, 21H04488 (K.O.), 20K11851, 20H01941, 20H00156 (H.R.T).
 This work was also supported in part by MEXT SPIRE, 
  MEXT as "Priority Issue on post-K computer" 
 (Elucidation of the Fundamental Laws and Evolution of the Universe) 
 and as “Program for Promoting Researches on the Supercomputer Fugaku” (Toward a unified view of the universe: from large scale structures to planets), and JICFuS.
\end{acknowledgements}

\vspace{5mm}





 \appendix

\section{Electron Distribution Function}

\subsection{Formalism}

\subsubsection{Maxwell-J\"uttner Distribution}

The distribution function of relativistic thermal electrons, i.e., the Maxwell-J\"uttner distribution, is described as:
\begin{eqnarray}
\frac{{\rm d} n_{\rm e, th}}{{\rm d} \gamma_{\rm e}}  &=& 
\frac{n_{\rm e, th}}{\Theta_{\rm e}} \frac{\gamma_{\rm e}^2 \sqrt{1 - \gamma_{\rm e}^{-2}}}{K_2(\Theta_{\rm e}^{-1})} \exp \left( - \frac{\gamma_{\rm e}}{\Theta_{\rm e}}\right), \label{eq:eDF_MJ}
\label{eq:MJdist}
\end{eqnarray}
where $n_{\rm e,  th}$ is the number density of thermal electrons, $\gamma_{\rm e}$ is the Lorentz factor of the thermal electron motions, $K_2$ is the modified Bessel function of the second kind, $\Theta_{\rm e} = k_{\rm B}T_{\rm e}/m_{\rm e} c^2$ is the dimensionless electron temperature, $T_{\rm e}$ is the electron temperature, $k_{\rm B}$ is the Boltzmann constant, $m_{\rm e}$ is the electron mass, and $c$ is the speed of light.

\subsubsection{Single Power Law Distribution}

The single power-law distribution ${\rm d}n_{ \rm e,  nth}/{\rm d}{\gamma}_{\rm e} ~{\propto}~ \gamma_{ e}^{-p}$ between $\gamma_{\rm e,  min} \le \gamma_{\rm e} \le \gamma_{\rm e,  max}$ is written as 
\begin{eqnarray}
\frac{{\rm d} n_{\rm e,nth}}{{\rm d} \gamma_{\rm e}} &=& \frac{n_{\rm e,nth}(1-p)}{\gamma_{\rm e, max}^{1-p} - \gamma_{\rm e, min}^{1-p}} \gamma_{\rm e}^{-p} ~~ ({\rm for}~~ p\ne 1), 
\label{eq:eDF_sPL_pgen} \\
\frac{{\rm d} n_{\rm e, nth}}{{\rm d} \gamma_{\rm e}} &=& \left[\ln \left(  \frac{\gamma_{\rm e, max}}{\gamma_{\rm e, min}} \right)
\right]^{-1} \gamma_{\rm e}^{-p} ~~ ({\rm for}~~ p = 1). \label{eq:eDF_sPL_p1} 
\end{eqnarray}

\subsubsection{Broken Power Law Distribution}

Alternatively, 
 one can use the broken power-law distribution function, in which the electrons with different power-law indices are distributed as ${\rm d} n_{\rm e, nth}/{\rm d}{\gamma}_{\rm e} ~{\propto}~ \gamma_{\rm e}^{-p_1}$ between the range $\gamma_{{\rm e, min}} \le \gamma_{\rm e} \le \gamma_{{\rm e, br}}$ and 
 ${\rm d} n_{\rm e,nth}/{\rm d}{\gamma}_{\rm e} ~{\propto}~ \gamma_{\rm e}^{-p_2}$ between the range $\gamma_{{\rm e, br}} < \gamma_{\rm e}
 \le \gamma_{{\rm e, max}}$.

 For $p_1 \ne 1$ and  $p_2 \ne 1$, 
 \begin{eqnarray}
 \frac{{\rm d} n_{\rm e,nth}}{{\rm d} \gamma_{\rm e}} &=& {\rm A} n_{\rm e,nth} \gamma_{\rm e}^{-p_1} 
 ~~({\rm for}~~ \gamma_{\rm e, min} \le \gamma_{\rm e} \le \gamma_{\rm e, br}), \label{eq:BrPL1} \\ 
 \frac{{\rm d} n_{\rm e,nth}}{{\rm d} \gamma_{\rm e}} &=& {\rm A} n_{\rm e,nth} \gamma_{\rm e, br}^{-p_1+p_2}\gamma_{\rm e}^{-p_2}
 ~ ~ ( {\rm for}~~ \gamma_{\rm e, br} < \gamma_{\rm e} \le \gamma_{\rm e, max}),  \label{eq:BrPL2}
 \end{eqnarray}
 where
 \begin{eqnarray}
 A \equiv \left[ \frac{\gamma_{\rm e, br}^{1-p_1}- \gamma_{\rm e, min}^{1-p_1}}{1-p_1}
 + \frac{\gamma_{\rm e, br}^{-p_1 + p_2} \left(\gamma_{\rm e, max}^{1-p_2}- \gamma_{\rm e, br}^{1-p_2} \right)}{1-p_2}\right]^{-1}. \label{eq:BrPL_A}
 \end{eqnarray}

\subsection{Monte Carlo Sampling}

\subsubsection{Maxwell-J\"utter Distribution} \label{sec:eDF_MJ_MC}

Using the same procedure as \cite{Canfield_etal_1987}, we sample the electrons with the Maxwell-J\"utter Distribution as following:
\begin{enumerate}
\item We generate a pseudo random number $\xi_1$ to select $i (=3,4,5,6)$ of the function $\pi_i$ in the following equation: 
\begin{eqnarray}
\pi_3 (\Theta_{\rm e}) &=& \frac{1}{S_3(\Theta_{\rm e})} \frac{\sqrt{\pi}}{4}, \\
\pi_4(\Theta_{\rm e}) &=&
\frac{1}{S_3(\Theta_{\rm e})}\frac{1}{2\sqrt{2}} \Theta_{\rm e}^{1/2}, \\
\pi_5(\Theta_{\rm e}) &=&
\frac{1}{S_3(\Theta_{\rm e})},
\frac{3\sqrt{\pi}}{8} \Theta_{\rm e} \\
\pi_6(\Theta_{\rm e}) &=&
\frac{1}{S_3(\Theta_{\rm e})}
 \frac{1}{\sqrt{2}} \Theta_{\rm e}^{3/2}, 
\end{eqnarray}
where $S_3 (\Theta_{\rm e})$ is described as follows.
\begin{eqnarray}
S_3(\Theta_{\rm e}) = \frac{\sqrt{\pi}}{4} + 
\frac{1}{2\sqrt{2}} \Theta_{\rm e}^{1/2}
+ \frac{3\sqrt{\pi}}{8} \Theta_{\rm e}
+ \frac{1}{\sqrt{2}} \Theta_{\rm e}^{3/2}.
\end{eqnarray}
 Obviously,  $\sum^{6}_{i=3} \pi_{i} (\Theta_{\rm e}) = 1$ and $i$ can be determined by using the pseudo random number $0 \le \xi_1 \le 1$.

\item Using the chosen $i$, we generate the random number $\xi_2$ reproducing $g_{i}$, which is the $\chi^2$ distribution of order $i$ 
\begin{eqnarray}
g_3 (\xi_2) &=& \frac{4}{\sqrt{\pi}} ({\xi_2})^2 \exp{[-({\xi_2})^2]}, \\
g_4 (\xi_2) &=& 2 ({\xi_2)}^3 \exp{[-({\xi_2})^2]}, \\
g_5 (\xi_2) &=& \frac{8}{3\sqrt{\pi}} ({\xi_2})^5 \exp{[-({\xi_2})^2]}, \\
g_6 (\xi_2) &=& ({\xi_2})^5 \exp{[-({\xi_2})^2]}, 
\end{eqnarray}
where $\xi_2$ distributes $-\infty < \xi_2 < \infty$.

\item
The generated $\xi_2$ is acceptable if it satisfies
\begin{eqnarray}
\frac{\sqrt{1 + ({\xi_2})^2 \Theta_{\rm e}}}{1 +  \xi_2 \sqrt{\Theta_{\rm e}/2}} \ge \xi_3, 
\end{eqnarray}
where $\xi_3$ ($0 \le \xi_3 \le 1$) is a newly generated random number  for the rejection method.
The Lorentz factor of electron is given by 
\begin{eqnarray}
\gamma_{\rm e} = 1 + ({\xi_2})^2.
\end{eqnarray}
If the above condition is not satisfied, then we go back to the step 1. The procedure repeats until the condition is satisfied.
\end{enumerate}

\subsubsection{Single Power-Law Distribution} \label{sec:eDF_sPL_MC}

On the other hand, for nonthermal electrons with a single/broken power law distribution, the  Lorentz factor of the scattering electrons can be sampled  much easier by the inverse function method.
For a single power law distribution, the Lorentz factor can be represented by generating a pseudo random number $\xi$:
\begin{eqnarray}
\gamma_{\rm e} &=& \gamma_{\rm min} \left[1 + \xi \left\{\left(\frac{\gamma_{\rm e, max}}{\gamma_{\rm e, min}}\right)^{1-p} -1\right\} \right]^{1/(1-p)}  ~~ ({\rm for}~~ p \ne 1), \\
\gamma_{\rm e} &=& \gamma_{\rm min} \left(\frac{\gamma_{\rm e, max}}{\gamma_{\rm e, min}}\right)^{\xi} 
~~ ({\rm for} ~~ p = 1).
\end{eqnarray}

\subsubsection{Broken Power-Law Distribution} \label{sec:eDF_brPL_MC}
For a broken power law distribution, the Lorentz factor can be written as follows for the case with  $p_1 \ne 1$ and $p_2 \ne 1$:
\begin{eqnarray}
\gamma_{\rm e} &=& \left[\gamma_{\rm e, min}^{1-p_1}  + A^{-1} 
\left(1-p_1 \right)  \xi \right]^{1/(1-p_1)} 
~~ ({\rm for} ~~ \xi \le \xi_{\rm br}), \\
\gamma_{\rm e} &=& \left[\gamma_{\rm e, br}^{1-p_2} - \left( 1-p_2 \right)\gamma_{\rm e, br}^{p_1-p_2} \left(\frac{\gamma_{\rm e, min}^{1-p_1} - \gamma_{\rm e, br}^{1-p_1}}{1-p_1} + A^{-1}\xi \right)  \right]^{1/(1-p_2)} 
~~ ({\rm for} ~~ \xi > \xi_{\rm br}), 
\end{eqnarray}
where 
\begin{eqnarray}
\xi_{\rm br} &\equiv& A \frac{\gamma_{\rm e, br}^{1-p_1} - \gamma_{\rm e, min}^{1-p_1}}{1-p_1}, \\
A &\equiv& \left[\frac{\gamma_{\rm e, br}^{1-p_1} - \gamma_{\rm e, min}^{1-p_1}}{1-p_1} + \gamma_{\rm e, br}^{-p_1+p_2} \frac{\gamma_{\rm e, max}^{1-p_2} - \gamma_{\rm e, br}^{1-p_2}}{1-p_2} \right]^{-1}.
\end{eqnarray}

For the case with $p_1 = 1$ and $p_2 \ne 1$, 
\begin{eqnarray}
\gamma_{\rm e} &=& \gamma_{\rm e, min} \exp\left(A^{-1}\xi \right)
~~ ({\rm for} ~~ \xi \le \xi_{\rm br}), \\
\gamma_{\rm e} &=& \left\{\gamma_{\rm e, br}^{1-p_2} - \left( 1-p_2 \right)\gamma_{\rm e, br}^{p_1-p_2} \left[\ln\left(\frac{\gamma_{\rm e, br}}{\gamma_{\rm e, min}}\right) - A^{-1}\xi \right]  \right\}^{1/(1-p_2)} 
~~ ({\rm for}~~ \xi > \xi_{\rm br}), 
\end{eqnarray}
where 
\begin{eqnarray}
\xi_{\rm br} &\equiv& A \ln\left(\frac{\gamma_{\rm e, br}}{\gamma_{\rm e, min}}\right), \\
A &\equiv& \left[\ln\left(\frac{\gamma_{\rm e, br}}{\gamma_{\rm e, min}}\right) + \gamma_{\rm e, br}^{-p_1+p_2} \frac{\gamma_{\rm e, max}^{1-p_2} - \gamma_{\rm e, br}^{1-p_2}}{1-p_2} \right]^{-1}.
\end{eqnarray}

For the case with $p_1 \ne 1$ and $p_2 = 1$, 
\begin{eqnarray}
\gamma_{\rm e} &=& \left[\gamma_{\rm e, min}^{1-p_1}  + A^{-1} 
\left(1-p_1 \right)  \xi \right]^{1/(1-p_1)} ({\rm for}~~ \xi \le \xi_{\rm br}), \\
\gamma_{\rm e} &=& \gamma_{\rm e, br} \exp\left[\gamma_{\rm e, br}^{p_1-p_2}\left( A^{-1}\xi -  \frac{\gamma_{\rm e, br}^{1-p_1} - \gamma_{\rm e, min}^{1-p_1}}{1-p_1} \right) \right] {\rm for}~~ (\xi > \xi_{\rm br}),  
\end{eqnarray}
where 
\begin{eqnarray}
\xi_{\rm br} &\equiv& A \frac{\gamma_{\rm e, br}^{1-p_1} - \gamma_{\rm e, min}^{1-p_1}}{1-p_1}, \\
A &\equiv& \left[\frac{\gamma_{\rm e, br}^{1-p_1} - \gamma_{\rm e, min}^{1-p_1}}{1-p_1}+ \gamma_{\rm e, br}^{-p_1+p_2} \ln\left(\frac{\gamma_{\rm e, max}}{\gamma_{\rm e, br}}\right)  \right]^{-1}.
\end{eqnarray}\\

\section{Emissivity}
\subsection{Formalism} \label{app:emission_form}
\subsubsection{Cyclo-Synchrotron or Synchrotron via Thermal Electron}

For plasma with mildly relativistic temperature, it is important to consider synchrotron emission taking into account the cyclotron emission.
It is convenient to use the fitting formula for the angle-averaded, cyclo-synchrotron emissivity \citep{Mahadevan_etal_1996}:
\begin{eqnarray}
j^{\rm (f)}_{\nu_{\rm f}{\rm(cs)}} &=& \frac{n_{\rm e,th} e^{2} \nu_{\rm f}}{\sqrt{3} c K_2({\Theta_{\rm e}^{-1}})} M(x_{\rm cs}), \\
M(x_{\rm cs}) &\equiv& \frac{4.0505}{x_{\rm cs}^{1/6}}
\left(1 + \frac{0.4}{x_{\rm cs}^{1/4}} + \frac{0.5316}{x_{\rm cs}^{1/2}} \right)      \exp(-1.8899x_{\rm cs}^{1/3}), \\
x_{\rm cs} &\equiv& \frac{\nu_{\rm f}}{\nu_{\rm c}}, \\
\nu_{\rm cs} &\equiv& \left(\frac{3eB}{4\pi m_{\rm e} c} \right) \Theta_{\rm e}^2.
\end{eqnarray}
where $K_2$ is the modified Bessel function of the second kind, $n_{\rm e,th}$ is the thermal electron density measured in the fluid-rest frame, 
$B$ is the magnetic field strength measured in the fluid-rest frame, 
$\Theta_{\rm e} (= k_{\rm B}T_{\rm e}/m_{\rm e}c^2)$ is the dimensionless electron temperature, $T_{\rm e}$ is the electron temperature, $k_B$ is the Boltzmann constant, $m_{\rm e}$ is the rest mass of electron, and $e$ is the elementary charge.

Alternatively, one can use a synchrotron formulation, which depends on the angle between the photon and magnetic field meashered in the fluid-rest frame $\theta_B$, for high temperature plasma ${\Theta} \gtrsim 0.5$ \citep{Leung_etal_2011}:
\begin{eqnarray}
j^{\rm (f)}_{\nu_{\rm f}{\rm(sy,th)}} & {\simeq} &  \frac{{\sqrt{2} \pi n_{\rm e,th} e^{2} \nu_{\rm s}}}
{3 c K_2({\Theta_{\rm e}^{-1}})}
\left(x_{\rm s}^{1/2} + 2^{11/12} x_{\rm s}^{1/6}\right)^2 \exp(-x_{\rm s}^{1/3}), \\
x_{\rm s} &\equiv& \frac{\nu_{\rm f}}{\nu_{\rm s}}, \\
\nu_{\rm s} &\equiv& \frac{2}{9} \left( \frac{eB}{2\pi m_{\rm e} c} \right)
\Theta_{\rm e}^2 \sin \theta_{B},
\end{eqnarray}
where 
\begin{eqnarray}
\cos \theta_{B} = -\frac{b^{\mu}p_{\mu}}{B p^{(f)}_0}.
\end{eqnarray}

\subsubsection{Synchrotron via Nonthermal Electron}

Synchrotron emission via the nonthermal electrons with single/broken power-law distribution functions are implemented.

The single power-law distribution ${\rm d}n_{\rm e,nth}/{\rm d}{\gamma}_e ~{\propto}~ \gamma_e^{-p}$ between the range $\gamma_{e,{\rm min}} \le \gamma_e \le \gamma_{e,{\rm max}}$ is written as 
\begin{eqnarray}
\frac{{\rm d} n_{\rm e,nth}}{{\rm d} \gamma_{\rm e}} &=& \frac{n_{\rm e,nth}(1-p)}{\gamma_{e,{\rm max}}^{1-p} - \gamma_{e,{\rm min}}^{1-p}} \gamma_{\rm e}^{-p} ~ ({\rm for}~~ p\ne 1), \\
\frac{{\rm d} n_{\rm e,nth}}{{\rm d} \gamma_{\rm e}} &=& \left[\ln \left(  \frac{\gamma_{e,{\rm max}}}{\gamma_{e,{\rm min}}} \right)
\right]^{-1} \gamma_{\rm e}^{-p} ~ ({\rm for}~~ p = 1).
\end{eqnarray}

The synchrotron emissivity is formulated as follows \citep[see, e.g.,][]{Dexter_etal_2012,Dexter_2016_GRTRANS}:
\begin{eqnarray}
j^{\rm (f)}_{\nu_{\rm f}{(\rm sy, nth)}} &=& \frac{(p-1)n_{{\rm e,  nth}}e^2 \nu_{\rm c}}{2\sqrt{3}c (\gamma_{e,{\rm min}}^{1-p} - \gamma_{e,{\rm max}}^{1-p})}
\left(\frac{\nu_{\rm f}}{\nu_{\rm c}}
\right)^{-(p-1)/2} 
\left[
G(x_{\rm max}) - G(x_{\rm min})
\right], \label{eq:sync_sPL}\\
G(x) &=& \int_x^{\infty} z^{(p-3)/2} F(z) {\rm d}z, \\
F(z) &=& z\int_z^{\infty} K_{5/3}(y) {\rm d}y,  
\end{eqnarray}
where $\nu_{\rm c} = 3eB \sin \theta_{B}/(4 \pi m_{\rm e} c)$,  $x = \nu/(\gamma_{\rm e}^2 \nu_{\rm c})$, $x_{\rm min(max)} =  \nu/(\gamma_{\rm e, min(max)}^2 \nu_{\rm c})$.

The double integral in the eq. (\ref{eq:sync_sPL}) can be reduced to the single integral \citep{Westfold_1959}:
\begin{eqnarray}
&&\int_x^{\infty} {\rm d} z z^{s-1} \int_z^{\infty}{\rm d}y K_{t+ 1}(y) 
= \frac{s+t}{s}
\int_x^{\infty}K_{t}(z) 
- \frac{x^s}{s} 
\left[
\int_x^{\infty} K_{t+1}(y) {\rm d}y - K_t(x)
\right].
\end{eqnarray}

 Alternatively, 
the broken power-law distribution function is also used to phenomenologically study the nonthermal emission, in which the electrons with different power-law index are distributed as $dn_{\rm e,nth}/d{\gamma}_e ~{\propto}~ \gamma_e^{-p_1}$ between the range $\gamma_{\rm e,min} \le \gamma_e \le \gamma_{\rm e, br}$ and 
 $dn_{\rm e,nth}/d{\gamma}_e ~{\propto}~ \gamma_e^{-p_2}$ between the range $\gamma_{\rm e, br} < \gamma_{\rm, e} \le \gamma_{\rm e max}$.

 For $p_1 \ne 1$ and  $p_2 \ne 1$, 
 \begin{eqnarray}
 \frac{{\rm d} n_{\rm e,nth}}{{\rm d} \gamma_{\rm e}} &=& {\rm A} n_{\rm e,nth} \gamma_{\rm e}^{-p_1} 
 ~~({\rm for}~~ \gamma_{\rm e, min} \le \gamma_{\rm e} \le \gamma_{\rm e, br}), \\
 \frac{{\rm d} n_{\rm e,nth}}{{\rm d} \gamma_{\rm e}} &=& {\rm A} n_{\rm e,nth} \gamma_{\rm e, br}^{-p_1+p_2}\gamma_{\rm e}^{-p_2}
 ~ ~ ({\rm for}~~ \gamma_{\rm e, br} < \gamma_{\rm e} \le \gamma_{\rm e, max}), 
 \end{eqnarray}
 where
 \begin{eqnarray}
 A \equiv \left[ \frac{\gamma_{\rm e, br}^{1-p_1}- \gamma_{\rm e, min}^{1-p_1}}{1-p_1}
 + \frac{\gamma_{\rm e, br}^{-p_1 + p_2} \left(\gamma_{\rm e, max}^{1-p_2}- \gamma_{\rm e, br}^{1-p_2} \right)}{1-p_2}\right]^{-1}. \label{eq:brPL_A}
 \end{eqnarray}

The synchrotron emissivity can be written as
\begin{eqnarray}
&&j^{\rm (f)}_{\nu_{\rm f}{(\rm sy, nth)}} =  A \frac{n_{{\rm e,  nth}}e^2 \nu_{\rm c}}{2\sqrt{3}c}
\left\{\left(\frac{\nu_{\rm f}}{\nu_{\rm c}}
\right)^{-(p_1-1)/2} 
\left[
G(x_{\rm br}) - G(x_{\rm min})\right] 
+ \gamma_{\rm e, br}^{-p_1+p_2}
\left(\frac{\nu}{\nu_{\rm c}}
\right)^{-(p_2-1)/2} \left[
G(x_{\rm max}) - G(x_{\rm br})\right]\right\}, \label{eq:sync_brPL}
\end{eqnarray}



\subsubsection{Bremsstrahlung  via Thermal Electron}
For relativistic temperature plasma, both of the electron-proton and electron-electron bremsstrahlung are important.
Here, we describe the bremsstrahlung emissivity via thermal electrons \citep[see, e.g.,][]{Svensson_1982,Stepney_Guilbert_1983,Narayan_Yi_1995, Manmoto_1997}.
The total bremsstrahlung emissivity $j^{\rm (f)}_{\nu_{\rm f}{(\rm brms)}}$ can be written as
\begin{eqnarray}
j^{\rm (f)}_{\nu_{\rm f} ({\rm brms})} 
= \frac{q^{-}_{\rm brms}f_{\rm Gaunt}}{4\pi} \exp\left(\frac{h \nu_{\rm f}}{k_{\rm B}T_{\rm e}}\right)
,
\end{eqnarray}
where $f_{\rm Gaunt}$  is the  Gaunt factor given by
\begin{eqnarray}
f_{\rm Gaunt} &=& \frac{h}{k_{\rm B}T_{\rm e}} 
\left(\frac{3}{\pi} \frac{k_{\rm B}T_{\rm e}}{h\nu_{\rm f}}\right)^{1/2}
~\left({\rm for}~~ \frac{k_{\rm B}T_{\rm e}}{h\nu_{\rm f}} < 1 \right), \\
f_{\rm Gaunt} &=& \frac{h}{k_{\rm B}T_{\rm e}} 
\frac{\sqrt{3}}{\pi}
\ln \left(\frac{4}{\zeta} \frac{k_{\rm B}T_{\rm e}}{h\nu_{\rm f}}\right) ~\left({\rm for}~~ \frac{k_{\rm B}T_{\rm e}}{h\nu_{\rm f}} \ge 1 \right),
\end{eqnarray}
and the bremsstrahlung cooling rate per unit volume $q^{-}_{\rm brms}$ can be written as
 \begin{eqnarray}
 q^{-}_{\rm brms} = q^{-}_{\rm brms, ep} + q^{-}_{\rm brms, ee}.
 \end{eqnarray}
 Here, $q^{-}_{\rm brms, ep} $ and $q^{-}_{\rm brms, ee}$ are the rate for  electron-proton and electron-electron bremmstrahlung, respectively. 
 These can be formulated as follows:
  \begin{eqnarray}
 q^{-}_{\rm brms, ep} = 1.25~ n_{\rm e, th}^2 \sigma_{\rm T} \alpha_{\rm fs} m_{\rm e} c^3 F_{\rm ep}(\Theta_{\rm e}),
 \end{eqnarray}
 where $\alpha_{\rm fs}$ and $\sigma_{\rm T}$ are the fine structure constant and the Thomsom scattering cross section, and  $F_{\rm ep}(\Theta_{\rm e})$ is
   \begin{eqnarray}
  F_{\rm ep}(\Theta_{\rm e}) = 4\left( \frac{2\Theta_{\rm e}}{\pi^3}\right)^{1/2} 
  \left( 1 + 1.781 \Theta_{\rm e}^{1.34} \right) 
  ~   \left({\rm for}~~ \Theta_{\rm e} < 1 \right), \\
  F_{\rm ep}(\Theta_{\rm e}) =  \frac{9\Theta_{\rm e}}{2\pi} 
  \left[ \ln \left( 1.123 \Theta_{\rm e} + 0,48\right)+1.5\right] 
  ~ \left({\rm for}~~ \Theta_{\rm e} \ge 1 \right),
 \end{eqnarray}
and 
  \begin{eqnarray}
 q^{-}_{\rm brms, ee} = n_{\rm e, th}^2 r_{\rm e}^2 \alpha_{\rm fs} c^3 
 \frac{20}{9\pi^{1/2}}\left(44 - 3\pi^2\right) \Theta_{\rm e}^{3/2}  \left(1 + 1.1\Theta_{\rm e}  + \Theta_{\rm e}^2 - 1.25\Theta_{\rm e}^{5/2}\right) ~ \left({\rm for}~~ \Theta_{\rm e} < 1 \right), \\
 q^{-}_{\rm brms, ee} = n_{\rm e, th}^2 r_{\rm e}^2 \alpha_{\rm fs} c^3 
 24 \Theta_{\rm e}^{3/2} \left[ \ln \left(1.1232 \Theta_{\rm e} \right)+ 1.28 \right] ~ \left({\rm for}~~ \Theta_{\rm e} \ge 1 \right),
 \end{eqnarray}
where $r_{\rm e}$ is the classical electron radius.


\subsection{MC sampling}

In general, there are mainly two ways to sample the photon emissions with MC methods: (i) superphotons with an uniform weight are generated in such a way that their number is proportional to the emissivity
of plasma among computational mesh, 
or (ii) the weight of superphotons is proportional to the emissivity with generating a uniform number of superphotons among computational meshes.

We employ the method (ii)  because we consider various emission processes, i.e., thermal/nonthermal synchrotorn and bremsstrahlung emission.
We can simply sum up the weight of superphotons when they are generated. The direction of the generated superphotons is sampled to be isotropic using uniform pseudo-random numbers. It should be emphasized that the dependence of the synchrotron emissivity on the direction are automatically incorporated by this method, because the weight of superphotons is set to be proportional to the emissivity.

Our procedure is summarized as follows:
\begin{enumerate}
    \item Superphotons are generated in such a way that their direction becomes isotropic in the fluid-rest frame, using uniform pseudo-random numbers.
    \item The weight of superphotons for each radiative process $w$, which is calculated by using eq. (\ref{eq:weight}), are summed up and the total weight of superphoton $w_{\rm tot}$ is comupted as follows:
    \begin{eqnarray}
        w_{\rm tot} = \sum w,
    \end{eqnarray}
    e.g., if we consider the synchrotron emission via thermal and nonthermal electrons and thermal bremsstrahlung emission, then
    \begin{eqnarray}
        w_{\rm tot} =  w_{\rm sy, th} + w_{\rm sy, nth} +  w_{\rm brms},
    \end{eqnarray}
        where $w_{\rm sy, th}$,  $w_{\rm sy, nth}$, and  $w_{\rm brms}$ are the weight of superphotons generated by the synchrotron emission via thermal electrons, and that via nonthermal electrons, and bremsstrahlung, respectively.
    \item The direction and energy of superphotons in the fluid-rest frame are transformed to those in the observer frame to integrate the geodesic equations with the radiative processes (i.e., the general relativistic radiative transfer equations).
\end{enumerate}





\section{Absorption Coefficient} \label{app:absorb_form}
For the radiative processes via thermal electrons, the absorption coefficients are obtained by the Kirchhoff's law, i.e.,
\begin{eqnarray}
\alpha^{\rm (f)}_{\nu_{\rm f}} = \frac{j^{\rm (f)}_{\nu_{\rm f}}}{B_{\nu_{\rm f}}}, \label{eq:alpha_th}
\end{eqnarray}
where $B_{\nu_{\rm f}}$ is the blackbody intensity of the plasma at which the radiative process takes place.

On the other hand, for the radiative processes via nonthermal electrons, the absorption coefficient are obtained by considering more generalized picture of the detailed balance between emission and absorption with the Einstein relation including continuum states
 \citep[see, e.g.,][]{Rybicki_Lightman_1979}.
The absorption coefficients for the nonthermal electrons with a single-power law distribution is 
\begin{eqnarray}
\alpha^{\rm (f)}_{\nu_{\rm f}({\rm sy, nth})} &=&
\frac{(p-1) (p+2) n_{{\rm e,  nth}}e^2 \nu_{\rm c}}{4\sqrt{3}c (\gamma_{e,{\rm min}}^{1-p} - \gamma_{e,{\rm max}}^{1-p})}
\left(\frac{\nu_{\rm f}}{\nu_{\rm c}}
\right)^{-(p+4)/2} 
\left[
G_a(x_{\rm max}) - G_a(x_{\rm min})
\right], \label{eq:alpha_sPL}\\
G_a(x) &=& \int_x^{\infty} z^{(p-2)/2} F(z) {\rm d}z. 
\end{eqnarray}

For the nonthermal electrons with a broken power-law distribution, the absorption coefficient is represented as follows:
\begin{eqnarray}
\alpha^{\rm (f)}_{\nu_{\rm f}({\rm sy, nth})} =
A \frac{n_{{\rm e,  nth}}e^2 }{4\sqrt{3}m_{\rm e}c \nu_{\rm c}}
&& \left\{
 (p_1+2) \left(\frac{\nu_{\rm f}}{\nu_{\rm c}}
\right)^{-(p_1+4)/2} 
\left[
G_a(x_{\rm br}) - G_a(x_{\rm min})\right] \right. \nonumber \\
&& \left. + (p_2+2) \gamma_{\rm e, br}^{-p_1+p_2}
\left(\frac{\nu_{\rm f}}{\nu_{\rm c}}
\right)^{-(p_2+4)/2} \left[
G_a(x_{\rm max}) - G_a(x_{\rm br})\right]\right\}
, \label{eq:alpha_brPL}
\end{eqnarray}
where $A$ is given in eq. (\ref{eq:brPL_A}).

\section{GRMHD simulation data} \label{appendix:simulation}

Three-dimensional GRMHD simulations are carried out by using GRRMHD simulation code \texttt{UWABAMI} in order to calculate the SEDs and images in \S \ref{sec:App}.
The radiation terms are not solved in this study since we are focusing on low mass accretion rate system, in which the two-temperature equations should be solved when we include radiation terms and this is beyond the scope of this paper.
The BH spin is set to be $a_* = 0.9375$.
The inner- and outer-outflow boundaries are located at $1.18 r_{\rm g}$ and $3.33 \times 10^3 r_{\rm g}$, respectively, where the former appears inside the event horizon of the black hole $r_{\rm H} = (1  + \sqrt{1 - a_{*}^2})r_{\rm g} \simeq 1.35 r_{\rm g}$.
The simulation domain is divided into $(N_r, N_\theta, N_\varphi) = (192, 120, 96)$ meshes, where $N_r$, $N_\theta$, $N_\varphi$ are the numbers of meshes in $r$, $\theta$, and $\varphi$ direction of the modified Kerr-Schild coordinate.

\begin{figure*}[ht!]
\centering
\includegraphics[width=0.99\textwidth]{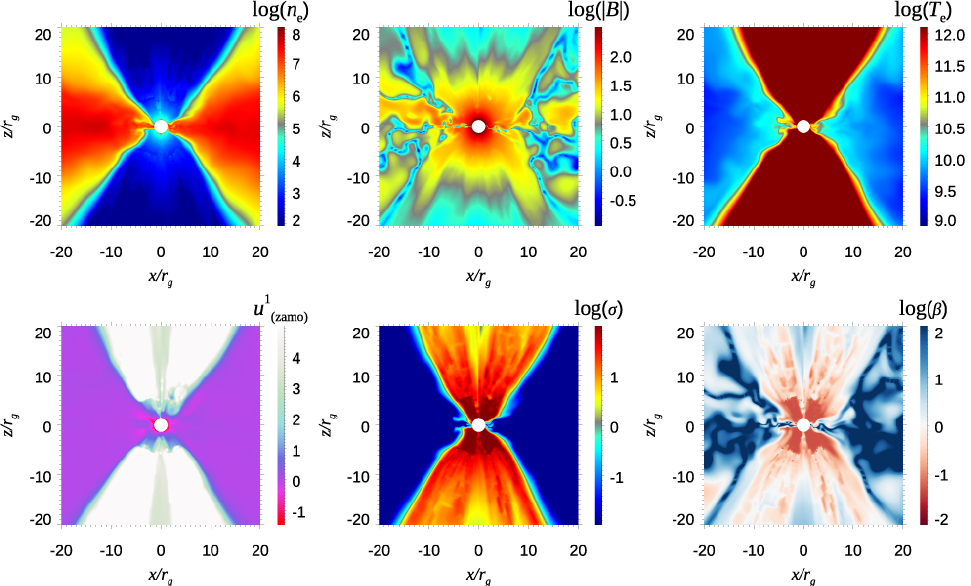}
\caption{Spatial distribution of electron number density (top-left), magnetic field strength (top-middle), electron temperature (top-right), radial component of  four-velocity in the ZAMO frame (bottom-left),  magnetization $\sigma = B^2/\rho c^2$ (bottom-middle), and plasma-$\beta$ (bottom-right).}
\label{fig:GRMHD_structure}
\end{figure*}

Initially, we set an isentropic hydroequilibrium torus rotating around the Kerr BH \citep{Fishbone_Moncrief_1976}, which is emmeded in a hot, static, uniform, and non-magnetized ambient gas.
The position of the inner edge and the pressure maximum of the torus are  set at $r = 20 r_{\rm g}$ and $r = 33 r_{\rm g}$ on the equatorial plane, respectively.
The specific heat ratio is assumed to be $\gamma_{\rm heat} = 13/9$.
After the growth of the magneto-rotational instability \citep{Balbus_Hawley_1991}, the accretion flow is formed via the angular momentum transport.
We use the snapshot at $t = 1.2 \times 10^4 r_{\rm g}/c$, in which the accretion flow in the quasi-steady state.
The structure of the accretion flow at $t = 1.2 \times 10^4 r_{\rm g}/c$ is displayed in  Fig.\ref{fig:GRMHD_structure}. 

The accretion flow is roughly categorized into two states by the magnitude of their magnetization: 
SANE (Standard And Normal Evolution)  and MAD (Magnetically Arrested Disk), which are weakly and strongly magnetized states, respectively \citep[e.g.,][ and references therein]{Narayan_etal_2012,Sadowski_etal_2013,Tchekhovskoy_2015,EHTC2019_5}.
The SANE and MAD are defined by 
the dimensionless magnetic flux threading the event horizon, which is described as
\begin{eqnarray}
\phi_{\rm BH} = \frac{\Phi_{\rm BH}}{\sqrt{{\dot M}_{\rm BH} r_{\rm g} c}},
\label{eq:phi_BH}
\end{eqnarray}
 where the magnetic flux $\Phi_{\rm BH}$ and the mass accretion rate onto the BH ${\dot M}_{\rm BH}$ is 
 \begin{eqnarray}
 \Phi_{\rm BH} = \frac{1}{2} \int_\theta \int_{\varphi} |B^r| dA_{\theta \varphi}, \\
 {\dot M}_{\rm BH} = \int_\theta \int_{\varphi} \rho u^r dA_{\theta \varphi}.
 \end{eqnarray}
 Here, $dA_{\theta \varphi} = \sqrt{-g} d\theta d\varphi$ is an area element in the $\theta$--$\varphi$ plane and $g$ is the determinant of the Kerr-Schild metric.
 In SANE and MAD states, $\Phi_{\rm BH} \sim 1$, and $\sim$ 15, respectively.
 The normalized magnetic flux is $\Phi_{\rm BH} \approx 5$ in our simulation, so that the accretion flow is categorized to the state between SANE and MAD, which is sometimes referred as semi-MAD \footnote{This corresponds to the normalized magnetic flux $\approx ~ 18$ in the formulation in \cite{Tchekhovskoy_etal_2011}, in which the magnetic flux is multiplied by $\sqrt{4\pi}$.} although the magnetic flux is slightly high.

\bibliography{GRRT.bib,GRRT2.bib,RHD.bib}{}

\begin{thebibliography}{}
\expandafter\ifx\csname natexlab\endcsname\relax\def\natexlab#1{#1}\fi
\providecommand{\url}[1]{\href{#1}{#1}}
\providecommand{\dodoi}[1]{doi:~\href{http://doi.org/#1}{\nolinkurl{#1}}}
\providecommand{\doeprint}[1]{\href{http://ascl.net/#1}{\nolinkurl{http://ascl.net/#1}}}
\providecommand{\doarXiv}[1]{\href{https://arxiv.org/abs/#1}{\nolinkurl{https://arxiv.org/abs/#1}}}

\bibitem[{{Abramowicz} {et~al.}(1988){Abramowicz}, {Czerny}, {Lasota}, \&
  {Szuszkiewicz}}]{Abramowicz_etal_1988}
{Abramowicz}, M.~A., {Czerny}, B., {Lasota}, J.~P., \& {Szuszkiewicz}, E. 1988,
  \apj, 332, 646, \dodoi{10.1086/166683}

\bibitem[{{Aleksi{\'c}} {et~al.}(2014){Aleksi{\'c}}, {Ansoldi}, {Antonelli},
  {Antoranz}, {Babic}, {Bangale}, {Barrio}, {Gonz{\'a}lez}, {Bednarek},
  {Bernardini}, {Biasuzzi}, {Biland}, {Blanch}, {Bonnefoy}, {Bonnoli},
  {Borracci}, {Bretz}, {Carmona}, {Carosi}, {Colin}, {Colombo}, {Contreras},
  {Cortina}, {Covino}, {Da Vela}, {Dazzi}, {De Angelis}, {De Caneva}, {De
  Lotto}, {Wilhelmi}, {Mendez}, {Prester}, {Dorner}, {Doro}, {Einecke},
  {Eisenacher}, {Elsaesser}, {Fonseca}, {Font}, {Frantzen}, {Fruck}, {Galindo},
  {L{\'o}pez}, {Garczarczyk}, {Terrats}, {Gaug}, {Godinovi{\'c}}, {Mu{\~n}oz},
  {Gozzini}, {Hadasch}, {Hanabata}, {Hayashida}, {Herrera}, {Hildebrand },
  {Hose}, {Hrupec}, {Idec}, {Kadenius}, {Kellermann}, {Kodani}, {Konno},
  {Krause}, {Kubo}, {Kushida}, {La Barbera}, {Lelas}, {Lewandowska},
  {Lindfors}, {Lombardi}, {Longo}, {L{\'o}pez}, {L{\'o}pez-Coto},
  {L{\'o}pez-Oramas}, {Lorenz}, {Lozano}, {Makariev}, {Mallot}, {Maneva},
  {Mankuzhiyil}, {Mannheim}, {Maraschi}, {Marcote}, {Mariotti},
  {Mart{\'\i}nez}, {Mazin}, {Menzel}, {Mirand a}, {Mirzoyan}, {Moralejo},
  {Munar-Adrover}, {Nakajima}, {Niedzwiecki}, {Nilsson}, {Nishijima}, {Noda},
  {Orito}, {Overkemping}, {Paiano}, {Palatiello}, {Paneque}, {Paoletti},
  {Paredes}, {Paredes-Fortuny}, {Persic}, {Poutanen}, {Moroni}, {Prandini},
  {Puljak}, {Reinthal}, {Rhode}, {Rib{\'o}}, {Rico}, {Garcia}, {R{\"u}gamer},
  {Saito}, {Saito}, {Satalecka}, {Scalzotto}, {Scapin}, {Schultz}, {Schweizer},
  {Shore}, {Sillanp{\"a}{\"a}}, {Sitarek}, {Snidaric}, {Sobczynska}, {Spanier},
  {Stamatescu}, {Stamerra}, {Steinbring}, {Storz}, {Strzys}, {Takalo},
  {Takami}, {Tavecchio}, {Temnikov}, {Terzi{\'c}}, {Tescaro}, {Teshima},
  {Thaele}, {Tibolla}, {Torres}, {Toyama}, {Treves}, {Uellenbeck}, {Vogler},
  {Zanin}, {Kadler}, {Schulz}, {Ros}, {Bach}, {Krau{\ss}}, \&
  {Wilms}}]{MAGIC_2014Science_IC310}
{Aleksi{\'c}}, J., {Ansoldi}, S., {Antonelli}, L.~A., {et~al.} 2014, Science,
  346, 1080, \dodoi{10.1126/science.1256183}

\bibitem[{{Balbus} \& {Hawley}(1991)}]{Balbus_Hawley_1991}
{Balbus}, S.~A., \& {Hawley}, J.~F. 1991, \apj, 376, 214,
  \dodoi{10.1086/170270}

\bibitem[{{Ball} {et~al.}(2016){Ball}, {{\"O}zel}, {Psaltis}, \&
  {Chan}}]{Ball_etal_2016}
{Ball}, D., {{\"O}zel}, F., {Psaltis}, D., \& {Chan}, C.-k. 2016, \apj, 826,
  77, \dodoi{10.3847/0004-637X/826/1/77}

\bibitem[{{Bardeen}(1973)}]{Bardeen_1973}
{Bardeen}, J.~M. 1973, in Black Holes (Les Astres Occlus), ed. C.~{Dewitt} \&
  B.~S. {Dewitt}, 241--289

\bibitem[{{Bardeen} {et~al.}(1972){Bardeen}, {Press}, \&
  {Teukolsky}}]{Bardeen_etal_1972}
{Bardeen}, J.~M., {Press}, W.~H., \& {Teukolsky}, S.~A. 1972, \apj, 178, 347,
  \dodoi{10.1086/151796}

\bibitem[{{Blandford} \& {Payne}(1982)}]{Blandford_Payne_1982}
{Blandford}, R.~D., \& {Payne}, D.~G. 1982, \mnras, 199, 883,
  \dodoi{10.1093/mnras/199.4.883}

\bibitem[{{Blandford} \& {Znajek}(1977)}]{Blandford_Znajek_1977}
{Blandford}, R.~D., \& {Znajek}, R.~L. 1977, \mnras, 179, 433,
  \dodoi{10.1093/mnras/179.3.433}

\bibitem[{{Broderick} \& {Loeb}(2006)}]{Broderick_Loeb_2006}
{Broderick}, A.~E., \& {Loeb}, A. 2006, \mnras, 367, 905,
  \dodoi{10.1111/j.1365-2966.2006.10152.x}

\bibitem[{{Canfield} {et~al.}(1987){Canfield}, {Howard}, \&
  {Liang}}]{Canfield_etal_1987}
{Canfield}, E., {Howard}, W.~M., \& {Liang}, E.~P. 1987, \apj, 323, 565,
  \dodoi{10.1086/165853}

\bibitem[{{Chael} {et~al.}(2021){Chael}, {Johnson}, \&
  {Lupsasca}}]{Chael_etal_2021_arXiv}
{Chael}, A., {Johnson}, M.~D., \& {Lupsasca}, A. 2021, arXiv e-prints,
  arXiv:2106.00683.
\newblock \doarXiv{2106.00683}

\bibitem[{{Chael} {et~al.}(2018){Chael}, {Johnson}, {Bouman}, {Blackburn},
  {Akiyama}, \& {Narayan}}]{Chael_etal_2018}
{Chael}, A.~A., {Johnson}, M.~D., {Bouman}, K.~L., {et~al.} 2018, \apj, 857,
  23, \dodoi{10.3847/1538-4357/aab6a8}

\bibitem[{{Chael} {et~al.}(2016){Chael}, {Johnson}, {Narayan}, {Doeleman},
  {Wardle}, \& {Bouman}}]{Chael_etal_2016}
{Chael}, A.~A., {Johnson}, M.~D., {Narayan}, R., {et~al.} 2016, \apj, 829, 11,
  \dodoi{10.3847/0004-637X/829/1/11}

\bibitem[{{Chael} {et~al.}(2017){Chael}, {Narayan}, \&
  {Sadowski}}]{Chael_etal_2017}
{Chael}, A.~A., {Narayan}, R., \& {Sadowski}, A. 2017, \mnras, 470, 2367,
  \dodoi{10.1093/mnras/stx1345}

\bibitem[{{Chandrasekhar}(1983)}]{Chandrasekahr_1983}
{Chandrasekhar}, S. 1983, in General Relativity and Gravitation, Volume 1, ed.
  B.~{Bertotti}, F.~{de Felice}, \& A.~{Pascolini}, Vol.~1, 6

\bibitem[{{Chatterjee} {et~al.}(2020){Chatterjee}, {Markoff}, {Neilsen},
  {Younsi}, {Witzel}, {Tchekhovskoy}, {Yoon}, {Ingram}, {van der Klis},
  {Boyce}, {Do}, {Haggard}, \& {Nowak}}]{Chatterjee_2020_arxiv}
{Chatterjee}, K., {Markoff}, S., {Neilsen}, J., {et~al.} 2020, arXiv e-prints,
  arXiv:2011.08904.
\newblock \doarXiv{2011.08904}

\bibitem[{{Cho} {et~al.}(2021){Cho}, {Zhao}, {Kawashima}, {Kino}, \& {et
  al.}}]{Cho_2021}
{Cho}, I., {Zhao}, G., {Kawashima}, T., {Kino}, M., \& {et al.} 2021, to be
  submitted to \apj

\bibitem[{{Curd} \& {Narayan}(2019)}]{Curd_Narayan_2019}
{Curd}, B., \& {Narayan}, R. 2019, \mnras, 483, 565,
  \dodoi{10.1093/mnras/sty3134}

\bibitem[{{Davelaar} {et~al.}(2018){Davelaar}, {Mo{\'s}cibrodzka}, {Bronzwaer},
  \& {Falcke}}]{Davelaar_etal_2018}
{Davelaar}, J., {Mo{\'s}cibrodzka}, M., {Bronzwaer}, T., \& {Falcke}, H. 2018,
  \aap, 612, A34, \dodoi{10.1051/0004-6361/201732025}

\bibitem[{{Davis} {et~al.}(2009){Davis}, {Blaes}, {Hirose}, \&
  {Krolik}}]{Davis_etal_2009}
{Davis}, S.~W., {Blaes}, O.~M., {Hirose}, S., \& {Krolik}, J.~H. 2009, \apj,
  703, 569, \dodoi{10.1088/0004-637X/703/1/569}

\bibitem[{{Dexter}(2016)}]{Dexter_2016_GRTRANS}
{Dexter}, J. 2016, \mnras, 462, 115, \dodoi{10.1093/mnras/stw1526}

\bibitem[{{Dexter} {et~al.}(2012){Dexter}, {McKinney}, \&
  {Agol}}]{Dexter_etal_2012}
{Dexter}, J., {McKinney}, J.~C., \& {Agol}, E. 2012, \mnras, 421, 1517,
  \dodoi{10.1111/j.1365-2966.2012.20409.x}

\bibitem[{{Dolence} {et~al.}(2009){Dolence}, {Gammie}, {Mo{\'s}cibrodzka}, \&
  {Leung}}]{Dolence_etal_2009}
{Dolence}, J.~C., {Gammie}, C.~F., {Mo{\'s}cibrodzka}, M., \& {Leung}, P.~K.
  2009, \apjs, 184, 387, \dodoi{10.1088/0067-0049/184/2/387}

\bibitem[{{EHT MWL Science Working Group} {et~al.}(2021){EHT MWL Science
  Working Group}, {Algaba}, {Anczarski}, {Asada}, {Balokovi{\'c}}, {Chandra},
  {Cui}, {Falcone}, {Giroletti}, {Goddi}, {Hada}, {Haggard}, {Jorstad}, {Kaur},
  {Kawashima}, {Keating}, {Kim}, {Kino}, {Komossa}, {Kravchenko}, {Krichbaum},
  {Lee}, {Lu}, {Lucchini}, {Markoff}, {Neilsen}, {Nowak}, {Park}, {Principe},
  {Ramakrishnan}, {Reynolds}, {Sasada}, {Savchenko}, {Williamson}, {Event
  Horizon Telescope Collaboration}, {Akiyama}, {Alberdi}, {Alef}, {Anantua},
  {Azulay}, {Baczko}, {Ball}, {Barrett}, {Bintley}, {Benson}, {Blackburn},
  {Blundell}, {Boland}, {Bouman}, {Bower}, {Boyce}, {Bremer}, {Brinkerink},
  {Brissenden}, {Britzen}, {Broderick}, {Broguiere}, {Bronzwaer}, {Byun},
  {Carlstrom}, {Chael}, {Chan}, {Chatterjee}, {Chatterjee}, {Chen}, {Chen},
  {Chesler}, {Cho}, {Christian}, {Conway}, {Cordes}, {Crawford}, {Crew},
  {Cruz-Osorio}, {Davelaar}, {de Laurentis}, {Deane}, {Dempsey}, {Desvignes},
  {Dexter}, {Doeleman}, {Eatough}, {Falcke}, {Farah}, {Fish}, {Fomalont},
  {Ford}, {Fraga-Encinas}, {Friberg}, {Fromm}, {Fuentes}, {Galison}, {Gammie},
  {Garc{\'\i}a}, {Gentaz}, {Georgiev}, {Gold}, {G{\'o}mez}, {G{\'o}mez-Ruiz},
  {Gu}, {Gurwell}, {Hecht}, {Hesper}, {Ho}, {Ho}, {Honma}, {Huang}, {Huang},
  {Hughes}, {Ikeda}, {Inoue}, {Issaoun}, {James}, {Jannuzi}, {Janssen},
  {Jeter}, {Jiang}, {Jim{\'e}nez-Rosales}, {Johnson}, {Jung}, {Karami},
  {Karuppusamy}, {Kettenis}, {Kim}, {Kim}, {Kim}, {Koay}, {Kofuji}, {Koch},
  {Koyama}, {Kramer}, {Kramer}, {Kuo}, {Lauer}, {Levis}, {Li}, {Li},
  {Lindqvist}, {Lico}, {Lindahl}, {Liu}, {Liu}, {Liuzzo}, {Lo}, {Lobanov},
  {Loinard}, {Lonsdale}, {MacDonald}, {Mao}, {Marchili}, {Marrone}, {Marscher},
  {Mart{\'\i}-Vidal}, {Matsushita}, {Matthews}, {Medeiros}, {Menten}, {Mizuno},
  {Mizuno}, {Moran}, {Moriyama}, {Moscibrodzka}, {M{\"u}ller}, {Musoke},
  {Mej{\'\i}as}, {Nagai}, {Nagar}, {Nakamura}, {Narayan}, {Narayanan},
  {Natarajan}, {Nathanail}, {Neri}, {Ni}, {Noutsos}, {Okino}, {Olivares},
  {Ortiz-Le{\'o}n}, {Oyama}, {{\"O}zel}, {Palumbo}, {Patel}, {Pen}, {Pesce},
  {Pi{\'e}tu}, {Plambeck}, {Popstefanija}, {Porth}, {P{\"o}tzl}, {Prather},
  {Preciado-L{\'o}pez}, {Psaltis}, {Pu}, {Rao}, {Rawlings}, {Raymond},
  {Rezzolla}, {Ricarte}, {Ripperda}, {Roelofs}, {Rogers}, {Ros}, {Rose},
  {Roshanineshat}, {Rottmann}, {Roy}, {Ruszczyk}, {Rygl}, {S{\'a}nchez},
  {S{\'a}nchez-Arguelles}, {Savolainen}, {Schloerb}, {Schuster}, {Shao},
  {Shen}, {Small}, {Sohn}, {Soohoo}, {Sun}, {Tazaki}, {Tetarenko}, {Tiede},
  {Tilanus}, {Titus}, {Toma}, {Torne}, {Trent}, {Traianou}, {Trippe}, {van
  Bemmel}, {van Langevelde}, {van Rossum}, {Wagner}, {Ward-Thompson}, {Wardle},
  {Weintroub}, {Wex}, {Wharton}, {Wielgus}, {Wong}, {Wu}, {Yoon}, {Young},
  {Young}, {Younsi}, {Yuan}, {Yuan}, {Zensus}, {Zhao}, {Zhao}, {Fermi Large
  Area Telescope Collaboration}, {Principe}, {Giroletti}, {D'Ammando},
  {Orienti}, {H.~E.~S.~S. Collaboration}, {Abdalla}, {Adam}, {Aharonian},
  {Benkhali}, {Ang{\"u}ner}, {Arcaro}, {Armand}, {Armstrong}, {Ashkar},
  {Backes}, {Baghmanyan}, {Barbosa Martins}, {Barnacka}, {Barnard},
  {Becherini}, {Berge}, {Bernl{\"o}hr}, {Bi}, {B{\"o}ttcher}, {Boisson},
  {Bolmont}, {de Lavergne}, {Breuhaus}, {Brun}, {Brun}, {Bryan}, {B{\"u}chele},
  {Bulik}, {Bylund}, {Caroff}, {Carosi}, {Casanova}, {Chand}, {Chen}, {Cotter},
  {Cury{\l}o}, {Damascene Mbarubucyeye}, {Davids}, {Davies}, {Deil}, {Devin},
  {Dewilt}, {Dirson}, {Djannati-Ata{\"\i}}, {Dmytriiev}, {Donath},
  {Doroshenko}, {Duffy}, {Dyks}, {Egberts}, {Eichhorn}, {Einecke}, {Emery},
  {Ernenwein}, {Feijen}, {Fegan}, {Fiasson}, {de Clairfontaine}, {Fontaine},
  {Funk}, {F{\"u}{\ss}ling}, {Gabici}, {Gallant}, {Giavitto}, {Giunti},
  {Glawion}, {Glicenstein}, {Gottschall}, {Grondin}, {Hahn}, {Haupt},
  {Hermann}, {Hinton}, {Hofmann}, {Hoischen}, {Holch}, {Holler}, {H{\"o}rbe},
  {Horns}, {Huber}, {Jamrozy}, {Jankowsky}, {Jankowsky}, {Jardin-Blicq},
  {Joshi}, {Jung-Richardt}, {Kasai}, {Kastendieck}, {Katarzy{\'n}ski}, {Katz},
  {Khangulyan}, {Kh{\'e}lifi}, {Klepser}, {Klu{\'z}niak}, {Komin}, {Konno},
  {Kosack}, {Kostunin}, {Kreter}, {Lamanna}, {Lemi{\`e}re}, {Lemoine-Goumard},
  {Lenain}, {Levy}, {Lohse}, {Lypova}, {Mackey}, {Majumdar}, {Malyshev},
  {Malyshev}, {Marandon}, {Marchegiani}, {Marcowith}, {Mares},
  {Mart{\'\i}-Devesa}, {Marx}, {Maurin}, {Meintjes}, {Meyer}, {Moderski},
  {Mohamed}, {Mohrmann}, {Montanari}, {Moore}, {Morris}, {Moulin}, {Muller},
  {Murach}, {Nakashima}, {Nayerhoda}, {de Naurois}, {Ndiyavala},
  {Niederwanger}, {Niemiec}, {Oakes}, {O'Brien}, {Odaka}, {Ohm},
  {Olivera-Nieto}, {de Ona Wilhelmi}, {Ostrowski}, {Panter}, {Panny},
  {Parsons}, {Peron}, {Peyaud}, {Piel}, {Pita}, {Poireau}, {Noel}, {Prokhorov},
  {Prokoph}, {P{\"u}hlhofer}, {Punch}, {Quirrenbach}, {Rauth}, {Reichherzer},
  {Reimer}, {Reimer}, {Remy}, {Renaud}, {Rieger}, {Rinchiuso}, {Romoli},
  {Rowell}, {Rudak}, {Ruiz-Velasco}, {Sahakian}, {Sailer}, {Sanchez},
  {Santangelo}, {Sasaki}, {Scalici}, {Schutte}, {Schwanke}, {Schwemmer},
  {Seglar-Arroyo}, {Senniappan}, {Seyffert}, {Shafi}, {Shiningayamwe},
  {Simoni}, {Sinha}, {Sol}, {Specovius}, {Spencer}, {Spir-Jacob}, {Stawarz},
  {Sun}, {Steenkamp}, {Stegmann}, {Steinmassl}, {Steppa}, {Takahashi},
  {Tavernier}, {Taylor}, {Terrier}, {Tiziani}, {Tluczykont}, {Tomankova},
  {Trichard}, {Tsirou}, {Tuffs}, {Uchiyama}, {van der Walt}, {van Eldik}, {van
  Rensburg}, {van Soelen}, {Vasileiadis}, {Veh}, {Venter}, {Vincent}, {Vink},
  {V{\"o}lk}, {Vuillaume}, {Wadiasingh}, {Wagner}, {Watson}, {Werner}, {White},
  {Wierzcholska}, {Wong}, {Yusafzai}, {Zacharias}, {Zanin}, {Zargaryan},
  {Zdziarski}, {Zech}, {Zhu}, {Zorn}, {Zouari}, {{\.Z}ywucka}, {MAGIC
  Collaboration}, {Acciari}, {Ansoldi}, {Antonelli}, {Engels}, {Artero},
  {Asano}, {Baack}, {Babi{\'c}}, {Baquero}, {de Almeida}, {Barrio}, {Becerra
  Gonz{\'a}lez}, {Bednarek}, {Bellizzi}, {Bernardini}, {Bernardos}, {Berti},
  {Besenrieder}, {Bhattacharyya}, {Bigongiari}, {Biland}, {Blanch}, {Bonnoli},
  {Bo{\v{s}}njak}, {Busetto}, {Carosi}, {Ceribella}, {Cerruti}, {Chai},
  {Chilingarian}, {Cikota}, {Colak}, {Colombo}, {Contreras}, {Cortina},
  {Covino}, {D'Amico}, {D'Elia}, {da Vela}, {Dazzi}, {de Angelis}, {de Lotto},
  {Delfino}, {Delgado}, {Delgado Mendez}, {Depaoli}, {di Pierro}, {di Venere},
  {Do Souto Espi{\~n}eira}, {Dominis Prester}, {Donini}, {Dorner}, {Doro},
  {Elsaesser}, {Ramazani}, {Fattorini}, {Ferrara}, {Fonseca}, {Font}, {Fruck},
  {Fukami}, {Garc{\'\i}a L{\'o}pez}, {Garczarczyk}, {Gasparyan}, {Gaug},
  {Giglietto}, {Giordano}, {Gliwny}, {Godinovi{\'c}}, {Green}, {Green},
  {Hadasch}, {Hahn}, {Heckmann}, {Herrera}, {Hoang}, {Hrupec}, {H{\"u}tten},
  {Inada}, {Inoue}, {Ishio}, {Iwamura}, {Jim{\'e}nez}, {Jormanainen}, {Jouvin},
  {Kajiwara}, {Karjalainen}, {Kerszberg}, {Kobayashi}, {Kubo}, {Kushida},
  {Lamastra}, {Lelas}, {Leone}, {Lindfors}, {Lombardi}, {Longo},
  {L{\'o}pez-Coto}, {L{\'o}pez-Moya}, {L{\'o}pez-Oramas}, {Loporchio}, {Machado
  de Oliveira Fraga}, {Maggio}, {Majumdar}, {Makariev}, {Mallamaci}, {Maneva},
  {Manganaro}, {Mannheim}, {Maraschi}, {Mariotti}, {Mart{\'\i}nez}, {Mazin},
  {Menchiari}, {Mender}, {Mi{\'c}anovi{\'c}}, {Miceli}, {Miener}, {Minev},
  {Miranda}, {Mirzoyan}, {Molina}, {Moralejo}, {Morcuende}, {Moreno},
  {Moretti}, {Neustroev}, {Nigro}, {Nilsson}, {Nishijima}, {Noda}, {Nozaki},
  {Ohtani}, {Oka}, {Otero-Santos}, {Paiano}, {Palatiello}, {Paneque},
  {Paoletti}, {Paredes}, {Pavleti{\'c}}, {Pe{\~n}il}, {Perennes}, {Persic},
  {Moroni}, {Prandini}, {Priyadarshi}, {Puljak}, {Rhode}, {Rib{\'o}}, {Rico},
  {Righi}, {Rugliancich}, {Saha}, {Sahakyan}, {Saito}, {Sakurai}, {Satalecka},
  {Saturni}, {Schleicher}, {Schmidt}, {Schweizer}, {Sitarek},
  {{\v{S}}nidari{\'c}}, {Sobczynska}, {Spolon}, {Stamerra}, {Strom}, {Strzys},
  {Suda}, {Suri{\'c}}, {Takahashi}, {Tavecchio}, {Temnikov}, {Terzi{\'c}},
  {Teshima}, {Tosti}, {Truzzi}, {Tutone}, {Ubach}, {van Scherpenberg}, {Vanzo},
  {Vazquez Acosta}, {Ventura}, {Verguilov}, {Vigorito}, {Vitale}, {Vovk},
  {Will}, {Wunderlich}, {Zari{\'c}}, {VERITAS Collaboration}, {Adams},
  {Benbow}, {Brill}, {Capasso}, {Christiansen}, {Chromey}, {Daniel}, {Errando},
  {Farrell}, {Feng}, {Finley}, {Fortson}, {Furniss}, {Gent}, {Giuri}, {Hassan},
  {Hervet}, {Holder}, {Hughes}, {Humensky}, {Jin}, {Kaaret}, {Kertzman},
  {Kieda}, {Kumar}, {Lang}, {Lundy}, {Maier}, {Moriarty}, {Mukherjee}, {Nieto},
  {Nievas-Rosillo}, {O'Brien}, {Ong}, {Otte}, {Patel}, {Pfrang}, {Pohl},
  {Prado}, {Pueschel}, {Quinn}, {Ragan}, {Reynolds}, {Ribeiro}, {Richards},
  {Roache}, {Rulten}, {Ryan}, {Santander}, {Sembroski}, {Shang}, {Weinstein},
  {Williams}, {Williamson}, {Eavn Collaboration}, {Hirota}, {Cui}, {Niinuma},
  {Ro}, {Sakai}, {Sawada-Satoh}, {Wajima}, {Wang}, {Liu}, \&
  {Yonekura}}]{EHT_MWL_2021}
{EHT MWL Science Working Group}, {Algaba}, J.~C., {Anczarski}, J., {et~al.}
  2021, \apjl, 911, L11, \dodoi{10.3847/2041-8213/abef71}

\bibitem[{{Event Horizon Telescope
  Collaboration}(2019{\natexlab{a}})}]{EHTC2019_1}
{Event Horizon Telescope Collaboration}. 2019{\natexlab{a}}, \apjl, 875, L1,
  \dodoi{10.3847/2041-8213/ab0ec7}

\bibitem[{{Event Horizon Telescope
  Collaboration}(2019{\natexlab{b}})}]{EHTC2019_2}
---. 2019{\natexlab{b}}, \apjl, 875, L2, \dodoi{10.3847/2041-8213/ab0c96}

\bibitem[{{Event Horizon Telescope
  Collaboration}(2019{\natexlab{c}})}]{EHTC2019_3}
---. 2019{\natexlab{c}}, \apjl, 875, L3, \dodoi{10.3847/2041-8213/ab0c57}

\bibitem[{{Event Horizon Telescope
  Collaboration}(2019{\natexlab{d}})}]{EHTC2019_4}
---. 2019{\natexlab{d}}, \apjl, 875, L4, \dodoi{10.3847/2041-8213/ab0e85}

\bibitem[{{Event Horizon Telescope
  Collaboration}(2019{\natexlab{e}})}]{EHTC2019_5}
---. 2019{\natexlab{e}}, \apjl, 875, L5, \dodoi{10.3847/2041-8213/ab0f43}

\bibitem[{{Event Horizon Telescope
  Collaboration}(2019{\natexlab{f}})}]{EHTC2019_6}
---. 2019{\natexlab{f}}, \apjl, 875, L6, \dodoi{10.3847/2041-8213/ab1141}

\bibitem[{{Event Horizon Telescope Collaboration} {et~al.}(2021){Event Horizon
  Telescope Collaboration}, {Akiyama}, {Algaba}, {Alberdi}, {Alef}, {Anantua},
  {Asada}, {Azulay}, {Baczko}, {Ball}, {Balokovi{\'c}}, {Barrett}, {Benson},
  {Bintley}, {Blackburn}, {Blundell}, {Boland}, {Bouman}, {Bower}, {Boyce},
  {Bremer}, {Brinkerink}, {Brissenden}, {Britzen}, {Broderick}, {Broguiere},
  {Bronzwaer}, {Byun}, {Carlstrom}, {Chael}, {Chan}, {Chatterjee},
  {Chatterjee}, {Chen}, {Chen}, {Chesler}, {Cho}, {Christian}, {Conway},
  {Cordes}, {Crawford}, {Crew}, {Cruz-Osorio}, {Cui}, {Davelaar}, {De
  Laurentis}, {Deane}, {Dempsey}, {Desvignes}, {Dexter}, {Doeleman}, {Eatough},
  {Falcke}, {Farah}, {Fish}, {Fomalont}, {Ford}, {Fraga-Encinas}, {Friberg},
  {Fromm}, {Fuentes}, {Galison}, {Gammie}, {Garc{\'\i}a}, {Gelles}, {Gentaz},
  {Georgiev}, {Goddi}, {Gold}, {G{\'o}mez}, {G{\'o}mez-Ruiz}, {Gu}, {Gurwell},
  {Hada}, {Haggard}, {Hecht}, {Hesper}, {Himwich}, {Ho}, {Ho}, {Honma},
  {Huang}, {Huang}, {Hughes}, {Ikeda}, {Inoue}, {Issaoun}, {James}, {Jannuzi},
  {Janssen}, {Jeter}, {Jiang}, {Jimenez-Rosales}, {Johnson}, {Jorstad}, {Jung},
  {Karami}, {Karuppusamy}, {Kawashima}, {Keating}, {Kettenis}, {Kim}, {Kim},
  {Kim}, {Kim}, {Kino}, {Koay}, {Kofuji}, {Koch}, {Koyama}, {Kramer}, {Kramer},
  {Krichbaum}, {Kuo}, {Lauer}, {Lee}, {Levis}, {Li}, {Li}, {Lindqvist}, {Lico},
  {Lindahl}, {Liu}, {Liu}, {Liuzzo}, {Lo}, {Lobanov}, {Loinard}, {Lonsdale},
  {Lu}, {MacDonald}, {Mao}, {Marchili}, {Markoff}, {Marrone}, {Marscher},
  {Mart{\'\i}-Vidal}, {Matsushita}, {Matthews}, {Medeiros}, {Menten}, {Mizuno},
  {Mizuno}, {Moran}, {Moriyama}, {Moscibrodzka}, {M{\"u}ller}, {Musoke}, {Mus
  Mej{\'\i}as}, {Michalik}, {Nadolski}, {Nagai}, {Nagar}, {Nakamura},
  {Narayan}, {Narayanan}, {Natarajan}, {Nathanail}, {Neilsen}, {Neri}, {Ni},
  {Noutsos}, {Nowak}, {Okino}, {Olivares}, {Ortiz-Le{\'o}n}, {Oyama},
  {{\"O}zel}, {Palumbo}, {Park}, {Patel}, {Pen}, {Pesce}, {Pi{\'e}tu},
  {Plambeck}, {PopStefanija}, {Porth}, {P{\"o}tzl}, {Prather},
  {Preciado-L{\'o}pez}, {Psaltis}, {Pu}, {Ramakrishnan}, {Rao}, {Rawlings},
  {Raymond}, {Rezzolla}, {Ricarte}, {Ripperda}, {Roelofs}, {Rogers}, {Ros},
  {Rose}, {Roshanineshat}, {Rottmann}, {Roy}, {Ruszczyk}, {Rygl},
  {S{\'a}nchez}, {S{\'a}nchez-Arguelles}, {Sasada}, {Savolainen}, {Schloerb},
  {Schuster}, {Shao}, {Shen}, {Small}, {Sohn}, {SooHoo}, {Sun}, {Tazaki},
  {Tetarenko}, {Tiede}, {Tilanus}, {Titus}, {Toma}, {Torne}, {Trent},
  {Traianou}, {Trippe}, {van Bemmel}, {van Langevelde}, {van Rossum}, {Wagner},
  {Ward-Thompson}, {Wardle}, {Weintroub}, {Wex}, {Wharton}, {Wielgus}, {Wong},
  {Wu}, {Yoon}, {Young}, {Young}, {Younsi}, {Yuan}, {Yuan}, {Zensus}, {Zhao},
  \& {Zhao}}]{EHTC_2021b}
{Event Horizon Telescope Collaboration}, {Akiyama}, K., {Algaba}, J.~C.,
  {et~al.} 2021, \apjl, 910, L13, \dodoi{10.3847/2041-8213/abe4de}

\bibitem[{{Falcke} {et~al.}(2000){Falcke}, {Melia}, \&
  {Agol}}]{Falcke_etal_2000}
{Falcke}, H., {Melia}, F., \& {Agol}, E. 2000, \apjl, 528, L13,
  \dodoi{10.1086/312423}

\bibitem[{{Fishbone} \& {Moncrief}(1976)}]{Fishbone_Moncrief_1976}
{Fishbone}, L.~G., \& {Moncrief}, V. 1976, \apj, 207, 962,
  \dodoi{10.1086/154565}

\bibitem[{{Gold} {et~al.}(2017){Gold}, {McKinney}, {Johnson}, \&
  {Doeleman}}]{Gold_etal_2017}
{Gold}, R., {McKinney}, J.~C., {Johnson}, M.~D., \& {Doeleman}, S.~S. 2017,
  \apj, 837, 180, \dodoi{10.3847/1538-4357/aa6193}

\bibitem[{{Gold} {et~al.}(2020){Gold}, {Broderick}, {Younsi}, {Fromm},
  {Gammie}, {Mo{\'s}cibrodzka}, {Pu}, {Bronzwaer}, {Davelaar}, {Dexter},
  {Ball}, {Chan}, {Kawashima}, {Mizuno}, {Ripperda}, {Akiyama}, {Alberdi},
  {Alef}, {Asada}, {Azulay}, {Baczko}, {Balokovi{\'c}}, {Barrett}, {Bintley},
  {Blackburn}, {Boland}, {Bouman}, {Bower}, {Bremer}, {Brinkerink},
  {Brissenden}, {Britzen}, {Broguiere}, {Byun}, {Carlstrom}, {Chael},
  {Chatterjee}, {Chatterjee}, {Chen}, {Chen}, {Cho}, {Christian}, {Conway},
  {Cordes}, {Crew}, {Cui}, {De Laurentis}, {Deane}, {Dempsey}, {Desvignes},
  {Doeleman}, {Eatough}, {Falcke}, {Fish}, {Fomalont}, {Fraga-Encinas},
  {Freeman}, {Friberg}, {G{\'o}mez}, {Galison}, {Garc{\'\i}a}, {Gentaz},
  {Georgiev}, {Goddi}, {Gu}, {Gurwell}, {Hada}, {Hecht}, {Hesper}, {Ho}, {Ho},
  {Honma}, {Huang}, {Huang}, {Hughes}, {Inoue}, {Issaoun}, {James}, {Jannuzi},
  {Janssen}, {Jeter}, {Jiang}, {Jimenez-Rosales}, {Johnson}, {Jorstad}, {Jung},
  {Karami}, {Karuppusamy}, {Keating}, {Kettenis}, {Kim}, {Kim}, {Kim}, {Kino},
  {Koay}, {Koch}, {Koyama}, {Kramer}, {Kramer}, {Krichbaum}, {Kuo}, {Lauer},
  {Lee}, {Li}, {Li}, {Lico}, {Lindqvist}, {Liu}, {Liuzzo}, {Lo}, {Lobanov},
  {Loinard}, {Lonsdale}, {Lu}, {MacDonald}, {Markoff}, {Mao}, {Marrone},
  {Marscher}, {Mart{\'\i}-Vidal}, {Matsushita}, {Matthews}, {Medeiros},
  {Menten}, {Mizuno}, {Moran}, {Moriyama}, {M{\"u}ller}, {Nagai}, {Nakamura},
  {Nagar}, {Narayan}, {Narayanan}, {Natarajan}, {Neri}, {Ni}, {Noutsos},
  {Okino}, {Ortiz-Le{\'o}n}, {Oyama}, {{\"O}zel}, {Palumbo}, {Park}, {Patel},
  {Pen}, {Pesce}, {Plambeck}, {Pi{\'e}tu}, {PopStefanija}, {Porth},
  {Preciado-L{\'o}pez}, {Psaltis}, {Ramakrishnan}, {Rao}, {Rawlings},
  {Raymond}, {Rezzolla}, {Roelofs}, {Rogers}, {Ros}, {Rose}, {Roshanineshat},
  {Rottmann}, {Roy}, {Ruszczyk}, {Rygl}, {S{\'a}nchez},
  {S{\'a}nchez-Arguelles}, {Sasada}, {Savolainen}, {Schuster}, {Schloerb},
  {Shao}, {Shen}, {Small}, {Sohn}, {SooHoo}, {Tiede}, {Tazaki}, {Tilanus},
  {Titus}, {Toma}, {Torne}, {Trent}, {Traianou}, {Trippe}, {Tsuda}, {van
  Langevelde}, {van Bemmel}, {van Rossum}, {Wagner}, {Wardle}, {Wex},
  {Weintroub}, {Wharton}, {Wielgus}, {Wong}, {Wu}, {Yoon}, {Young}, {Young},
  {Yuan}, {Yuan}, {Zensus}, {Zhao}, {Zhao}, {Zhu}, \& {Event Horizon Telescope
  Collaboration}}]{GRRT_code_comparison_2020}
{Gold}, R., {Broderick}, A.~E., {Younsi}, Z., {et~al.} 2020, \apj, 897, 148,
  \dodoi{10.3847/1538-4357/ab96c6}

\bibitem[{{Gorecki} \& {Wilczewski}(1984)}]{Gorecki_Wilczewski_1984}
{Gorecki}, A., \& {Wilczewski}, W. 1984, \actaa, 34, 141

\bibitem[{{Johannsen}(2013)}]{Johannsen_2013}
{Johannsen}, T. 2013, \apj, 777, 170, \dodoi{10.1088/0004-637X/777/2/170}

\bibitem[{{Johnson} {et~al.}(2020){Johnson}, {Lupsasca}, {Strominger}, {Wong},
  {Hadar}, {Kapec}, {Narayan}, {Chael}, {Gammie}, {Galison}, {Palumbo},
  {Doeleman}, {Blackburn}, {Wielgus}, {Pesce}, {Farah}, \&
  {Moran}}]{Johnson_etal_2020_ScienceAdv}
{Johnson}, M.~D., {Lupsasca}, A., {Strominger}, A., {et~al.} 2020, Science
  Advances, 6, eaaz1310, \dodoi{10.1126/sciadv.aaz1310}

\bibitem[{{Kawanaka} {et~al.}(2008){Kawanaka}, {Kato}, \&
  {Mineshige}}]{Kawanaka_etal_2008}
{Kawanaka}, N., {Kato}, Y., \& {Mineshige}, S. 2008, \pasj, 60, 399,
  \dodoi{10.1093/pasj/60.2.399}

\bibitem[{{Kawanaka} \& {Mineshige}(2021)}]{Kawanaka_Mineshige_2021}
{Kawanaka}, N., \& {Mineshige}, S. 2021, \pasj, 73, 630,
  \dodoi{10.1093/pasj/psab023}

\bibitem[{{Kawashima} {et~al.}(2019){Kawashima}, {Kino}, \&
  {Akiyama}}]{Kawashima_etal_2019}
{Kawashima}, T., {Kino}, M., \& {Akiyama}, K. 2019, \apj, 878, 27,
  \dodoi{10.3847/1538-4357/ab19c0}

\bibitem[{{Kawashima} {et~al.}(2012){Kawashima}, {Ohsuga}, {Mineshige},
  {Yoshida}, {Heinzeller}, \& {Matsumoto}}]{Kawashima_etal_2012}
{Kawashima}, T., {Ohsuga}, K., {Mineshige}, S., {et~al.} 2012, \apj, 752, 18,
  \dodoi{10.1088/0004-637X/752/1/18}

\bibitem[{{Kawashima} {et~al.}(2021){Kawashima}, {Toma}, {Kino}, {Akiyama},
  {Nakamura}, \& {Moriyama}}]{Kawashima_etal_2021}
{Kawashima}, T., {Toma}, K., {Kino}, M., {et~al.} 2021, \apj, 909, 168,
  \dodoi{10.3847/1538-4357/abd5bb}

\bibitem[{{Kitaki} {et~al.}(2017){Kitaki}, {Mineshige}, {Ohsuga}, \&
  {Kawashima}}]{Kitaki_etal_2017}
{Kitaki}, T., {Mineshige}, S., {Ohsuga}, K., \& {Kawashima}, T. 2017, \pasj,
  69, 92, \dodoi{10.1093/pasj/psx101}

\bibitem[{{Kompaneets}(1957)}]{Kompaneets_1957}
{Kompaneets}, A.~S. 1957, Soviet Journal of Experimental and Theoretical
  Physics, 4, 730

\bibitem[{{Leung} {et~al.}(2011){Leung}, {Gammie}, \&
  {Noble}}]{Leung_etal_2011}
{Leung}, P.~K., {Gammie}, C.~F., \& {Noble}, S.~C. 2011, \apj, 737, 21,
  \dodoi{10.1088/0004-637X/737/1/21}

\bibitem[{{Luminet}(1979)}]{Luminet_1979}
{Luminet}, J.-P. 1979, \aap, 75, 228

\bibitem[{{Lynden-Bell}(1996)}]{Lynden-Bell_1996}
{Lynden-Bell}, D. 1996, \mnras, 279, 389, \dodoi{10.1093/mnras/279.2.389}

\bibitem[{{MAGIC Collaboration} {et~al.}(2008){MAGIC Collaboration}, {Albert},
  {Aliu}, {Anderhub}, {Antonelli}, {Antoranz}, {Backes}, {Baixeras}, {Barrio},
  {Bartko}, {Bastieri}, {Becker}, {Bednarek}, {Berger}, {Bernardini},
  {Bigongiari}, {Biland}, {Bock}, {Bonnoli}, {Bordas}, {Bosch-Ramon}, {Bretz},
  {Britvitch}, {Camara}, {Carmona}, {Chilingarian}, {Commichau}, {Contreras},
  {Cortina}, {Costado}, {Covino}, {Curtef}, {Dazzi}, {De Angelis}, {de Cea del
  Pozo}, {de los Reyes}, {De Lotto}, {De Maria}, {De Sabata}, {Delgado Mendez},
  {Dominguez}, {Dorner}, {Doro}, {Errando}, {Fagiolini}, {Ferenc},
  {Fern{\'a}ndez}, {Firpo}, {Fonseca}, {Font}, {Galante}, {Garc{\'\i}a
  L{\'o}pez}, {Garczarczyk}, {Gaug}, {Goebel}, {Hayashida}, {Herrero},
  {H{\"o}hne}, {Hose}, {Hsu}, {Huber}, {Jogler}, {Kneiske}, {Kranich}, {La
  Barbera}, {Laille}, {Leonardo}, {Lindfors}, {Lombardi}, {Longo}, {L{\'o}pez},
  {Lorenz}, {Majumdar}, {Maneva}, {Mankuzhiyil}, {Mannheim}, {Maraschi},
  {Mariotti}, {Mart{\'\i}nez}, {Mazin}, {Meucci}, {Meyer}, {Miranda},
  {Mirzoyan}, {Mizobuchi}, {Moles}, {Moralejo}, {Nieto}, {Nilsson}, {Ninkovic},
  {Otte}, {Oya}, {Panniello}, {Paoletti}, {Paredes}, {Pasanen}, {Pascoli},
  {Pauss}, {Pegna}, {Perez-Torres}, {Persic}, {Peruzzo}, {Piccioli}, {Prada},
  {Prandini}, {Puchades}, {Raymers}, {Rhode}, {Rib{\'o}}, {Rico}, {Rissi},
  {Robert}, {R{\"u}gamer}, {Saggion}, {Saito}, {Salvati}, {Sanchez-Conde},
  {Sartori}, {Satalecka}, {Scalzotto}, {Scapin}, {Schmitt}, {Schweizer},
  {Shayduk}, {Shinozaki}, {Shore}, {Sidro}, {Sierpowska-Bartosik},
  {Sillanp{\"a}{\"a}}, {Sobczynska}, {Spanier}, {Stamerra}, {Stark}, {Takalo},
  {Tavecchio}, {Temnikov}, {Tescaro}, {Teshima}, {Tluczykont}, {Torres},
  {Turini}, {Vankov}, {Venturini}, {Vitale}, {Wagner}, {Wittek}, {Zabalza},
  {Zandanel}, {Zanin}, \& {Zapatero}}]{MAGIC_EBL_2008_Science}
{MAGIC Collaboration}, {Albert}, J., {Aliu}, E., {et~al.} 2008, Science, 320,
  1752, \dodoi{10.1126/science.1157087}

\bibitem[{{MAGIC Collaboration} {et~al.}(2020){MAGIC Collaboration}, {Acciari},
  {Ansoldi}, {Antonelli}, {Arbet Engels}, {Arcaro}, {Baack}, {Babi{\'c}},
  {Banerjee}, {Bangale}, {Barres de Almeida}, {Barrio}, {Becerra Gonz{\'a}lez},
  {Bednarek}, {Bellizzi}, {Bernardini}, {Berti}, {Besenrieder},
  {Bhattacharyya}, {Bigongiari}, {Biland}, {Blanch}, {Bonnoli},
  {Bo{\v{s}}njak}, {Busetto}, {Carosi}, {Ceribella}, {Chai}, {Chilingaryan},
  {Cikota}, {Colak}, {Colin}, {Colombo}, {Contreras}, {Cortina}, {Covino},
  {D'Elia}, {da Vela}, {Dazzi}, {de Angelis}, {de Lotto}, {Delfino}, {Delgado},
  {Depaoli}, {di Pierro}, {di Venere}, {Do Souto Espi{\~n}eira}, {Dominis
  Prester}, {Donini}, {Dorner}, {Doro}, {Elsaesser}, {Fallah Ramazani},
  {Fattorini}, {Fern{\'a}ndez-Barral}, {Ferrara}, {Fidalgo}, {Foffano},
  {Fonseca}, {Font}, {Fruck}, {Fukami}, {Garc{\'\i}a L{\'o}pez}, {Garczarczyk},
  {Gasparyan}, {Gaug}, {Giglietto}, {Giordano}, {Godinovi{\'c}}, {Green},
  {Guberman}, {Hadasch}, {Hahn}, {Herrera}, {Hoang}, {Hrupec}, {H{\"u}tten},
  {Inada}, {Inoue}, {Ishio}, {Iwamura}, {Jouvin}, {Kerszberg}, {Kubo},
  {Kushida}, {Lamastra}, {Lelas}, {Leone}, {Lindfors}, {Lombardi}, {Longo},
  {L{\'o}pez}, {L{\'o}pez-Coto}, {L{\'o}pez-Oramas}, {Loporchio}, {Machado de
  Oliveira Fraga}, {Maggio}, {Majumdar}, {Makariev}, {Mallamaci}, {Maneva},
  {Manganaro}, {Mannheim}, {Maraschi}, {Mariotti}, {Mart{\'\i}nez}, {Masuda},
  {Mazin}, {Mi{\'c}anovi{\'c}}, {Miceli}, {Minev}, {Miranda}, {Mirzoyan},
  {Molina}, {Moralejo}, {Morcuende}, {Moreno}, {Moretti}, {Munar-Adrover},
  {Neustroev}, {Nigro}, {Nilsson}, {Ninci}, {Nishijima}, {Noda}, {Nogu{\'e}s},
  {N{\"o}the}, {Nozaki}, {Paiano}, {Palacio}, {Palatiello}, {Paneque},
  {Paoletti}, {Paredes}, {Pe{\~n}il}, {Peresano}, {Persic}, {Prada Moroni},
  {Prand ini}, {Puljak}, {Rhode}, {Rib{\'o}}, {Rico}, {Righi}, {Rugliancich},
  {Saha}, {Sahakyan}, {Saito}, {Sakurai}, {Satalecka}, {Schmidt}, {Schweizer},
  {Sitarek}, {{\v{S}}nidari{\'c}}, {Sobczynska}, {Somero}, {Stamerra}, {Strom},
  {Strzys}, {Suda}, {Suri{\'c}}, {Takahashi}, {Tavecchio}, {Temnikov},
  {Terzi{\'c}}, {Teshima}, {Torres-Alb{\`a}}, {Tosti}, {Tsujimoto}, {Vagelli},
  {van Scherpenberg}, {Vanzo}, {Acosta}, {Vigorito}, {Vitale}, {Vovk}, {Will},
  {Zari{\'c}}, {Asano}, {Hada}, {Harris}, {Giroletti}, {Jermak}, {Madrid},
  {Massaro}, {Richter}, {Spanier}, {Steele}, \& {Walker}}]{MAGIC_2020_M87}
{MAGIC Collaboration}, {Acciari}, V.~A., {Ansoldi}, S., {et~al.} 2020, \mnras,
  492, 5354, \dodoi{10.1093/mnras/staa014}

\bibitem[{{Mahadevan} {et~al.}(1996){Mahadevan}, {Narayan}, \&
  {Yi}}]{Mahadevan_etal_1996}
{Mahadevan}, R., {Narayan}, R., \& {Yi}, I. 1996, \apj, 465, 327,
  \dodoi{10.1086/177422}

\bibitem[{{Manmoto} {et~al.}(1997){Manmoto}, {Mineshige}, \&
  {Kusunose}}]{Manmoto_1997}
{Manmoto}, T., {Mineshige}, S., \& {Kusunose}, M. 1997, \apj, 489, 791,
  \dodoi{10.1086/304817}

\bibitem[{{Mao} {et~al.}(2017){Mao}, {Dexter}, \& {Quataert}}]{Mao_etal_2017}
{Mao}, S.~A., {Dexter}, J., \& {Quataert}, E. 2017, \mnras, 466, 4307,
  \dodoi{10.1093/mnras/stw3324}

\bibitem[{{Massaro} {et~al.}(2006){Massaro}, {Tramacere}, {Perri}, {Giommi}, \&
  {Tosti}}]{Massaro_etal_2006}
{Massaro}, E., {Tramacere}, A., {Perri}, M., {Giommi}, P., \& {Tosti}, G. 2006,
  \aap, 448, 861, \dodoi{10.1051/0004-6361:20053644}

\bibitem[{{Mo{\'s}cibrodzka} {et~al.}(2017){Mo{\'s}cibrodzka}, {Dexter},
  {Davelaar}, \& {Falcke}}]{Moscibrodzka_etal_2017}
{Mo{\'s}cibrodzka}, M., {Dexter}, J., {Davelaar}, J., \& {Falcke}, H. 2017,
  \mnras, 468, 2214, \dodoi{10.1093/mnras/stx587}

\bibitem[{{Mo{\'s}cibrodzka} {et~al.}(2016){Mo{\'s}cibrodzka}, {Falcke}, \&
  {Shiokawa}}]{Moscibrodzka_etal_2016}
{Mo{\'s}cibrodzka}, M., {Falcke}, H., \& {Shiokawa}, H. 2016, \aap, 586, A38,
  \dodoi{10.1051/0004-6361/201526630}

\bibitem[{{Mo{\'s}cibrodzka} {et~al.}(2009){Mo{\'s}cibrodzka}, {Gammie},
  {Dolence}, {Shiokawa}, \& {Leung}}]{Moscibrodzka_etal_2009}
{Mo{\'s}cibrodzka}, M., {Gammie}, C.~F., {Dolence}, J.~C., {Shiokawa}, H., \&
  {Leung}, P.~K. 2009, \apj, 706, 497, \dodoi{10.1088/0004-637X/706/1/497}

\bibitem[{{Narayan} {et~al.}(2012){Narayan}, {S{\"A} dowski}, {Penna}, \&
  {Kulkarni}}]{Narayan_etal_2012}
{Narayan}, R., {S{\"A} dowski}, A., {Penna}, R.~F., \& {Kulkarni}, A.~K. 2012,
  \mnras, 426, 3241, \dodoi{10.1111/j.1365-2966.2012.22002.x}

\bibitem[{{Narayan} {et~al.}(2017){Narayan}, {Sa{\`I}{\textsection}dowski}, \&
  {Soria}}]{Narayan_etal_2017}
{Narayan}, R., {Sa{\`I}{\textsection}dowski}, A., \& {Soria}, R. 2017, \mnras,
  469, 2997, \dodoi{10.1093/mnras/stx1027}

\bibitem[{{Narayan} \& {Yi}(1994)}]{Narayan_Yi_1994}
{Narayan}, R., \& {Yi}, I. 1994, \apjl, 428, L13, \dodoi{10.1086/187381}

\bibitem[{{Narayan} \& {Yi}(1995)}]{Narayan_Yi_1995}
---. 1995, \apj, 452, 710, \dodoi{10.1086/176343}

\bibitem[{{Noble} {et~al.}(2007){Noble}, {Leung}, {Gammie}, \&
  {Book}}]{Noble_etal_2007}
{Noble}, S.~C., {Leung}, P.~K., {Gammie}, C.~F., \& {Book}, L.~G. 2007,
  Classical and Quantum Gravity, 24, S259, \dodoi{10.1088/0264-9381/24/12/S17}

\bibitem[{{Ohsuga} {et~al.}(2005){Ohsuga}, {Kato}, \&
  {Mineshige}}]{Ohsuga_etal_2005_SED}
{Ohsuga}, K., {Kato}, Y., \& {Mineshige}, S. 2005, \apj, 627, 782,
  \dodoi{10.1086/430432}

\bibitem[{{{\"O}zel} {et~al.}(2000){{\"O}zel}, {Psaltis}, \&
  {Narayan}}]{Ozel_2000}
{{\"O}zel}, F., {Psaltis}, D., \& {Narayan}, R. 2000, \apj, 541, 234,
  \dodoi{10.1086/309396}

\bibitem[{{Pozdnyakov} {et~al.}(1977){Pozdnyakov}, {Sobol}, \&
  {Syunyaev}}]{Pozdnyakov_etal_1977}
{Pozdnyakov}, L.~A., {Sobol}, I.~M., \& {Syunyaev}, R.~A. 1977, \sovast, 21,
  708

\bibitem[{{Pozdnyakov} {et~al.}(1983){Pozdnyakov}, {Sobol}, \&
  {Syunyaev}}]{Pozdnyakov_etal_1983}
---. 1983, \apspr, 2, 189

\bibitem[{{Press} {et~al.}(2002){Press}, {Teukolsky}, {Vetterling}, \&
  {Flannery}}]{Numerical_Recipes}
{Press}, W.~H., {Teukolsky}, S.~A., {Vetterling}, W.~T., \& {Flannery}, B.~P.
  2002, {Numerical recipes in C++ : the art of scientific computing}

\bibitem[{{Psaltis} {et~al.}(2015){Psaltis}, {{\"O}zel}, {Chan}, \&
  {Marrone}}]{Psaltis_etal_2015}
{Psaltis}, D., {{\"O}zel}, F., {Chan}, C.-K., \& {Marrone}, D.~P. 2015, \apj,
  814, 115, \dodoi{10.1088/0004-637X/814/2/115}

\bibitem[{{Pu} {et~al.}(2016){Pu}, {Yun}, {Younsi}, \&
  {Yoon}}]{Pu_etal_2016_Odyssey}
{Pu}, H.-Y., {Yun}, K., {Younsi}, Z., \& {Yoon}, S.-J. 2016, \apj, 820, 105,
  \dodoi{10.3847/0004-637X/820/2/105}

\bibitem[{{Rees}(1984)}]{Rees_1984review}
{Rees}, M.~J. 1984, \araa, 22, 471, \dodoi{10.1146/annurev.aa.22.090184.002351}

\bibitem[{{Rybicki} \& {Lightman}(1979)}]{Rybicki_Lightman_1979}
{Rybicki}, G.~B., \& {Lightman}, A.~P. 1979, {Radiative processes in
  astrophysics}

\bibitem[{{Scepi} {et~al.}(2021){Scepi}, {Dexter}, \&
  {Begelman}}]{Scepi_etal_2021_arxiv}
{Scepi}, N., {Dexter}, J., \& {Begelman}, M.~C. 2021, arXiv e-prints,
  arXiv:2107.08056.
\newblock \doarXiv{2107.08056}

\bibitem[{{Schnittman} \& {Krolik}(2013)}]{Schnittman_Krolik_2013}
{Schnittman}, J.~D., \& {Krolik}, J.~H. 2013, \apj, 777, 11,
  \dodoi{10.1088/0004-637X/777/1/11}

\bibitem[{{Schnittman} {et~al.}(2013){Schnittman}, {Krolik}, \&
  {Noble}}]{Schnittman_etal_2013}
{Schnittman}, J.~D., {Krolik}, J.~H., \& {Noble}, S.~C. 2013, \apj, 769, 156,
  \dodoi{10.1088/0004-637X/769/2/156}

\bibitem[{{Shakura} \& {Sunyaev}(1973)}]{Shakura_Sunyaev_1973}
{Shakura}, N.~I., \& {Sunyaev}, R.~A. 1973, \aap, 500, 33

\bibitem[{{Shcherbakov} {et~al.}(2012){Shcherbakov}, {Penna}, \&
  {McKinney}}]{Shcherbakov_etal_2012}
{Shcherbakov}, R.~V., {Penna}, R.~F., \& {McKinney}, J.~C. 2012, \apj, 755,
  133, \dodoi{10.1088/0004-637X/755/2/133}

\bibitem[{{Sikora}(1981)}]{Sikora_1981}
{Sikora}, M. 1981, \mnras, 196, 257, \dodoi{10.1093/mnras/196.2.257}

\bibitem[{{S{\"A}dowski} {et~al.}(2013){S{\"A}dowski}, {Narayan}, {Penna},
  \& {Zhu}}]{Sadowski_etal_2013}
{S{\"A}dowski}, A., {Narayan}, R., {Penna}, R., \& {Zhu}, Y. 2013, \mnras,
  436, 3856, \dodoi{10.1093/mnras/stt1881}

\bibitem[{{Stepney} \& {Guilbert}(1983)}]{Stepney_Guilbert_1983}
{Stepney}, S., \& {Guilbert}, P.~W. 1983, \mnras, 204, 1269,
  \dodoi{10.1093/mnras/204.4.1269}

\bibitem[{{Svensson}(1982)}]{Svensson_1982}
{Svensson}, R. 1982, \apj, 258, 335, \dodoi{10.1086/160082}

\bibitem[{{Takahashi} {et~al.}(2016){Takahashi}, {Ohsuga}, {Kawashima}, \&
  {Sekiguchi}}]{Takahashi_etal_2016}
{Takahashi}, H.~R., {Ohsuga}, K., {Kawashima}, T., \& {Sekiguchi}, Y. 2016,
  \apj, 826, 23, \dodoi{10.3847/0004-637X/826/1/23}

\bibitem[{{Takahashi}(2004)}]{Takahashi_2004}
{Takahashi}, R. 2004, \apj, 611, 996, \dodoi{10.1086/422403}

\bibitem[{{Tchekhovskoy}(2015)}]{Tchekhovskoy_2015}
{Tchekhovskoy}, A. 2015, {Launching of Active Galactic Nuclei Jets}, ed.
  I.~{Contopoulos}, D.~{Gabuzda}, \& N.~{Kylafis}, Vol. 414, 45,
  \dodoi{10.1007/978-3-319-10356-3_3}

\bibitem[{{Tchekhovskoy} {et~al.}(2011){Tchekhovskoy}, {Narayan}, \&
  {McKinney}}]{Tchekhovskoy_etal_2011}
{Tchekhovskoy}, A., {Narayan}, R., \& {McKinney}, J.~C. 2011, \mnras, 418, L79,
  \dodoi{10.1111/j.1745-3933.2011.01147.x}

\bibitem[{{Tramacere} {et~al.}(2009){Tramacere}, {Giommi}, {Perri},
  {Verrecchia}, \& {Tosti}}]{Tramacere_etal_2009}
{Tramacere}, A., {Giommi}, P., {Perri}, M., {Verrecchia}, F., \& {Tosti}, G.
  2009, \aap, 501, 879, \dodoi{10.1051/0004-6361/200810865}

\bibitem[{{Tramacere} {et~al.}(2011){Tramacere}, {Massaro}, \&
  {Taylor}}]{Tramacere_etal_2011}
{Tramacere}, A., {Massaro}, E., \& {Taylor}, A.~M. 2011, \apj, 739, 66,
  \dodoi{10.1088/0004-637X/739/2/66}

\bibitem[{{Tsunetoe} {et~al.}(2020){Tsunetoe}, {Mineshige}, {Ohsuga},
  {Kawashima}, \& {Akiyama}}]{Tsunetoe_etal_2020}
{Tsunetoe}, Y., {Mineshige}, S., {Ohsuga}, K., {Kawashima}, T., \& {Akiyama},
  K. 2020, \pasj, 72, 32, \dodoi{10.1093/pasj/psaa008}

\bibitem[{{Tsunetoe} {et~al.}(2021){Tsunetoe}, {Mineshige}, {Ohsuga},
  {Kawashima}, \& {Akiyama}}]{Tsunetoe_etal_2021}
---. 2021, \pasj, \dodoi{10.1093/pasj/psab054}

\bibitem[{{Volonteri}(2012)}]{Volonteri_2012Science}
{Volonteri}, M. 2012, Science, 337, 544, \dodoi{10.1126/science.1220843}

\bibitem[{{Westfold}(1959)}]{Westfold_1959}
{Westfold}, K.~C. 1959, \apj, 130, 241, \dodoi{10.1086/146713}

\bibitem[{{Wong}(2021)}]{Wong_2021}
{Wong}, G.~N. 2021, \apj, 909, 217, \dodoi{10.3847/1538-4357/abdd2d}

\bibitem[{{Wong} {et~al.}(2021){Wong}, {Ryan}, \& {Gammie}}]{Wong_etal_2021}
{Wong}, G.~N., {Ryan}, B.~R., \& {Gammie}, C.~F. 2021, \apj, 907, 73,
  \dodoi{10.3847/1538-4357/abd0f9}

\bibitem[{{Younsi} {et~al.}(2012){Younsi}, {Wu}, \&
  {Fuerst}}]{Younsi_etal_2012}
{Younsi}, Z., {Wu}, K., \& {Fuerst}, S.~V. 2012, \aap, 545, A13,
  \dodoi{10.1051/0004-6361/201219599}

\bibitem[{{Yuan} {et~al.}(2003){Yuan}, {Quataert}, \&
  {Narayan}}]{Yuan_etal_2003}
{Yuan}, F., {Quataert}, E., \& {Narayan}, R. 2003, \apj, 598, 301,
  \dodoi{10.1086/378716}

\end{thebibliography}
\bibliographystyle{aasjournal}




\end{CJK*}
\end{document}